\documentclass[preprint,amsmath,amssymb,aps,a4]{revtex4-2}
\usepackage{graphicx}
\usepackage{bm}
\usepackage{txfonts}
\usepackage{hyperref}
\usepackage{caption}
\usepackage{amsmath}
\usepackage{amssymb}
\usepackage{braket}
\usepackage{subfig}
\usepackage{rotating}
\date{\today}
\newcommand{\be}{\begin{eqnarray}}
\newcommand{\ee}{\end{eqnarray}}

\newcommand{\bfk}{{\bf k}_{\perp}}

\usepackage{xcolor}
\begin{document}
\title{T-even TMDs for the spin-0 pseudo-scalar mesons upto twist-4 using light-front formalism}
\author{Satyajit Puhan$^{1}$}
\email{puhansatyajit@gmail.com}
 \author{Shubham Sharma$^{1}$}
\email{s.sharma.hep@gmail.com}
  \author{Navpreet Kaur$^{1}$}
\email{knavpreet.hep@gmail.com}
   \author{Narinder Kumar$^{1,2}$}
\email{narinderhep@gmail.com} 
  \author{Harleen Dahiya$^{1}$}
\email{dahiyah@nitj.ac.in}
\affiliation{$^1$ Department of Physics, Dr. B.R. Ambedkar National
	Institute of Technology, Jalandhar, 144008, India}
 \affiliation{$^2$ Computational Theoretical High Energy Physics Lab,
Department of Physics, Doaba College, Jalandhar 144004, India}

\date{\today}%
\begin{abstract}
We have investigated the pseudo-scalar meson structure in the form of transverse momentum-dependent parton distribution functions (TMDs) in the light-front based holographic model and quark model. Starting from leading order, we have calculated all the time-reversal even TMDs for pion and kaon up to twist-$4$ in these models. We have shown the 3-dimensional structure as well as the 2-dimensional structure of these particles along with their average quark transverse momenta. The parton distribution functions (PDFs) of pseudo-scalar mesons have been compared with the results of other models. The sum rules, TMD transverse dependence, inverse moments and Gaussian transverse dependence ratio in these models have also been studied. 
Further, the transverse quark densities have also been analyzed in the momentum space plane for these particles. The higher twist kaon properties in light-front dynamics have been predicted for the first time in this work.

 \vspace{0.1cm}
\end{abstract}
%
\maketitle
%
%
\section{Introduction\label{secintro}}
The ultimate goal of theoretical and experimental physicists is to revealed, how the internal structure of hadrons is composed of constituent quarks, gluons and sea quarks.
Understanding the multidimensional structure of the hadrons using the distribution functions has been an interesting topic of study. The study of $3$-dimensional ($3$D) transverse momentum-dependent parton distribution functions (TMDs) \cite{Diehl:2015uka,Angeles-Martinez:2015sea,Pasquini:2008ax} along with $1$-dimensional ($1$D) parton distribution functions (PDFs) \cite{Martin:1998sq} describe the properties of the transverse and longitudinal section of the quarks inside the hadron. The PDFs carry the information only about the longitudinal momentum fraction ($x$) carried by the quark from the hadron and the TMDs have information about the transverse momenta ($\bfk$) along with longitudinal momentum fraction. TMDs are the extended form of collinear PDFs. One can get the information about the spatial distributions of the quarks by studying the generalized parton distributions (GPDs) \cite{Diehl:2003ny}. Experimentally, PDFs can be extracted from the deep inelastic scattering (DIS) processes \cite{Polchinski:2002jw}, whereas TMDs from the Drell-Yan (DY) processes \cite{Zhou:2009jm}, semi-inclusive DIS scattering (SIDIS) processes \cite{Bacchetta:2017gcc} and $Z^0/ W^\pm $ production \cite{Catani:2015vma}. 
\par The pion beams impinging on nuclear targets in DY process provide data about the leading twist PDF $f_1^q (x)$ \cite{Aicher:2010cb,Gluck:1999xe,Hecht:2000xa}. The unpolarized DY differential cross section, given in Collins-Soper frame \cite{Lorce:2016ugb,Peng:2008sp}, provides the information about TMDs and is expressed as 
\begin{eqnarray*}
     \frac{d\sigma}{d \Omega} \propto 
 \bigg( 1 + \lambda \cos^2 \theta
      + \mu \sin 2\theta \cos \phi
      + \frac{\nu}{2} \sin^2 \theta \cos 2\phi \bigg) \; .
\end{eqnarray*}
The coefficient $\lambda$ in the above equation corresponds to the twist-$2$ unpolarized $f_1^q (x,\bfk^2)$ TMD. Here, the coefficient $\mu$ arises from the contribution of twist-$3$ TMDs \cite{Arnold:2008kf} and time-reversal odd (T-odd) Boer-Mulders function is linked to $\nu$ coefficient \cite{Boer:1999mm}. Therefore, for the sake of completeness, one must study the leading twist as well as the sub-leading twist distributions of the quark inside the hadron. Considering its importance, there have been some recent developments in the model computation of higher twist nucleon TMDs \cite{Sharma:2023wha,Sharma:2023azu}, which is observed to be in sync with CLAS experiment data for $e^q(x)$ PDF \cite{Courtoy:2014ixa}. In the present work, we mainly target on the pseudo-scalar mesons like pions ($\pi$) and kaons ($K$). The leading twist TMDs for $\pi$ have been studied in various theoretical models like Nambu-Jona-Lasinio (NJL) model \cite{Noguera:2015iia}, MIT Bag model \cite{Lu:2012hh} and in light-front (LF) dynamics \cite{Ahmady:2019yvo,Kaur:2020vkq}. 
Few higher order $\pi$ TMDs has been calculated \cite{Lorce:2016ugb,Zhu:2023lst}, however, no work has been reported for higher order $K$ TMDs. For the case of spin-$0$ pseudo-scalar mesons, there are total $8$ TMDs up to twist-$4$ \cite{Meissner:2008ay}. Out of them,  $4$ are T-even and remaining are of T-odd nature. The DY experiments at J-PARC and Fermilab \cite{Peng:2008sp}, the pion-induced DY process and $J/\psi$ production \cite{Bourrely:2023yzi} have provided the twist-2 $f_1^q (x)$ PDF information for $\pi$ and $K$. The COMPASS (CERN) aims to provide the leading and sub-leading twist in the $\pi$ impinging on polarized proton targets experiments. Higher twist PDFs and TMDs, providing a doorway to quark-gluon dynamics, can be decomposed by an equation of motions (EOM) of quantum chromodynamics (QCD) \cite{Lorce:2016ugb}.
\par In this work, we have considered two LF dynamics models, the light-front holographic model (LFHM) \cite{Brodsky:2011xx,Liu:2015jna}and the light-cone quark model (LCQM) \cite{Belyaev:1997iu,Tawfiq:1998nk,Choi:2017uos,Geng:2016pyr} to provide a wider perspective on desired distributions. We have taken the lower Fock-state $|M \rangle = \sum | q \bar{q} \rangle \psi_{q \bar{q}}$ for this work. The exchange of gluons within the quarks plays an important role in the higher twist distribution function calculations. Gluon contributions have not been included in this work. The LFHM connects the $5$-dimensional Anti-de Sitter (AdS) space-time of gravitational theory to the quantum field theory in LF formalism for strongly interacting particles. Hadrons are described by string-like objects in the higher dimensional AdS space and the behavior of these strings corresponds to the confinement of quarks and gluons within the hadrons \cite{Brodsky:2005en}. LFHM has been widely used to study the  $B\to \rho \gamma$ \cite{Ahmady:2012dy}, $B\to K^*$ \cite{Ahmady:2013cva}, $B\to \rho$ \cite{Ahmady:2013cga}, $B\to K^*\mu^-\mu^+$ \cite{Ahmady:2014cpa}, $\rho$ meson TMDs \cite{Kaur:2020emh}, $\rho$ meson electroproduction at HERA \cite{Forshaw:2012im} etc. LCQM is a non-perturbative approach framework describing the structure and properties of hadrons as well as the internal structure of
quarks within them \cite{Belyaev:1997iu}. It is gauge invariant and relativistic by nature. LCQM focuses on the valence quarks of the hadrons as they are the primary constituents responsible for the overall structure and properties of hadrons. LCQM cannot describe higher Fock-state contributions and the effect of confinement which play an important role in describing the complete dynamics of hadrons. LCQM has been widely used for the last two decades and provides better results in a non-perturbative regime. It has been used to study of $B_{\textit{i}{(\frac{1}{2}^+)}}\to B_{\textit{f}{(\frac{3}{2}^+)}}$ transitions \cite{Lu:2023rmq}, $ \Xi_{cc} \to \Xi_c$ weak decay \cite{Ke:2019lcf}, $V' \to V''$ transitions \cite{Chang:2019obq}, $(\pi^0, \eta, \eta') \to \gamma^* \gamma^*$ transtions \cite{Choi:2019wqx} etc. Both models have different spin structures and provide good results in understanding the valence quark distributions of hadrons.
\par In this work, we have calculated and studied the valence quark longitudinal and transverse distribution properties of $\pi$ and $K$ in both the models. As a first step, we have solved the quark-quark correlation function for spin-$0$ mesons to compute all the  T-even TMDs. Following which, the explicit forms of $f_1^q(x,\bfk^2)$, $e^q(x,\bfk^2)$, $f^{\perp q}(x,\bfk^2)$ and $f_3^q(x,\bfk^2)$ T-even TMDs have been written for both the models in terms of the LF wave functions. We have discussed the relations between these TMDs and terms of the positivity inequalities between them. We have also calculated the PDFs corresponding to these TMDs and discussed the sum rules of the PDFs. We have compared the behavior of TMDs and PDFs in both the models through $3$D and $2$D plots with respect to longitudinal momentum fraction $x$ and transverse momenta of quark $\bfk$ for \textit{u}-quark and $\bar s$-quark for both the particles. We have also compared the behavior of $2$D  PDF plots with experimental data as well as with other model's predictions. The inverse moments and transverse dependence of TMDs have been calculated in both the models. The transverse momenta values of our models have been compared with other models. At last, we have discussed the unpolarized quark densities of T-even TMDs in both models for both the particles.
\par This paper is arranged as follows. In Section \ref{kanha}, we have discussed the spin-$0$ T-even TMDs coming from the quark-quark correlation function at lower Fock-state. The PDFs related to each T-even TMD with the corresponding sum rules given in Section \ref{krishna}. The overlap form of T-even TMDs in the LF wave function form has been given for the LCQM and LFHM in Section \ref{satya} and \ref{kanhu} respectively. In these sections, we have discussed the TMDs and PDFs behavior through plots concerning different variables. The numerical results of the work have been discussed in Section \ref{puhan}. A detailed discussion on inverse moments, transverse dependence and Gaussian transverse dependence ratio of TMDs has also been presented in this section. Finally, we conclude in Section \ref{conli}.


 \section{Methodology}
\subsection{Transverse momentum-dependent parton distributions}\label{kanha}
For spin-$0$ pseudo-scalar mesons, the valence quark TMDs can be expressed through the quark-quark correlation function, which is defined as \cite{Meissner:2008ay,Kaur:2020vkq}
\begin{align}\label{coor}
      \Phi_q^{[\Gamma]}(x,\textbf{k}_\perp)=\frac{1}{2}\int\frac{\mathrm{d}z^-\mathrm{d}^2\vec{z}_\perp}{2(2\pi)^3}e^{i \textbf{k}\cdot z} \langle \pi(K)|\bar\psi(0)\Gamma\mathcal{W}(0,z)\psi(z)|\pi(K)\rangle|_{z^+=0} ,
\end{align}
where $z=(z^+,z^-,z^\perp)$ is the position four-vector. Momentum four-vector of the quark is denoted by $\textbf{k}$ which is given by $\textbf{k} \equiv (\textbf{k}^+,\textbf{k}^-,\textbf{k}^\perp)$, where $\textbf{k}^\perp$ and $\textbf{k}^+$ are the transverse and longitudinal momentum of the quark respectively. $|\pi/K\rangle$ is the LF bound state of pion/kaon with masses and momentum $M_{\pi(K)}$ and $(P^+, P_{\perp})$ respectively. $x= \frac{\textbf{k}^+}{P^+}$ is the longitudinal momentum fraction carried by the quarks from the target meson. $\mathcal{W}(0,z)$ is the Wilson line which preserves the gauge invariance of the bilocal quark field operators in the correlation functions \cite{Bacchetta:2020vty} which has been taken as $1$ here. $\psi(z)$ is the quark field operator, $\Gamma$ is the Dirac matrix which determines the Lorentz structure of the correlator $\Phi_q^{[\Gamma]}$. There are a total of $4$ T-even TMDs at twist-$2,3$ and $4$ case. These TMDs can be expressed as a projection of correlator in Eq. (\ref{coor}) under different cases of Dirac matrix structure $(\Gamma)$. In case of the leading twist (twist-$2$), the T-even TMD $f_1^q(x,\bfk^2)$ can be expressed as 
\begin{align}
  \Phi_q^{[\gamma^{+}]}(x,\textbf{k}_\perp) &=f^q_{1}(x,\textbf{k}^2_\perp). \label{tw-2TMDs1}
\end{align}
For subleading twist (twist-$3$) case, there are $2$ T-even TMDs, which can be expressed as
\begin{align}
    \Phi^{[1]}_q (x,\textbf{k}_\perp)&=\frac{M_{\pi(K)}}{P^{+}}e^q(x,\textbf{k}^2_\perp),     \label{tw-3TMDs-e}\\
    \Phi_q^{[\gamma^{j}]}(x,\textbf{k}_\perp) &=\frac{k_{\perp}^{j}}{P^{+}}f^{\perp q}(x,\textbf{k}^2_\perp), \label{tw-3TMDs-fperp}
\end{align}
with $j=1$, $2$.
The twist-$4$ T-even TMD $f_3^q(x,\bfk^2)$ can be expressed as
\begin{align}
  \Phi_q^{[\gamma^{-}]}(x,\textbf{k}_\perp) = \frac{M^2_{\pi(K)}}{(P^+)^2}f^q_{3}(x,\textbf{k}^2_\perp). \label{tw-4TMDs1}
\end{align}
Twist-$2$ $(f^q_{1}(x,\bfk^2))$ and
twist-$3$ ($e^q(x,\textbf{k}^2_\perp)$ and $f^{\perp q}(x,\textbf{k}^2_\perp)$) TMDs can be accessed through DY \cite{Arnold:2008kf}, but twist-$4$ ($f^q_{3}(x,\textbf{k}^2_\perp)$) cannot be accessed through current known experiments, therefore it has to be considered as an
academic object. Nevertheless, $f^q_{3}(x,\textbf{k}^2_\perp)$ completes the description of the quark TMD correlator through twist-4 \cite{Goeke:2005hb}. All these T-even TMDs in Eqs. (\ref{tw-2TMDs1}) to (\ref{tw-4TMDs1}) are related to each other through the following relations \cite{Lorce:2016ugb,Zhu:2023lst}
\begin{align}
	x\,e^q(x,\textbf{k}^2_\perp) & = 
	x\,\tilde{e}^q(x,\textbf{k}^2_\perp) + \frac{m_{q(\bar q)}}{M_{\pi(K)}}\,f_1^q(x,\textbf{k}^2_\perp),\label{Eq:eom-e}\\
 	x\,f^{\perp q}(x,\textbf{k}^2_\perp) & = 
	x\,\tilde{f}^{\perp q}(x,\textbf{k}^2_\perp) + f_1^q(x,\textbf{k}^2_\perp),\label{Eq:eom-fperp}\\
	x^2f^q_3(x,\textbf{k}^2_\perp) & = 
	x^2\tilde{f}_3^q(x,\textbf{k}^2_\perp) + \frac{\textbf{k}^2_\perp+m^2_{q(\bar q)}}{M^{2}_{\pi(K)}}\;f_1^q(x,\textbf{k}^2_\perp).\label{Eq:eom-f4}
\end{align}
Here, $m_{q(\bar q)}$ is the mass of the constituent quark (anti-quark) in meson. In QCD, the tilde-terms are expressed in terms of the projection of the quark-gluon-quark correlator \cite{Accardi:2009au}. Since, we have not included the gluonic Fock-state of $\pi(K)$ in the present work, the tilde terms vanish. The results presented in this work correspond to Wandzura-Wilczek \cite{Accardi:2009au} where the above relations can be rewritten as
\begin{align}
	x\,e^q(x,\textbf{k}^2_\perp) & = 
  \frac{m_{q(\bar q)}}{M_{\pi(K)}}\,f_1^q(x,\textbf{k}^2_\perp),\label{Eq:eom-e1}\\
 	x\,f^{\perp q}(x,\textbf{k}^2_\perp) & = 
 f_1^q(x,\textbf{k}^2_\perp),\label{Eq:eom-fperp1}\\
	x^2f^q_3(x,\textbf{k}^2_\perp) & = 
	 \frac{\textbf{k}^2_\perp+m^2_{q(\bar q)}}{M^{2}_{\pi(K)}}\;f_1^q(x,\textbf{k}^2_\perp).\label{Eq:eom-f41}
\end{align}
Positivity inequalities provide an important test for the validation of the TMDs \cite{Lorce:2014hxa,Lorce:2016ugb} which are given as 
\begin{align}
	f_1^q(x,\textbf{k}^2_\perp) \ge 0, \label{Eq:f1-inequality} &  \\
    	f_3^q(x,\textbf{k}^2_\perp) \ge 0. \label{Eq:f4-inequality} &
\end{align}
\subsection{Parton distribution functions}\label{krishna}
By integrating the T-even TMDs over parton's transverse momentum $\textbf{k}_\perp$, we can find their respective PDFs. Since there is no PDF for the twist-$3$ $f^{\perp q}(x,\textbf{k}^2_\perp)$ TMD due to the presence of explicit $\textbf{k}_{\perp}^{j}$ factor in it, one can formally define $f^{\perp q}(x)$ and other PDFs as 
\begin{align}
	f^{\perp q}(x) & =\int d^2 \textbf{k}_\perp f^{\perp q}(x,\textbf{k}^2_\perp) \label{fper} \\
   f_1^q(x) & =\int d^2 \bfk f_1^q(x,\textbf{k}^2_\perp),\label{fpdf}
   \\
   e^q(x) & =\int d^2 \bfk e^q(x,\textbf{k}^2_\perp),\label{epdf}
   \\
   f^q_3(x) & =\int d^2 \bfk f^q_3(x,\textbf{k}^2_\perp).
   \label{f3pdf}
 \end{align} 
Sum rules of PDF play an important role in  testing the consistency of any model. If the valence number of quark with flavor $q$ is $N_q$ (which is for instance
$N_u=N_{\bar d}=N_{\bar s}=1$ in $\pi^+$ and $K^+$), the sum rules are then expressed as \cite{Lorce:2016ugb,Zhu:2023lst}
\begin{align}
  \int d x f_1^q(x) & = N_q\,, 	\label{Eq:f1-sum-rule} \\
  \sum_q\int d x x f_1^q(x)& = 1  \,, \label{Eq:mom-sum-rule} \\
  \sum_q\int d x e^q(x)& = \frac{\sigma_{\pi(K)}}{m_{q(\bar q)}}\,,
\label{Eq:sigma-sum-rule} \\
  \int d x x e^q(x) 	& = \frac{m_{q(\bar q)}}{M_{\pi(K)}}\;N_q\,,
\label{Eq:Jaffe-Ji-sum-rule} \\
  2\int d x\,f_3^q(x) 	& = N_q. 	\label{Eq:f4-sum-rule}
\end{align}
The $\sigma_{\pi(K)}$ term in the right-hand side of Eq. (\ref{Eq:sigma-sum-rule}) refers to the scalar form factor at zero-momentum transfer \cite{Efremov:2002qh}. In case of $\pi$, the term $\sigma_{\pi}$ is equal to $\frac{1}{2 M_{\pi}}$. According to Refs. \cite{Tangerman:1994bb,Lorce:2016ugb,Lorce:2014hxa}, the $f^q_3(x)$ PDF can be written in the form of $f_1^q(x)$ and $f^{\perp q}(x)$ as 
\begin{align}
    f^q_3(x) =\frac{f_1^q(x)}{2} + \frac{d}{dx} \int {d^2 \textbf{k}_{\perp} \frac{\textbf{k}^2_{\perp}}{2 M^2_{\pi(K)}} f^{\perp q}(x, \bfk^2)}.
    \label{qLIR}
\end{align}
The above equation is referred to as “quark-model Lorentz-invariance relations (qLIRs)" for spin-$0$ PDFs.

\subsection{Light-cone quark model}\label{satya}
The hadron wave function based on the  light-cone quantization of QCD  using multi-particle Fock-state expansion is expressed as \cite{lcqm,lcqm1.1,lcqm1.2,lcqm2,Kaur:2020vkq,Qian:2008px}
\begin{eqnarray}
|M (P^+, \mathbf{P}_\perp, S_z) \rangle
   &=&\sum_{n,\lambda_i}\int\prod_{i=1}^n \frac{\mathrm{d} x_i \mathrm{d}^2
        \mathbf{k}_{\perp i}}{\sqrt{x_i}~16\pi^3}
 16\pi^3  \delta\Big(1-\sum_{i=1}^n x_i\Big)\nonumber\\
 &&\delta^{(2)}\Big(\sum_{i=1}^n \mathbf{k}_{\perp i}\Big) \psi_{n/M}(x_i,\mathbf{k}_{\perp i},\lambda_i)   | n ; x_i P^+, x_i \mathbf{P}_\perp+\mathbf{k}_{\perp i},
        \lambda_i \rangle
        .\nonumber\\
\end{eqnarray}
Here $|M (P^+, \mathbf{P}_\perp, S_z) \rangle$ is the hadron eigenstate, $P=(P^+,P^-,P_{\perp})$ is the meson's total momentum, $ \lambda_i$ is the helicity of the $i$-th constituent and $S_z$ is the longitudinal spin projection of the target. To simplify the calculations, we have taken the minimal Fock-state description of mesons in the form of quark-antiquark, which is given by
\begin{eqnarray}
|M(P, S_Z)\rangle &=& \sum_{\lambda_1,\lambda_2}\int
\frac{\mathrm{d} x \mathrm{d}^2
        \mathbf{k}_{\perp}}{\sqrt{x(1-x)}16\pi^3}
           \psi^{(1)M}_{S_Z}(x,\mathbf{k}_{\perp},\lambda_1,\lambda_2)|x,\mathbf{k}_{\perp},
        \lambda_1,\lambda_2 \rangle
        .
        \label{meson}
\end{eqnarray}
The momenta of the meson ($P$), constituent quark ($k_1$) and anti-quark ($k_2$) in the LF frame are respectively defined as 
\begin{eqnarray}
P&\equiv&\bigg(P^+,\frac{M_{\pi(K)}^2}{P^+},\textbf{0}_\perp \bigg),\\
k_1&\equiv&\bigg(x P^+, \frac{\textbf{k}_\perp^2+m_q^2}{x P^+},\textbf{k}_\perp \bigg),\\
k_2&\equiv&\bigg((1-x) P^+, \frac{\textbf{k}_\perp^2+m_{\bar q}^2}{(1-x) P^+},-\textbf{k}_\perp \bigg).
\end{eqnarray} 
The LF meson wave function in LCQM is expressed as 
\begin{eqnarray}
\psi_{S_z}^{(1)F}(x,\textbf{k}_\perp, \lambda_1, \lambda_2)= \chi_{S_z}^F(x,\textbf{k}_\perp, \lambda_1, \lambda_2) \varphi^{\pi(K)}(x, \textbf{k}_\perp).\
\label{space}
\end{eqnarray}
Here $\chi_{S_z}^F$ and $\varphi^{\pi(K)}(x,\textbf{k}_\perp)$ corresponds to the spin wave function and momentum space wave function of the pion (kaon) respectively, where superscript $F$ in the above equation stands for the front-form. The superscript $1$ in $\psi_{S_z}^{(1)F}(x,\textbf{k}_\perp, \lambda_1, \lambda_2)$ is used as a notation for LCQM. The front-form spin wave function for spin-$0$ mesons has been derived from the instant-form spin through the Melosh-Wigner rotation method \cite{Kaur:2020vkq}. Understanding the Melosh-Wigner rotation, which is essentially a relativistic phenomenon brought on by the transversal motion of quarks inside the hadron, is crucial for solving the proton ``spin puzzle" in the nucleon scenario \cite{Qian:2008px,Xiao:2002iv}. This transformation is given as 
\begin{eqnarray}
\chi_i^\uparrow(T)&=&-\frac{[q_i^R \chi_i^\downarrow(F)-(q_i^+ +m_{q(\bar q)})\chi_i^\uparrow(F)]}{\omega_i},\label{instant-front1}\\
\chi_i^\downarrow(T)&=&\frac{[q_i^L\chi_i^\uparrow(F)+(q_i^+ +m_{q(\bar q)})\chi_i^\downarrow(F)]}{\omega_i}.
\label{instant-front}
\end{eqnarray}
The instant-form notations for four vector momenta for the quarks are $q_1^\mu =(q_1^0,\textbf{q})$, $q_2^\mu=(q_2^0,-\textbf{q})$ with $q_i^0=(\textbf{q}^2+m_i^2)^{1/2}$ and $q_i^{R,L}=q_i^1 \pm i q_i^2$ for $i=1$ and $2$. The function $\omega_i$ is given as
\begin{eqnarray}
     \omega_1  &=&  [(x M_{\pi(K)} + m_q)^2 + \textbf{k}^2_\perp]^{\frac{1}{2}},\label{l2}\\ 
    \omega_2 &=&  [((1-x) M_{\pi(K)} + m_{\bar q})^2+ \textbf{k}^2_\perp]^\frac{1}{2}.
    \label{l1}
\end{eqnarray}
In the case of pion, we are introducing  $m_q=m_{\bar q }=m$ and for kaon $m_q \neq m_{\bar q }$. Hence for pion and kaon, we have different definitions of masses which are as follows 
\begin{eqnarray}
     M_{\pi}^2 &=& \frac{m^2+\textbf{k}^2_{\perp}}{x (1-x)},
   \\
   M_{K}^2 &=& \frac{m^2_{q}+\textbf{k}^2_{\perp}}{x} + \frac{m^2_{\bar q}+\textbf{k}^2_{\perp}}{1-x}.
\end{eqnarray}
For different particles, we can write different forms of Eq. (\ref{l2}) and Eq. (\ref{l1}). Now we can get the spin wave function for spin-$0$ particles depending on their compositions for $S_z=0$ as
\begin{eqnarray}
\chi^\mathcal{P}(x,\textbf{k}_\perp)=\sum_{\lambda_1, \lambda_2}\kappa_{0}^F(x,\textbf{k}_\perp, \lambda_1, \lambda_2) \chi_1^{\lambda_1}(F) \chi_2^{\lambda_2}(F),
\end{eqnarray}
with $\mathcal{P}$ being the pseudo-scalar meson notation and $\lambda$ is quark helicity of pion (kaon). $\kappa_{0}^F(x,\textbf{k}_\perp, \lambda_1, \lambda_2)$ is the coefficient of spin wave function. Depending upon the different possibilities of parton's helicity, this coefficient is given by 
\begin{eqnarray}
\kappa_0^F (x, \textbf{k}_\perp, \uparrow, \downarrow)=  [(x M_{\pi(K)} +m_q)((1-x)M_{\pi(K)} + m_{\bar q})-k_\perp^2]/\sqrt{2}\, \omega_1 \omega_2, \nonumber\\
\kappa_0^F (x, \textbf{k}_\perp, \downarrow, \uparrow)=  - [(x M_{\pi(K)} +m_q)((1-x)M_{\pi(K)} + m_{\bar q})-k_\perp^2]/\sqrt{2}\, \omega_1 \omega_2, \nonumber\\
\kappa_0^F (x, \textbf{k}_\perp, \uparrow, \uparrow)=   [(x M_{\pi(K)} +m_q)q_2^L-((1-x)M_{\pi(K)} + m_{\bar q})q_1^L]/\sqrt{2}\, \omega_1 \omega_2, \nonumber\\
\kappa_0^F (x, \textbf{k}_\perp, \downarrow, \downarrow)= [(x M_{\pi(K)} +m_q)q_2^R-((1-x)M_{\pi(K)} + m_{\bar q})q_1^R]/\sqrt{2}\, \omega_1 \omega_2. \label{coeff}
\end{eqnarray}
These coefficients satisfy the following normalization relation
\begin{eqnarray}
\sum_{\lambda_1,\lambda_2} \kappa_0^{F*}(x, \textbf{k}_\perp, \lambda_1, \lambda_2)\kappa_0^F(x, \textbf{k}_\perp, \lambda_1, \lambda_2)=1.
\end{eqnarray}
Further, the momentum space wave function of Eq. (\ref{space}) can be given using Brodsky-Huang-Lepage (BHL) prescription for pion as \cite{Qian:2008px,Xiao:2002iv}
\begin{eqnarray}
\varphi^\pi(x,\textbf{k}_\perp)=A^\pi \ {\rm exp}\bigg[-\frac{1}{8 \beta_\pi^2} \frac{{\bf k}^2_\perp+m^2} {x(1-x)}+ \frac{m^2}{2 \beta_\pi^2} \bigg].
\label{bhl-pi}
\end{eqnarray}
Similarly, for kaon we have
\begin{eqnarray}
\varphi^K(x,\textbf{k}_\perp)= A^K \ {\rm exp} \Bigg[-\frac{\frac{\textbf{k}^2_\perp+m_q^2}{x}+\frac{\textbf{k}^2_\perp+m^2_{\bar q}}{1-x}}{8 \beta_K^2}
-\frac{(m_q^2-m_{\bar q}^2)^2}{8 \beta_K^2 \bigg(\frac{\textbf{k}^2_\perp+m_q^2}{x}+\frac{\textbf{k}^2_\perp+m_{\bar q}^2}{1-x}\bigg)} + \frac{m^2_q +m^2_{\bar q}}{4 \beta_K^2}\Bigg],
\label{bhl-k}
\end{eqnarray}
where $ A^{\pi(K)}$ and $\beta_{\pi (K)}$ are the normalization constants and harmonic scale parameters for pion (kaon) respectively.
\begin{figure}[!htb]
     \centering
     {\includegraphics[width=.5\linewidth]{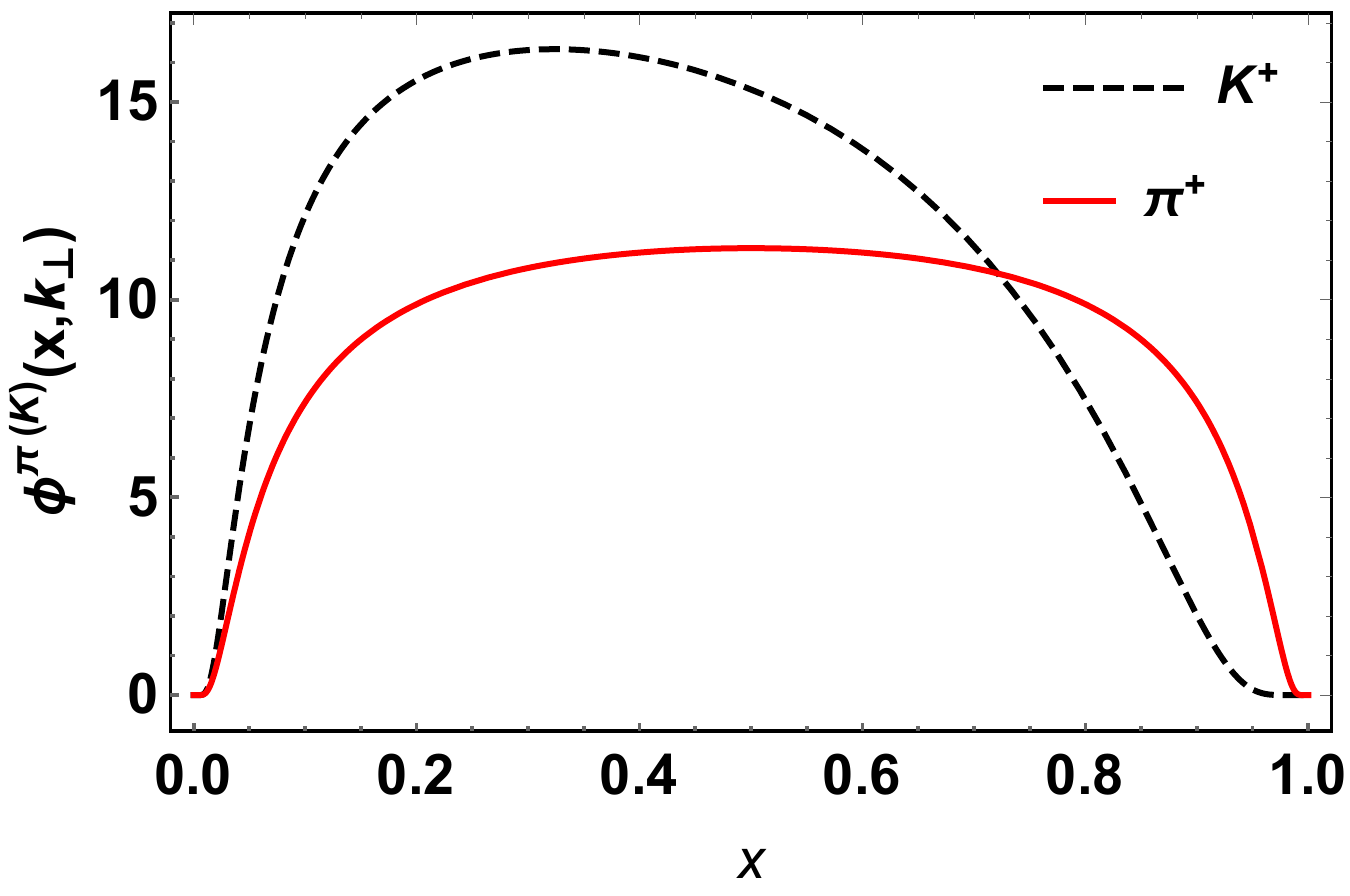}}
     \caption{Momentum space wave function in the LCQM at $\textbf{k}^2_{\perp} = 0.2$ GeV$^2$ for pion (solid red line) and kaon (dashed black line).}\label{Tom}
\end{figure}
\begin{figure}
\centering
(a){\label{4figs-a1} \includegraphics[width=0.45\textwidth]{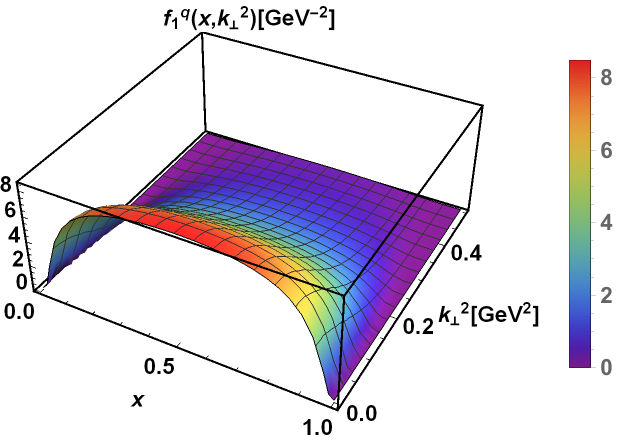}}
\hfill
(b){\label{4figs-b2} \includegraphics[width=0.47\textwidth]{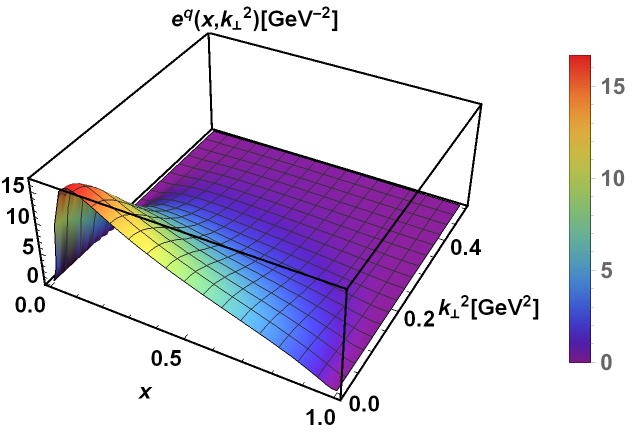}}%
\hfill \\
(c){\label{4figs-c3} \includegraphics[width=0.45\textwidth]{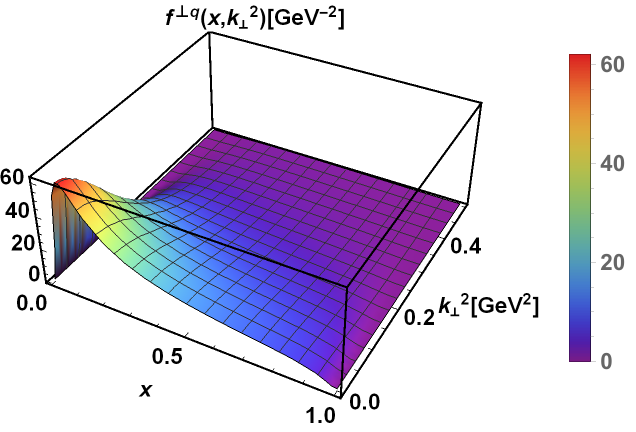}}%
\hfill
(d){\label{4figs-d4} \includegraphics[width=0.45\textwidth]{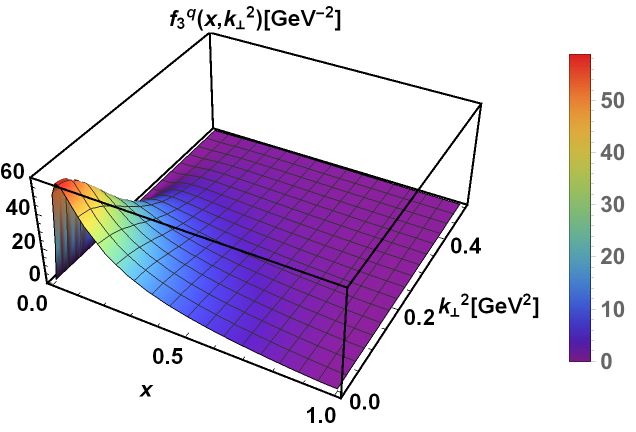}}%
\caption{3D Plots of T-even TMDs for \textit{u}-quark of pion with respect to $x$ and $\textbf{k}^2_\perp$ in the LCQM.}
\label{4figs}
\end{figure}
\begin{figure}
\centering
(a){\label{4figs-a5} \includegraphics[width=0.45\textwidth]{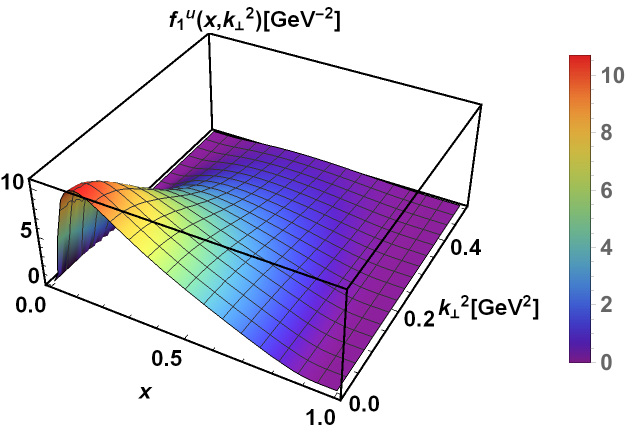}}
\hfill
(b){\label{4figs-b6} \includegraphics[width=0.47\textwidth]{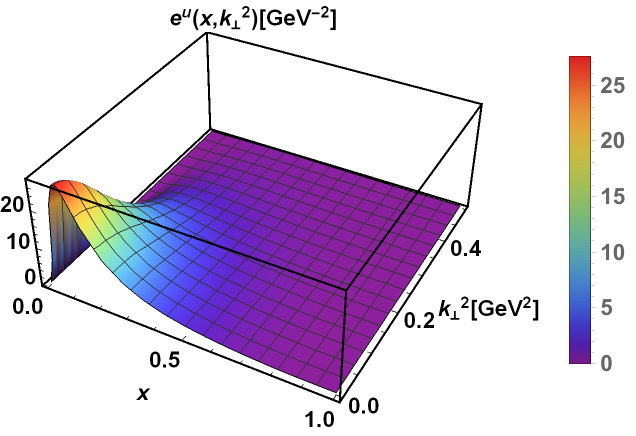}}%
\hfill \\
(c){\label{4figs-c7} \includegraphics[width=0.45\textwidth]{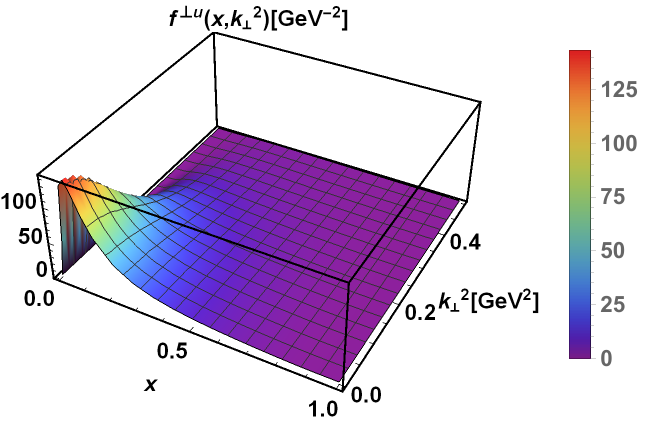}}%
\hfill
(d){\label{4figs-d8} \includegraphics[width=0.45\textwidth]{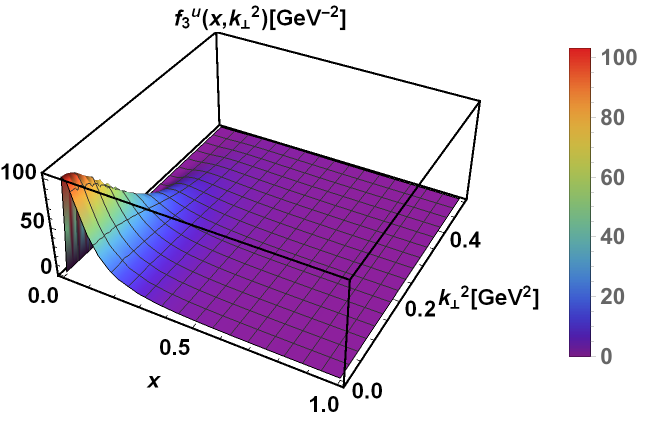}}%
\caption{3D Plots of T-even TMDs for \textit{u}-quark of kaon with respect to $x$ and $\textbf{k}^2_\perp$ in the LCQM.}
\label{4figs1}
\end{figure}
\begin{figure}
\centering
(a){\label{4figs-a9} \includegraphics[width=0.45\textwidth]{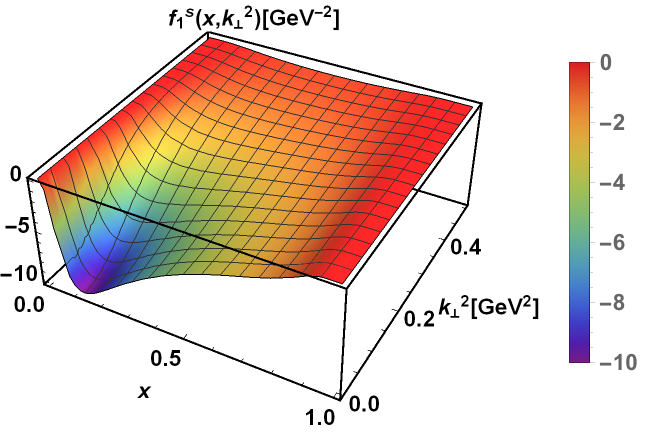}}
\hfill
(b){\label{4figs-b10} \includegraphics[width=0.45\textwidth]{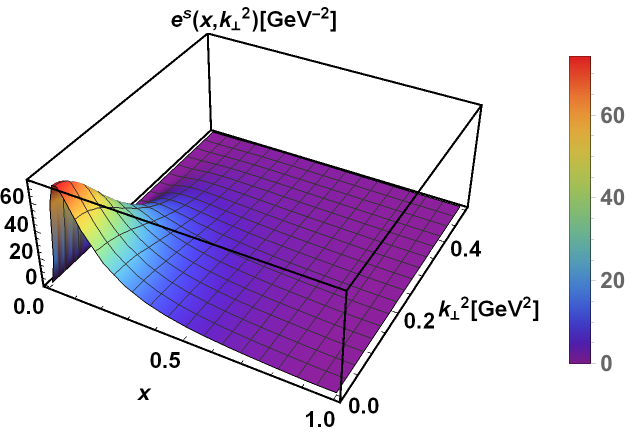}}%
\hfill \\
(c){\label{4figs-c11} \includegraphics[width=0.45\textwidth]{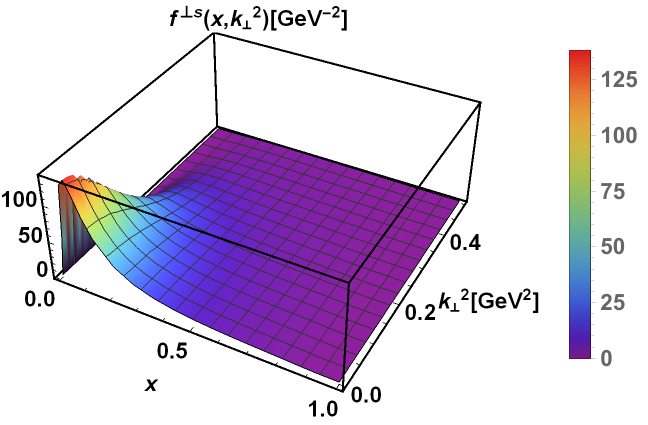}}%
\hfill
(d){\label{4figs-d12} \includegraphics[width=0.45\textwidth]{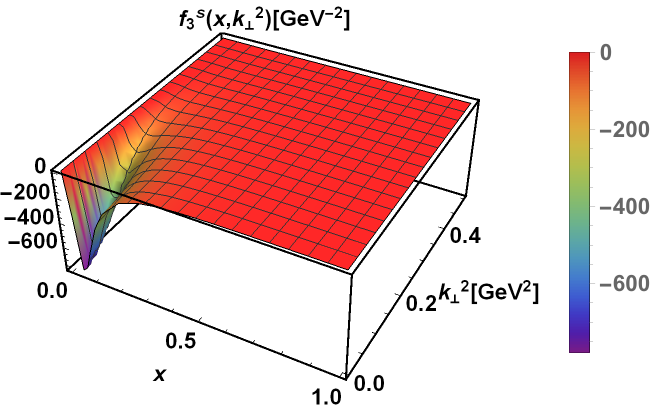}}%
\caption{3D plots of T-even TMDs for $\bar s$-quark of kaon with respect to $x$ and $\textbf{k}^2_\perp$ in the LCQM}
\label{4figs2}
\end{figure}
\par The two-particle Fock-state in Eq. (\ref{meson}) can be written in the LF wave functions with all possible of helicities of its constituent quark and anti-quark as
\begin{eqnarray}
\ket{\pi(K)(P^+,\textbf{P}_\perp,S_z)}&=&\int \frac{{\rm d}^2\textbf{k}_\perp  {\rm d}x}{\sqrt{x(1-x)}2 (2 \pi)^3}\big[\psi^{(1)\pi(K)}_{S_z}(x,\textbf{k}_\perp, \uparrow, \uparrow)\ket{x P^+, \textbf{k}_\perp, \uparrow, \uparrow} \nonumber\\
&&+\psi^{(1)\pi(K)}_{S_z}(x,\textbf{k}_\perp, \downarrow, \downarrow)\ket{x P^+, \textbf{k}_\perp, \downarrow, \downarrow}+\psi^{(1)\pi(K)}_{S_z}(x,\textbf{k}_\perp, \downarrow, \uparrow)\ket{x P^+, \textbf{k}_\perp, \downarrow, \uparrow}\nonumber\\
&&+\psi^{(1)\pi(K)}_{S_z}(x,\textbf{k}_\perp, \uparrow, \downarrow)\ket{x P^+, \textbf{k}_\perp, \uparrow, \downarrow}\big].
\label{overlap}
\end{eqnarray}
The twist-$2$ unpolarized TMD $f^{q}_{1}(x,k^2_{\perp})$ in Eq. (\ref{tw-2TMDs1}) for both kaon and pion in the form of LF wave functions is found to be 
\begin{eqnarray}
f_1^{q (\pi(K))}(x,\textbf{k}^2_\perp)&=&\frac{1}{16 \pi^3} \big[\mid{ \psi_0^{(1)\pi(K)}(x,\textbf{k}_\perp, \uparrow, \uparrow )}\mid^2 +\mid \psi_0^{(1)\pi(K)}(x,\textbf{k}_\perp, \downarrow, \downarrow )\mid^2 + \mid \psi_0^{(1)\pi(K)} (x,\textbf{k}_\perp, \downarrow, \uparrow )\mid^2 \nonumber\\
&& + \mid \psi_0^{(1)\pi(K)}(x,\textbf{k}_\perp, \uparrow, \downarrow)\mid^2\big].
\end{eqnarray}
Using the states, all the T-even TMDs for pion can be written as
\begin{eqnarray}
f_1^{q (\pi)}(x,\textbf{k}^2_\perp)&=&\frac{1}{16 \pi^3} \bigg[\big((x {M}_\pi+m)((1-x){M}_\pi+m)-\textbf{k}^2_\perp\big)^2
+\big({M}_\pi+2 m\big)^2\bigg]\frac{\mid \varphi^{\pi}(x,\textbf{k}_\perp)\mid^2}{\omega^2_1 \omega^2_2},\label{f1}
\\
e^{q (\pi)}(x,\textbf{k}^2_{\perp}) &=& \frac{m}{x M_{\pi}}f_1^{q (\pi)}(x,\textbf{k}^2_\perp) \nonumber\\
&=&\frac{1}{16 \pi^3} \bigg[ \frac{m}{x M_{\pi}} \big((x {M}_\pi+m)((1-x){M}_\pi+m)-\textbf{k}^2_\perp\big)^2
+\big({M}_\pi+2 m\big)^2\bigg]\frac{\mid \varphi^{\pi}(x,\textbf{k}_\perp)\mid^2}{\omega^2_1 \omega^2_2},\label{e}
\\
f^{\perp q (\pi)}(x,\textbf{k}^2_{\perp})&=& \frac{f_1^{q (\pi)}(x,\textbf{k}^2_\perp)}{x}\nonumber\\
&=&\frac{1}{16 \pi^3} \bigg[ \frac{1}{x} \big((x {M}_\pi+m)((1-x){M}_\pi+m)-\textbf{k}^2_\perp\big)^2
+\big({M}_\pi+2 m\big)^2\bigg]\frac{\mid \varphi^{\pi}(x,\textbf{k}_\perp)\mid^2}{\omega^2_1 \omega^2_2},\label{fperp}
\end{eqnarray}
\begin{eqnarray}
f^{q (\pi)}_{3}(x,\textbf{k}^2_{\perp})&=& \frac{\textbf{k}^2_{\perp}+m^2}{x^2 M^2_{\pi}} f_1^{q (\pi)}(x,\textbf{k}^2_\perp)\nonumber\\
&=&\frac{1}{16 \pi^3} \bigg[ \frac{\textbf{k}^2_{\perp}+m^2}{x^2 M^2_{\pi}} \big((x {M}_\pi+m)((1-x){M}_\pi+m)-\textbf{k}^2_\perp\big)^2
+\big({M}_\pi+2 m\big)^2\bigg]\frac{\mid \varphi^{\pi}(x,\textbf{k}_\perp)\mid^2}{\omega^2_1 \omega^2_2}.\nonumber\\ \label{f3}
\end{eqnarray}
Similarly, for the case of kaon, the TMDs are expressed as 
\begin{eqnarray}
f_1^{q (K)}(x,\textbf{k}^2_\perp)&=&\frac{1}{16 \pi^3} \bigg[\big((x {M}_K+m_q)((1-x){M}_K+m_{\bar q})-\textbf{k}^2_\perp\big)^2
+\big({M}_K+ m_q+m_{\bar q}\big)^2\bigg]  \frac{\mid \varphi^{K}(x,\textbf{k}_\perp)\mid^2}{\omega^2_1 \omega^2_2},\label{f1k}
\\
e^{q (K)}(x,\textbf{k}^2_{\perp}) &=& \frac{m_{q}}{x M_{K}}f_1^{q (K)}(x,\textbf{k}^2_\perp)\nonumber\\
&=& \frac{1}{16 \pi^3} \bigg[ \frac{m_{q}}{x M_{K}}\big((x {M}_K+m_q)((1-x){M}_K+m_{\bar q})-\textbf{k}^2_\perp\big)^2
+\big({M}_K+ m_q+m_{\bar q}\big)^2\bigg]  \frac{\mid \varphi^{K}(x,\textbf{k}_\perp)\mid^2}{\omega^2_1 \omega^2_2},\nonumber\\ \label{ek}
\\
f^{\perp q (K)}(x,\textbf{k}^2_{\perp})&=& \frac{f_1^{q (K)}(x,\textbf{k}^2_\perp)}{x}\nonumber\\
&=&\frac{1}{16 \pi^3} \bigg[ \frac{1}{x} \big((x {M}_K+m_q)((1-x){M}_K+m_{\bar q})-\textbf{k}^2_\perp\big)^2
+\big({M}_K+ m_q+m_{\bar q}\big)^2\bigg]  \frac{\mid \varphi^{K}(x,\textbf{k}_\perp)\mid^2}{\omega^2_1 \omega^2_2},\nonumber \label{fperpk}
\\
\end{eqnarray}
\begin{eqnarray}
f^{q (K)}_{3}(x,\textbf{k}^2_{\perp})&=& \frac{\textbf{k}^2_{\perp}+m_{q}^2}{x^2 M^2_{K}} f_1^{q (K)}(x,\textbf{k}^2_\perp)\nonumber\\
&=& \frac{1}{16 \pi^3} \bigg[\frac{\textbf{k}^2_{\perp}+m_{q}^2}{x^2 M^2_{K}} \big((x {M}_K+m_q)((1-x){M}_K+m_{\bar q})-\textbf{k}^2_\perp\big)^2
+\big({M}_K+ m_q+m_{\bar q}\big)^2\bigg]  \frac{\mid \varphi^{K}(x,\textbf{k}_\perp)\mid^2}{\omega^2_1 \omega^2_2}. \nonumber\\ \label{f3k}
\end{eqnarray}
The above equations are the relations of T-even TMDs, when the struck parton is quark. For anti-quark distributions, we have adopted the following conversion \cite{Diehl:2003ny,Kaur:2019jow}
\begin{eqnarray}
    F^q (x, \textbf{k}^2_{\perp},m_q,m_{\bar q}) = -F^{\bar q} (-x, -\textbf{k}^2_{\perp},m_{\bar q},m_q).
\end{eqnarray}

\subsection{Light-front holographic model}\label{kanhu}
LFHM has been widely used to describe the internal hadron structure and their microscopic properties like radius, charge, distribution amplitudes (DAs) and quark-gluon interaction \cite{Vega:2009zb,Brodsky:2011xx,Chabysheva:2012fe,Liu:2015jna}. The LF wave function of meson for quark-anti-quark Fock-state $|M(P)\rangle$ can be written as the superposition as the meson wave functions of $0$ and $1$ quark orbital angular momentum (QOAM) [$L_z$] with momentum $P$ as \cite{Pasquini:2023aaf,Kaur:2019jfa,Ahmady:2019hag}
\begin{eqnarray}
    |M(P)\rangle &=& |M(P)\rangle_{L_z=0} +
|M(P)\rangle_{|L_z|=1}.
\end{eqnarray}
 With different QOAM, the meson states are expressed as \cite{Kaur:2019jfa,Pasquini:2023aaf}
\begin{figure}[!htb]
\centering
\includegraphics[width=.5\linewidth]{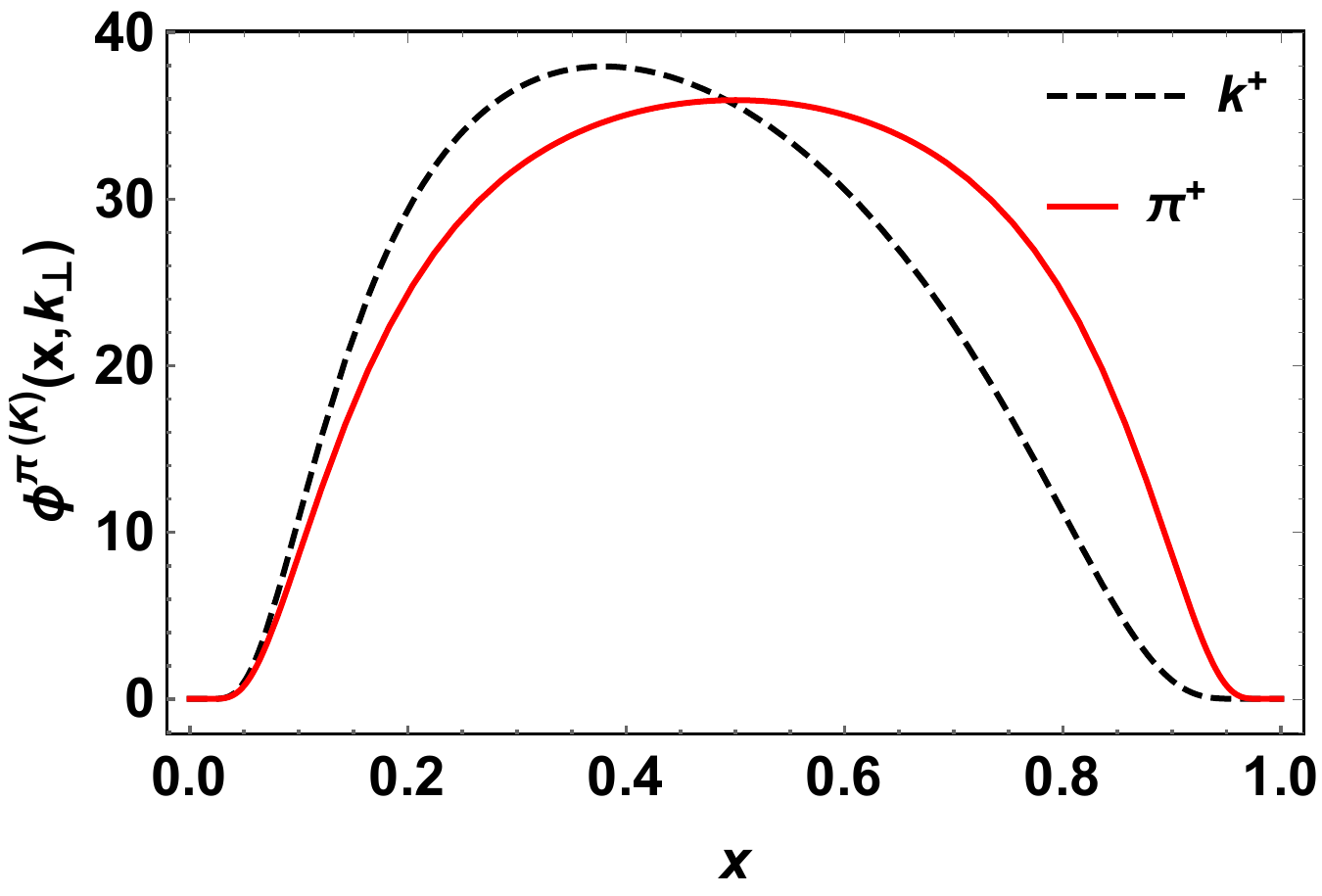}
    \caption{Momentum space wave function in LFHM at $\textbf{k}^2_{\perp} = 0.2$ GeV$^2$ for pion (solid red line) and kaon (dashed black line). }\label{jerry}
\end{figure}
\begin{figure}
\centering
(a){\label{4figs-a13} \includegraphics[width=0.45\textwidth]{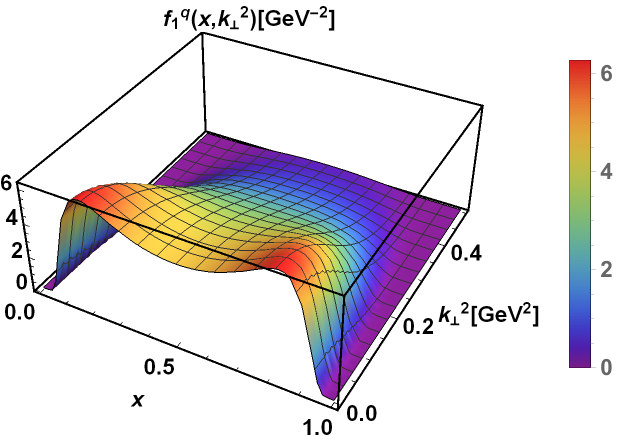}}
\hfill
(b){\label{4figs-b14} \includegraphics[width=0.45\textwidth]{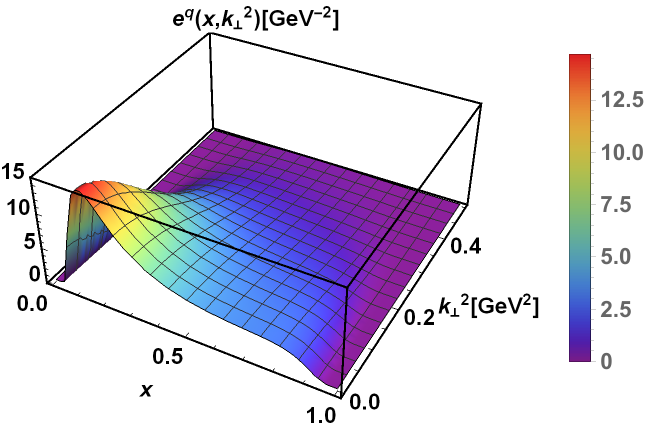}}%
\hfill \\
(c){\label{4figs-c15} \includegraphics[width=0.45\textwidth]{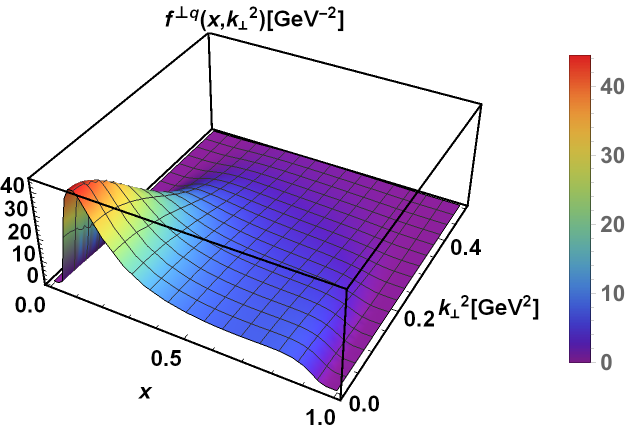}}%
\hfill
(d){\label{4figs-d16} \includegraphics[width=0.45\textwidth]{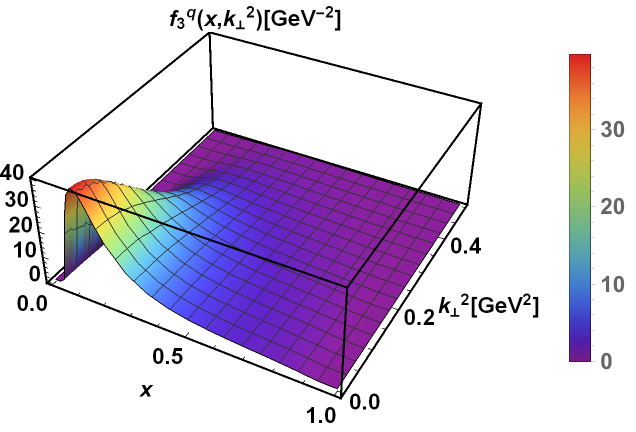}}%
\caption{3D Plots of T-even TMDs for \textit{u}-quark of pion with respect to $x$ and $\textbf{k}^2_\perp$ in the LFHM.}
\label{4figs3}
\end{figure}
\begin{figure}
\centering
(a){\label{4figs-a17} \includegraphics[width=0.45\textwidth]{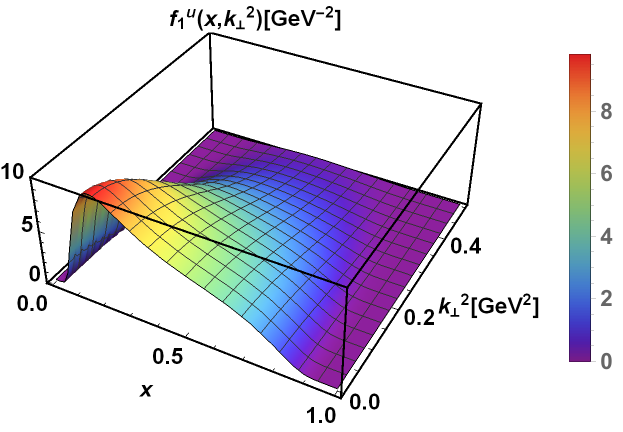}}
\hfill
(b){\label{4figs-b18} \includegraphics[width=0.45\textwidth]{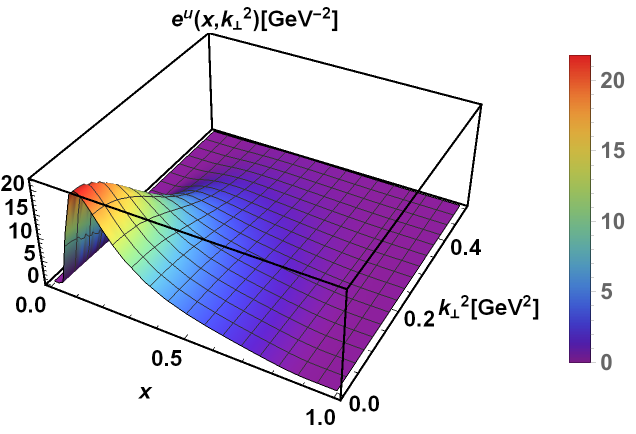}}%
\hfill \\
(c){\label{4figs-c19} \includegraphics[width=0.45\textwidth]{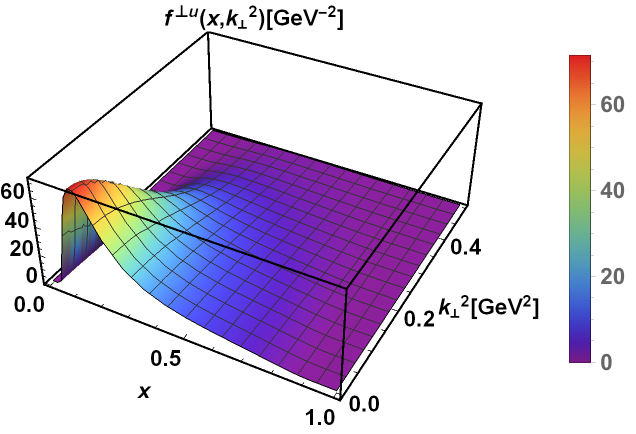}}%
\hfill
(d){\label{4figs-d20} \includegraphics[width=0.45\textwidth]{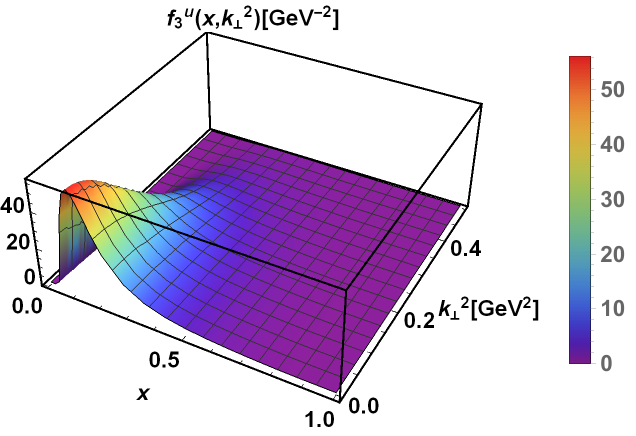}}%
\caption{3D Plots of T-even TMDs for \textit{u}-quark of Kaon with respect to $x$ and $\textbf{k}^2_\perp$ in the LFHM.}
\label{4figs4}
\end{figure}
\begin{figure}
\centering
(a){\label{4figs-a21} \includegraphics[width=0.45\textwidth]{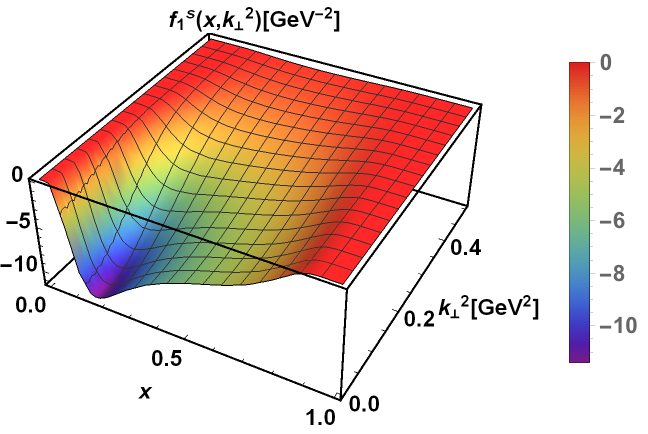}}
\hfill
(b){\label{4figs-b22} \includegraphics[width=0.46\textwidth]{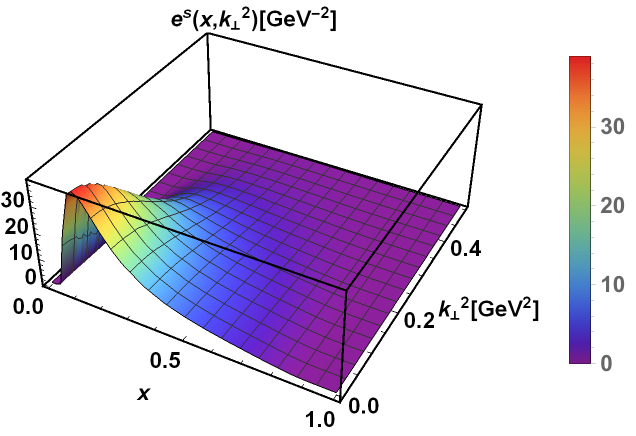}}%
\hfill \\
(c){\label{4figs-c23} \includegraphics[width=0.45\textwidth]{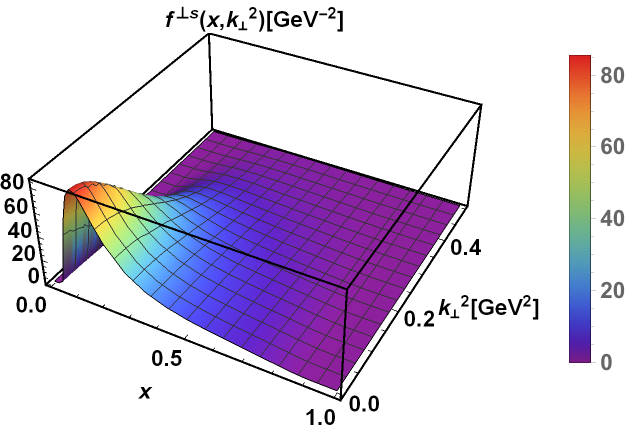}}%
\hfill
(d){\label{4figs-d24} \includegraphics[width=0.46\textwidth]{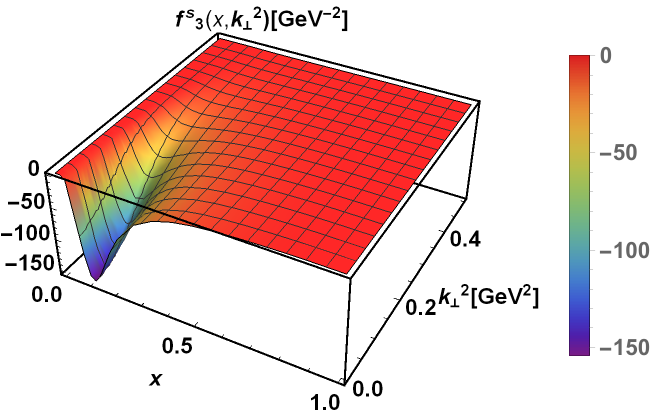}}%
\caption{3D Plots of T-even TMDs for $\bar s$-quark of Kaon with respect to $x$ and $\textbf{k}^2_\perp$ in the LFHM.}
\label{4figs5}
\end{figure}
\begin{figure}
\centering
(a){\label{4figs-a25} \includegraphics[width=0.45\textwidth]{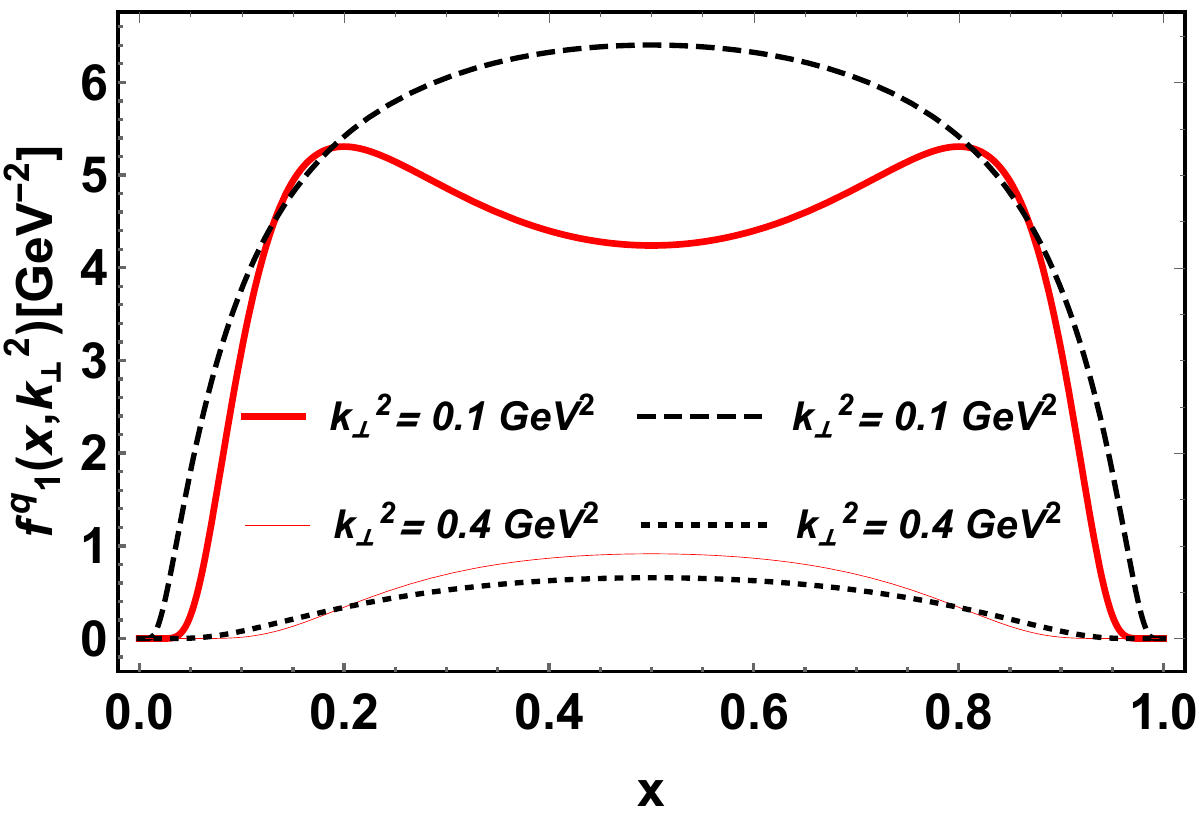}}
\hfill
(b){\label{4figs-b26} \includegraphics[width=0.46\textwidth]{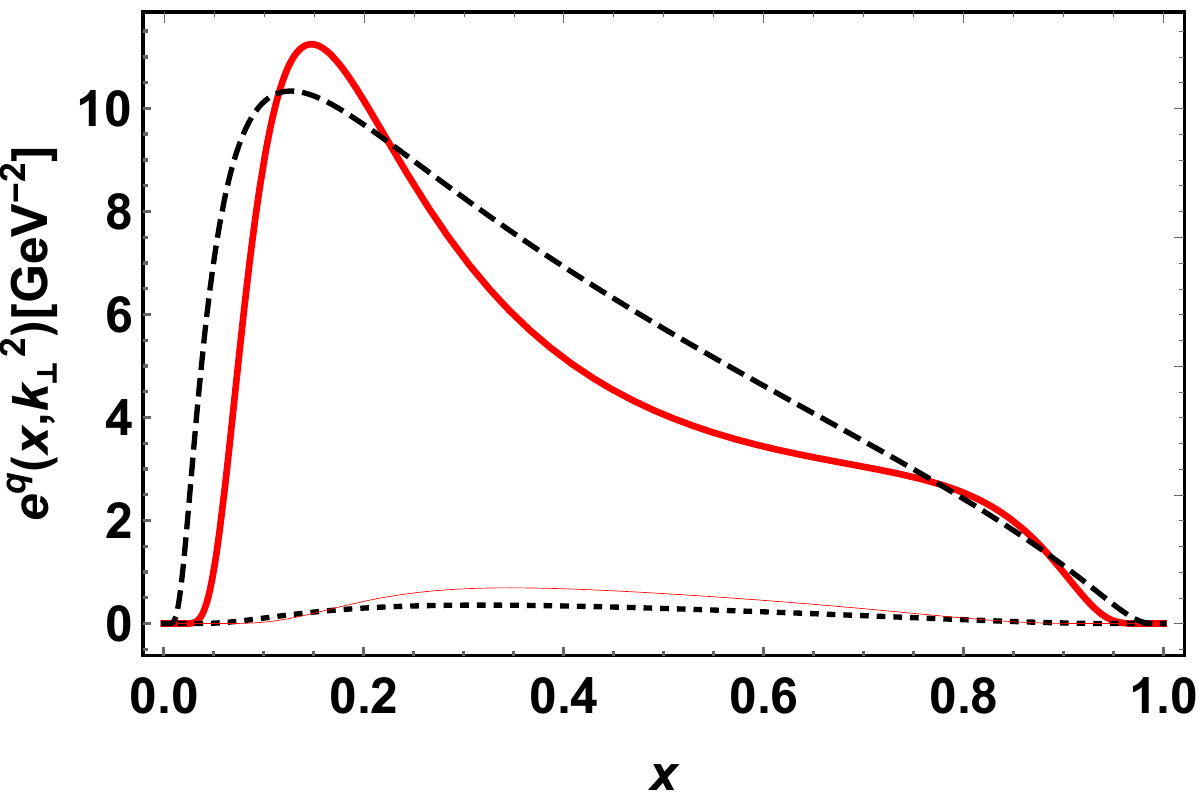}}%
\hfill \\
(c){\label{4figs-c27} \includegraphics[width=0.45\textwidth]{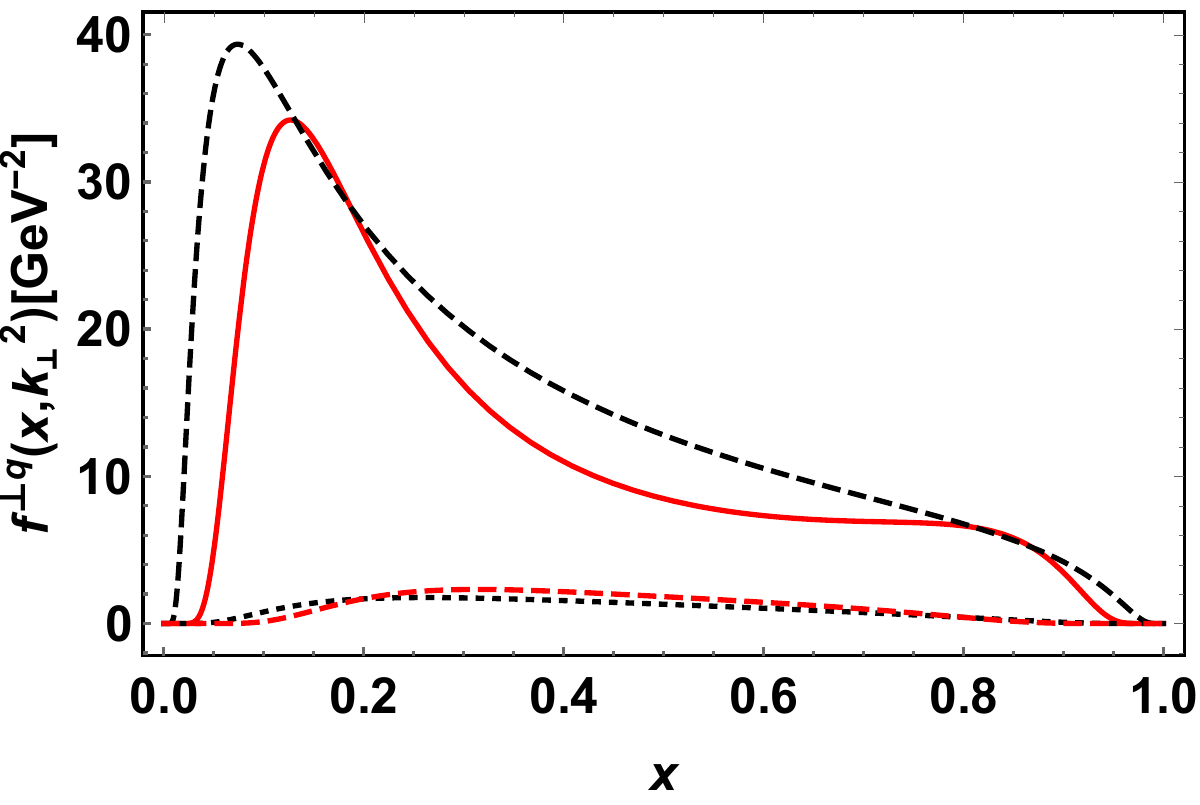}}%
\hfill
(d){\label{4figs-d28} \includegraphics[width=0.45\textwidth]{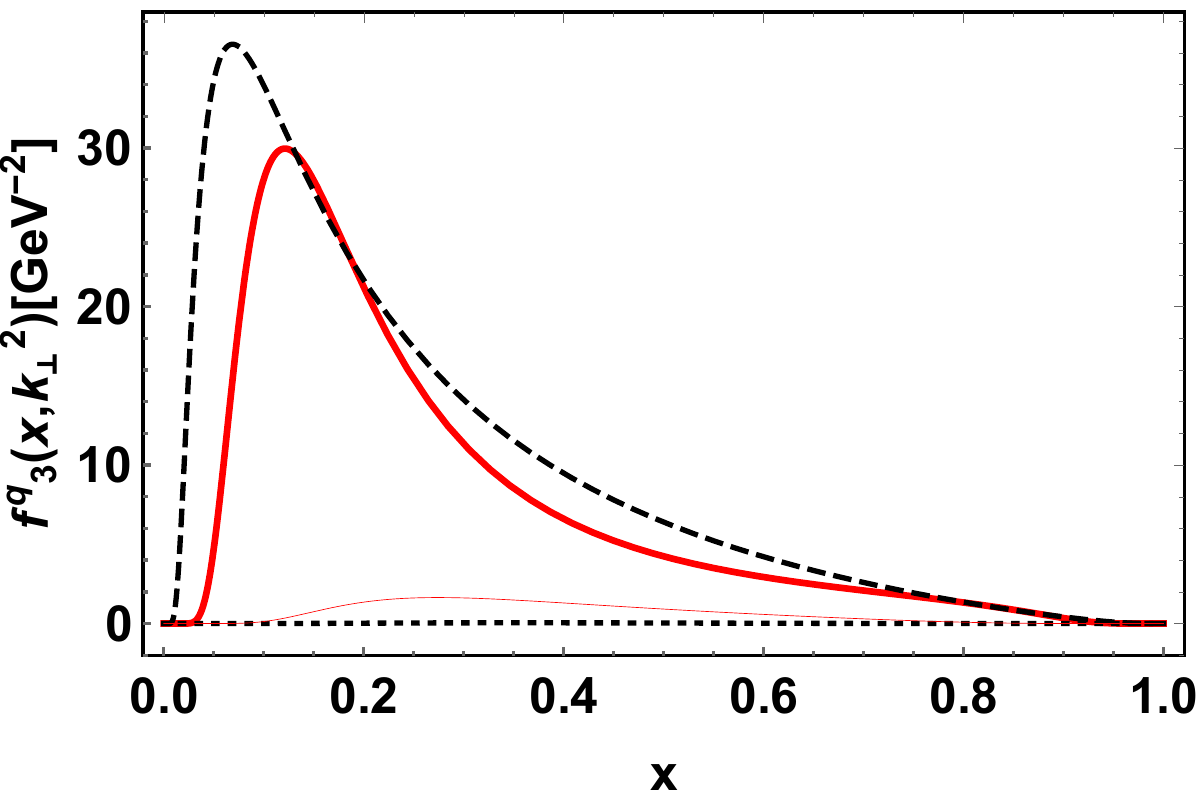}}%
\caption{Comparison of T-even TMDs with respect to longitudinal momentum fraction $x$ in both the LCQM and LFHM for \textit{u}-quark of pion. The solid thick and thin red lines are for the case in LFHM, while dashed and dotted black lines are in the LCQM.}
\label{4figs6}
\end{figure}
\begin{figure}
\centering
(a){\label{4figs-a29} \includegraphics[width=0.45\textwidth]{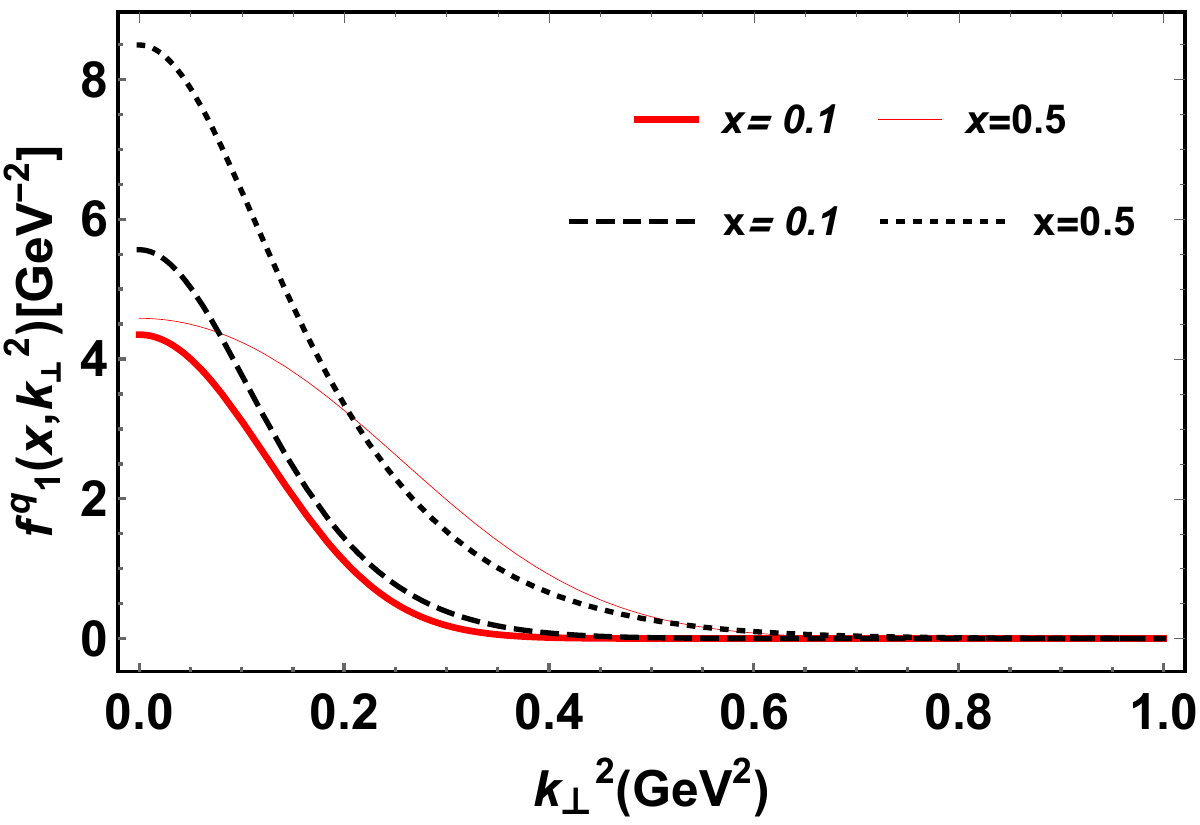}}
\hfill
(b){\label{4figs-b30} \includegraphics[width=0.47\textwidth]{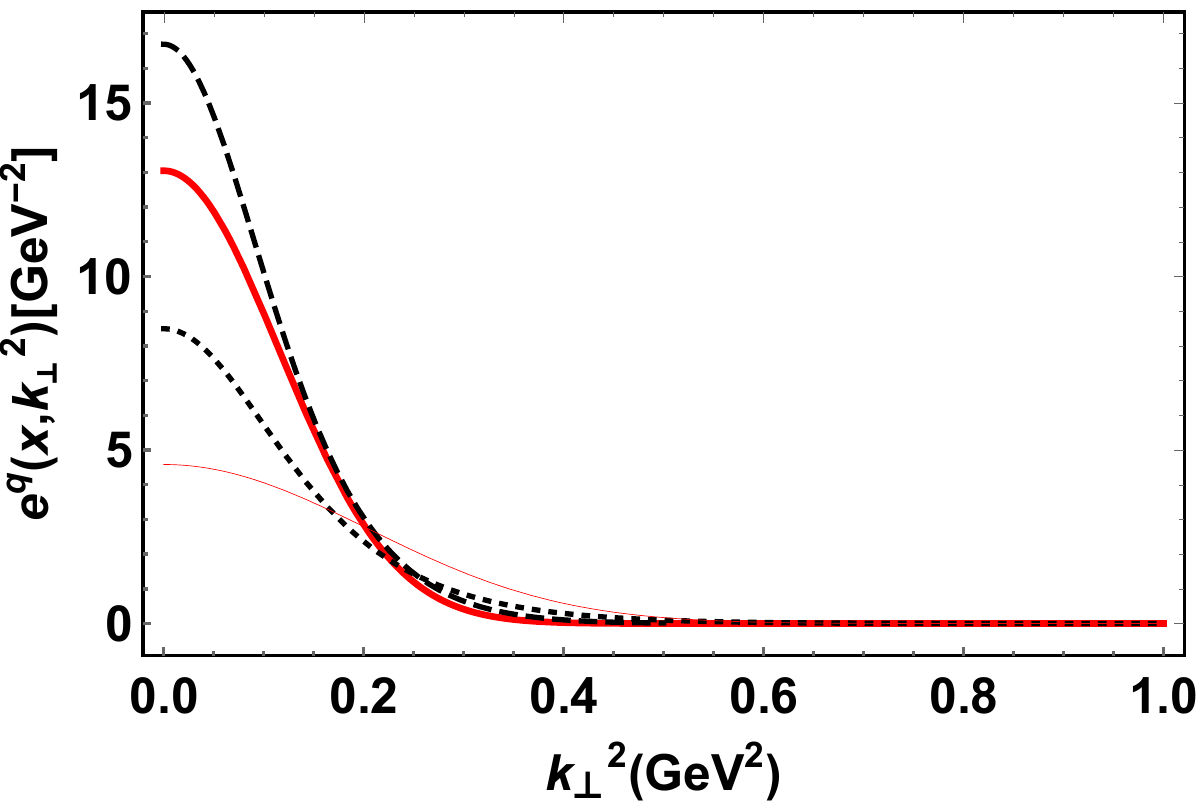}}%
\hfill \\
(c){\label{4figs-c31} \includegraphics[width=0.45\textwidth]{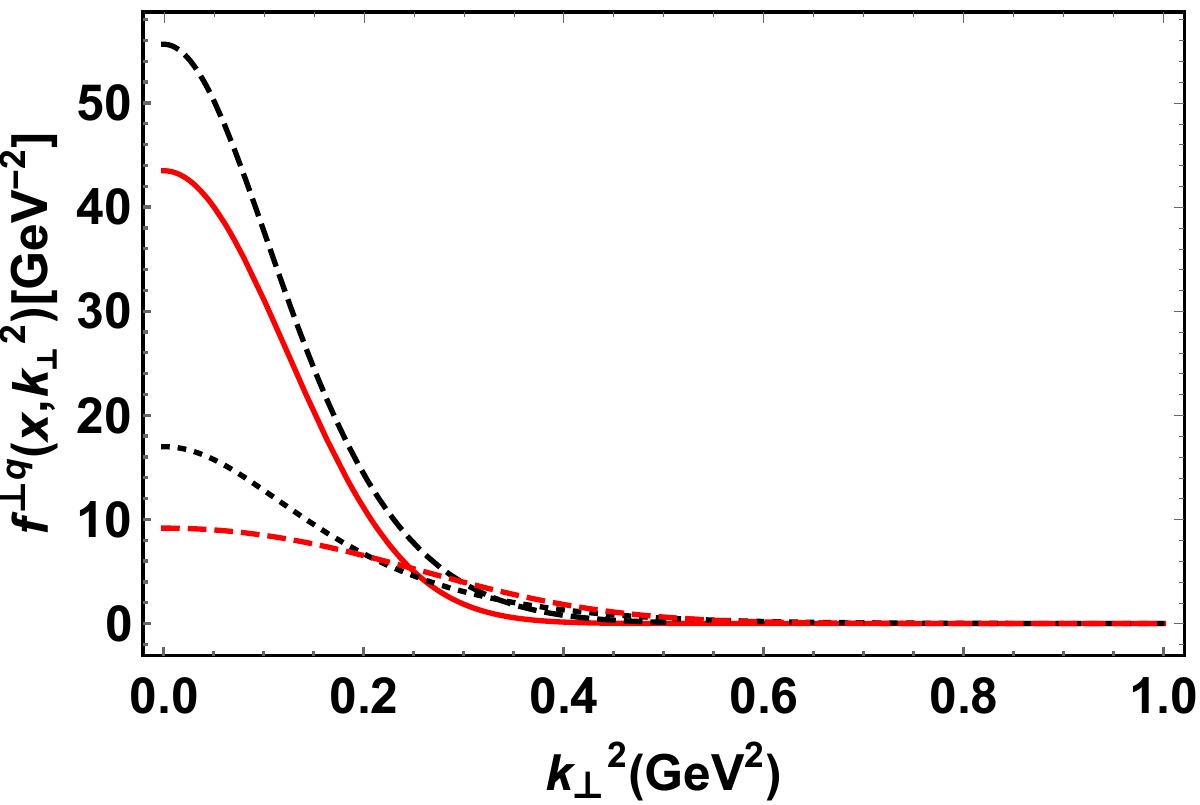}}%
\hfill
(d){\label{4figs-d32} \includegraphics[width=0.46\textwidth]{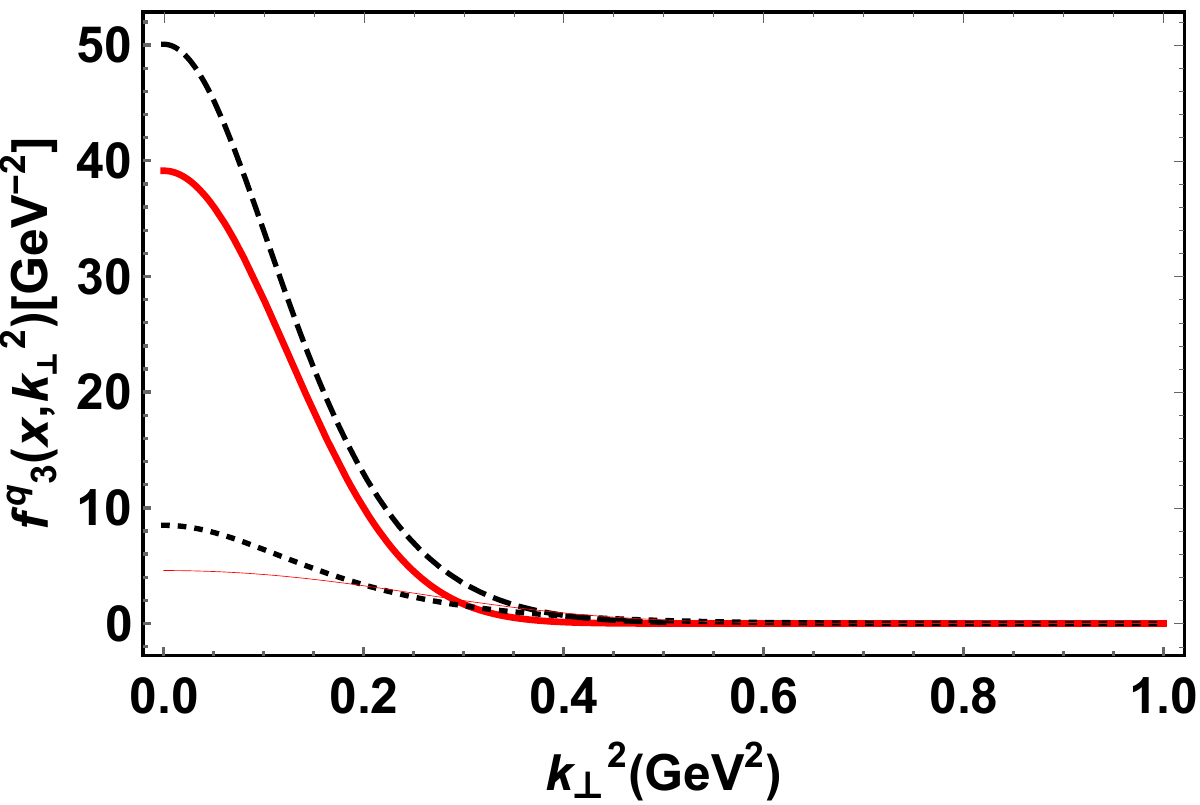}}%
\caption{Comparison of T-even TMDs with respect to $\bfk^2(GeV^2)$ in both the LCQM and LFHM for \textit{u}-quark of pion. The solid thick and thin red lines are for the case in LFHM, while dashed and dotted black lines are in the LCQM.}
\label{4figs7}
\end{figure}
\begin{figure}
\centering
(a){\label{4figs-a33} \includegraphics[width=0.45\textwidth]{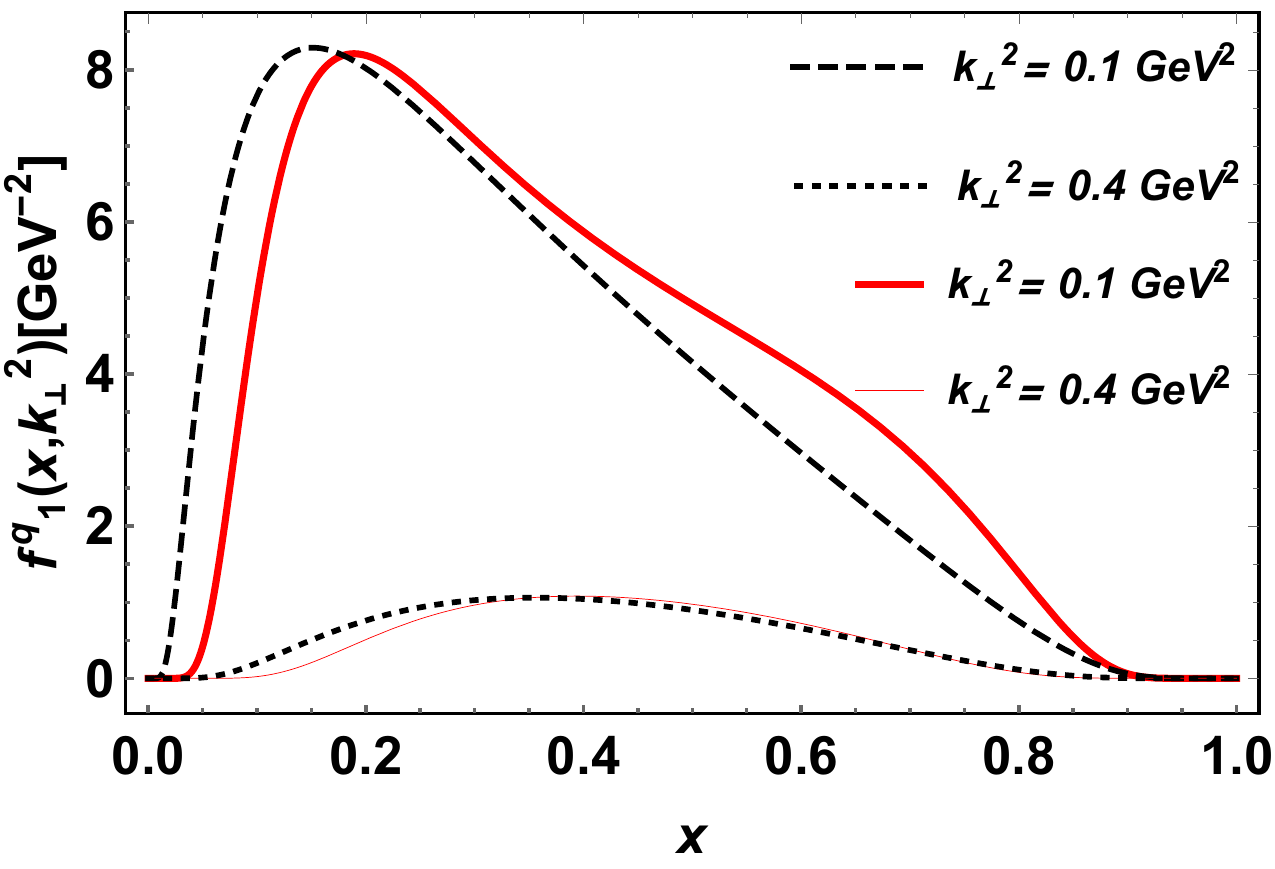}}
\hfill
(b){\label{4figs-b34} \includegraphics[width=0.46\textwidth]{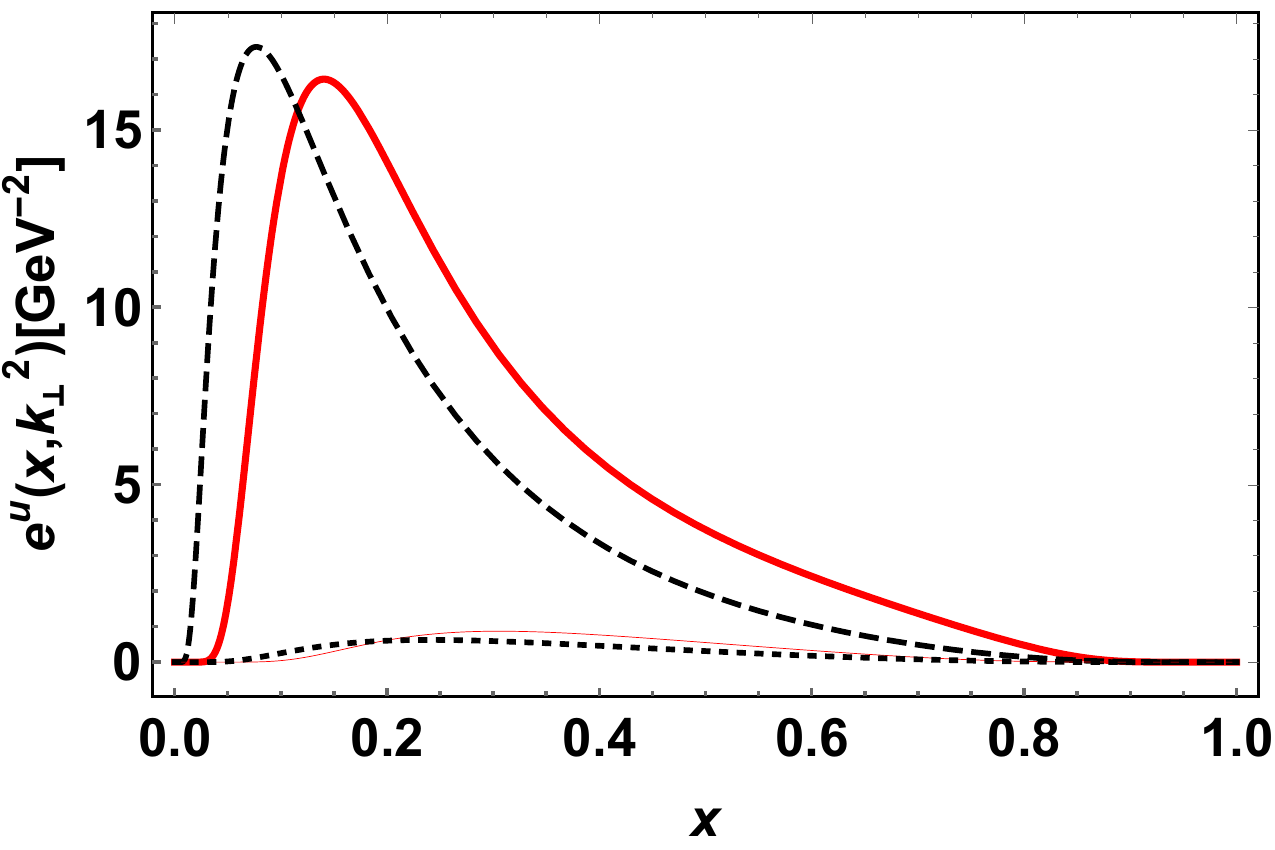}}%
\hfill \\
(c){\label{4figs-c35} \includegraphics[width=0.45\textwidth]{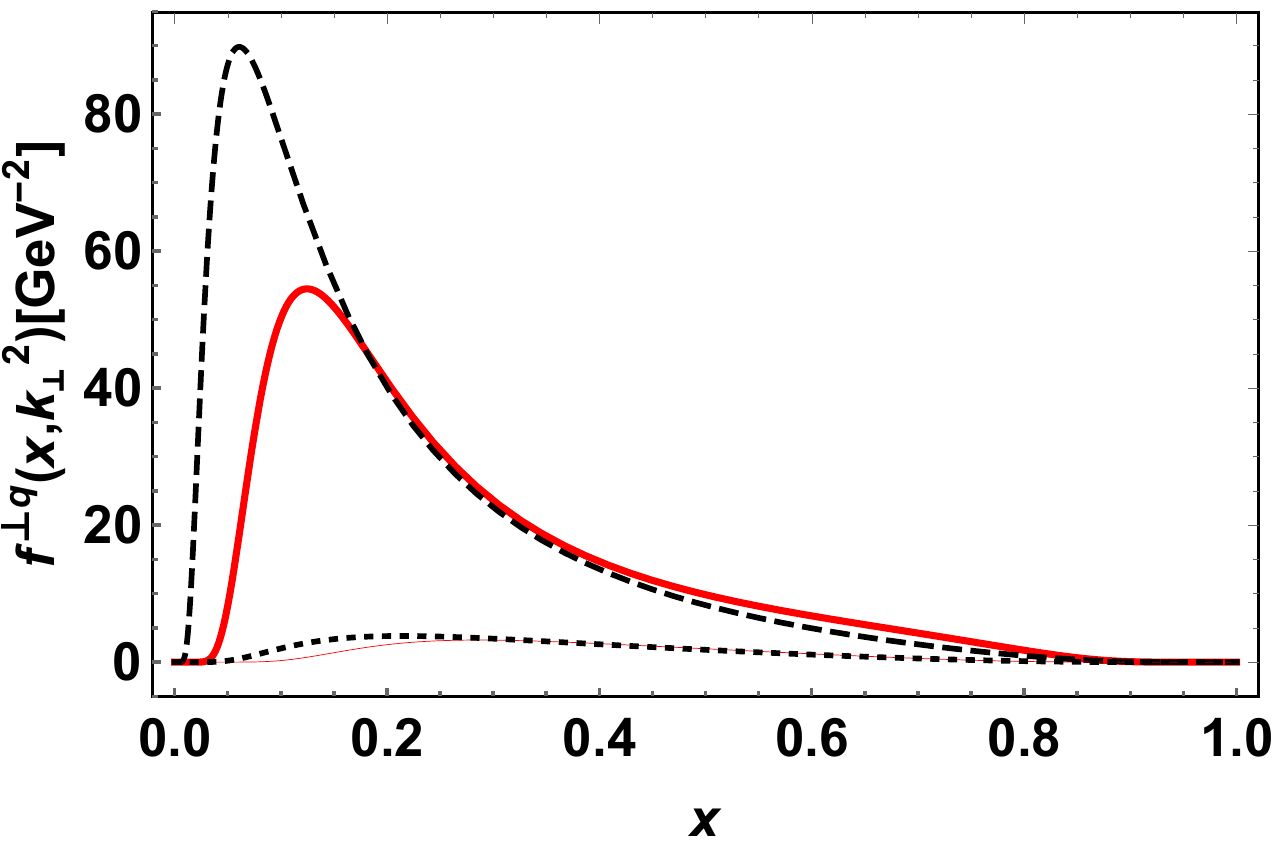}}%
\hfill
(d){\label{4figs-d36} \includegraphics[width=0.46\textwidth]{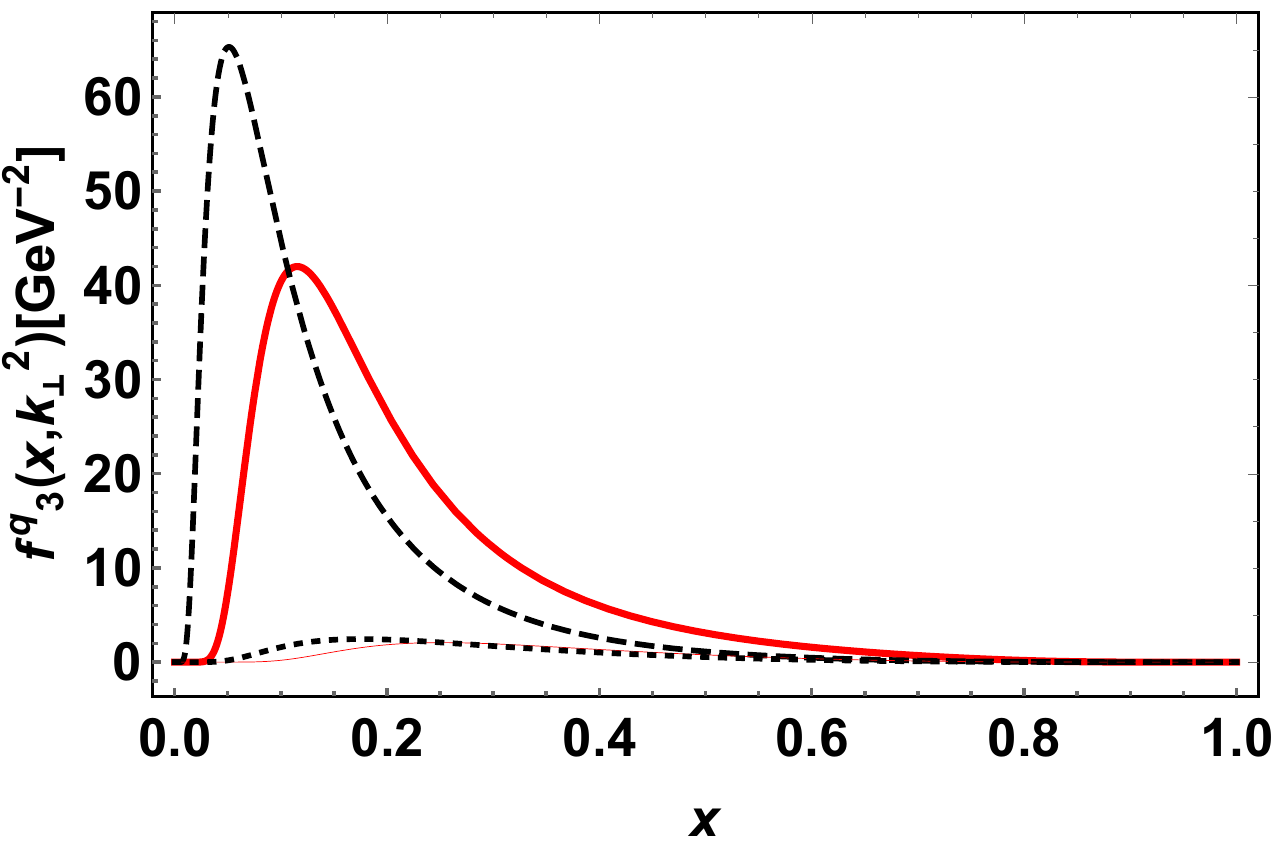}}%
\caption{Comparison of T-even TMDs with respect to longitudinal momentum fraction $(x)$ in both the LCQM and LFHM for \textit{u}-quark of kaon. The solid thick and thin red lines are for the case in LFHM, while dashed and dotted black lines are in the LCQM respectively.}
\label{4figs8}
\end{figure}
\begin{figure}
\centering
(a){\label{4figs-a37} \includegraphics[width=0.45\textwidth]{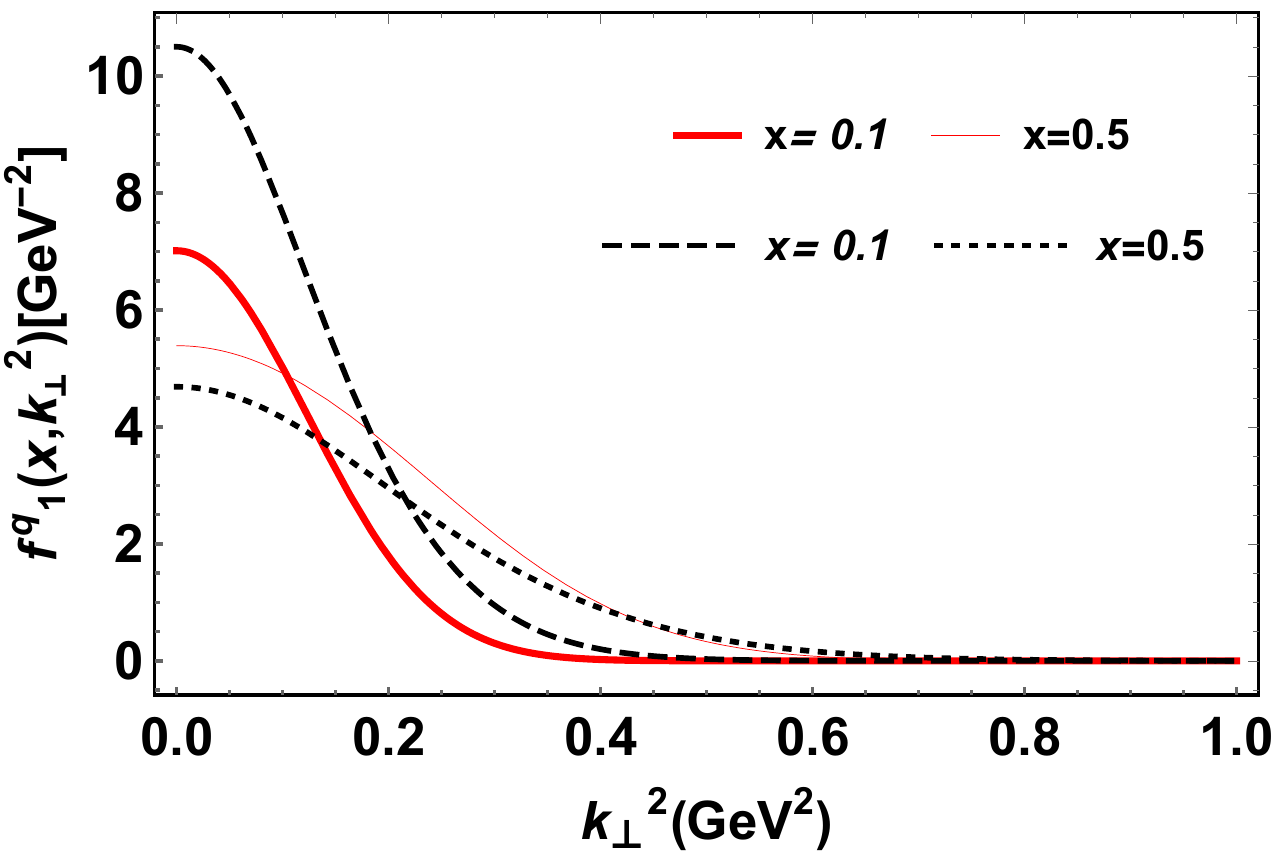}}
\hfill
(b){\label{4figs-b38} \includegraphics[width=0.46\textwidth]{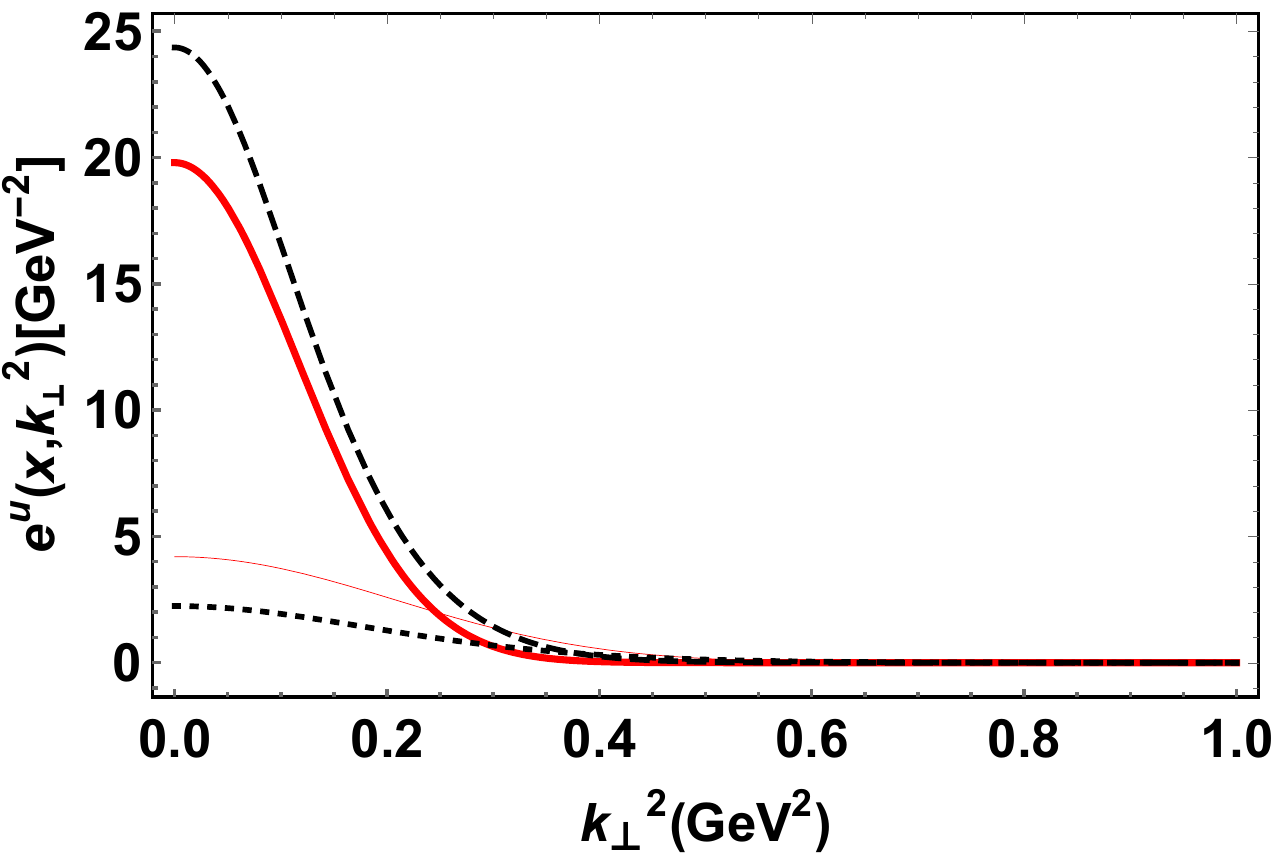}}%
\hfill \\
(c){\label{4figs-c39} \includegraphics[width=0.45\textwidth]{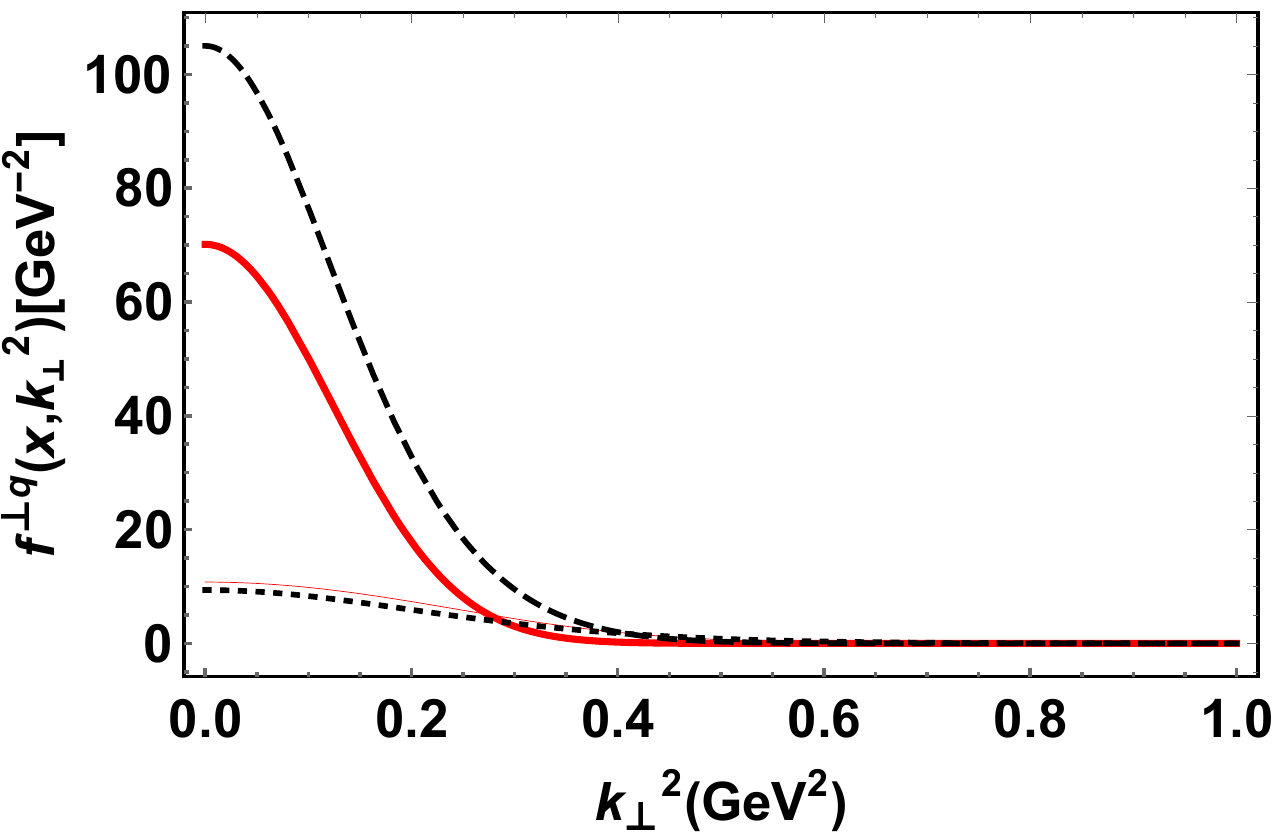}}%
\hfill
(d){\label{4figs-d40} \includegraphics[width=0.45\textwidth]{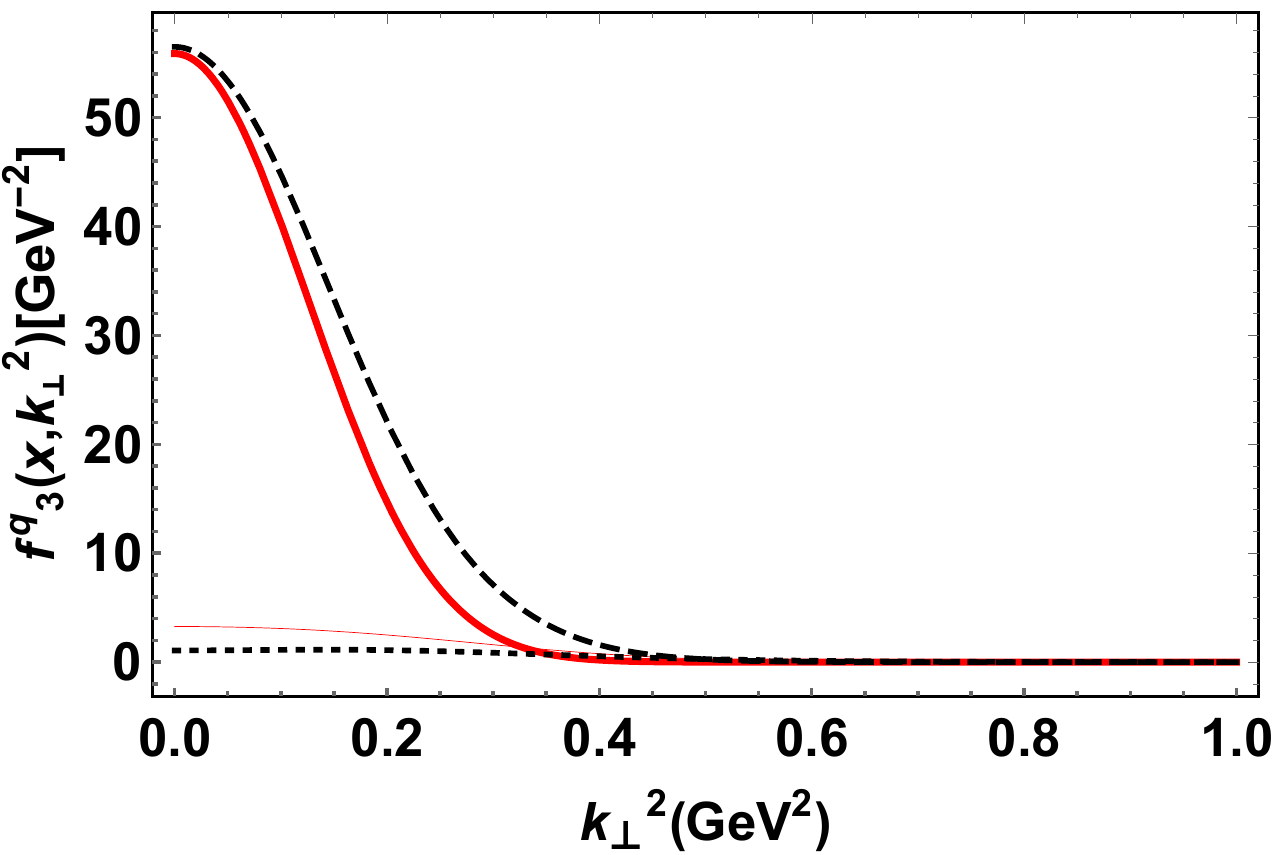}}%
\caption{Comparison of T-even TMDs with respect to $\bfk^2(GeV^2)$ in both the LCQM and LFHM for \textit{u}-quark of kaon. The solid thick and thin red lines are for the case in LFHM, while dashed and dotted black lines are in the LCQM.}
\label{4figs9}
\end{figure}
\begin{figure}
\centering
(a){\label{4figs-a41} \includegraphics[width=0.45\textwidth]{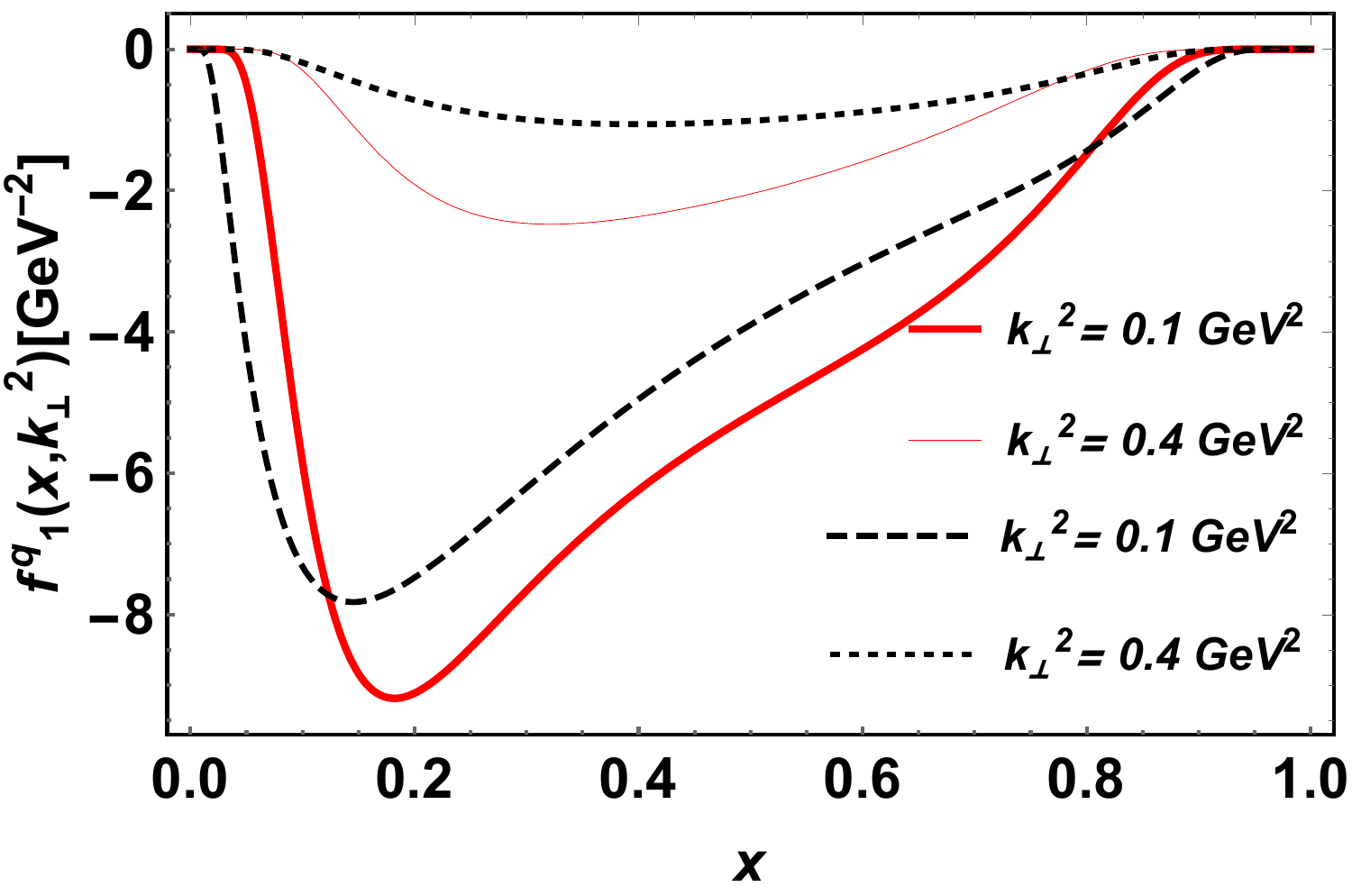}}
\hfill
(b){\label{4figs-b42} \includegraphics[width=0.46\textwidth]{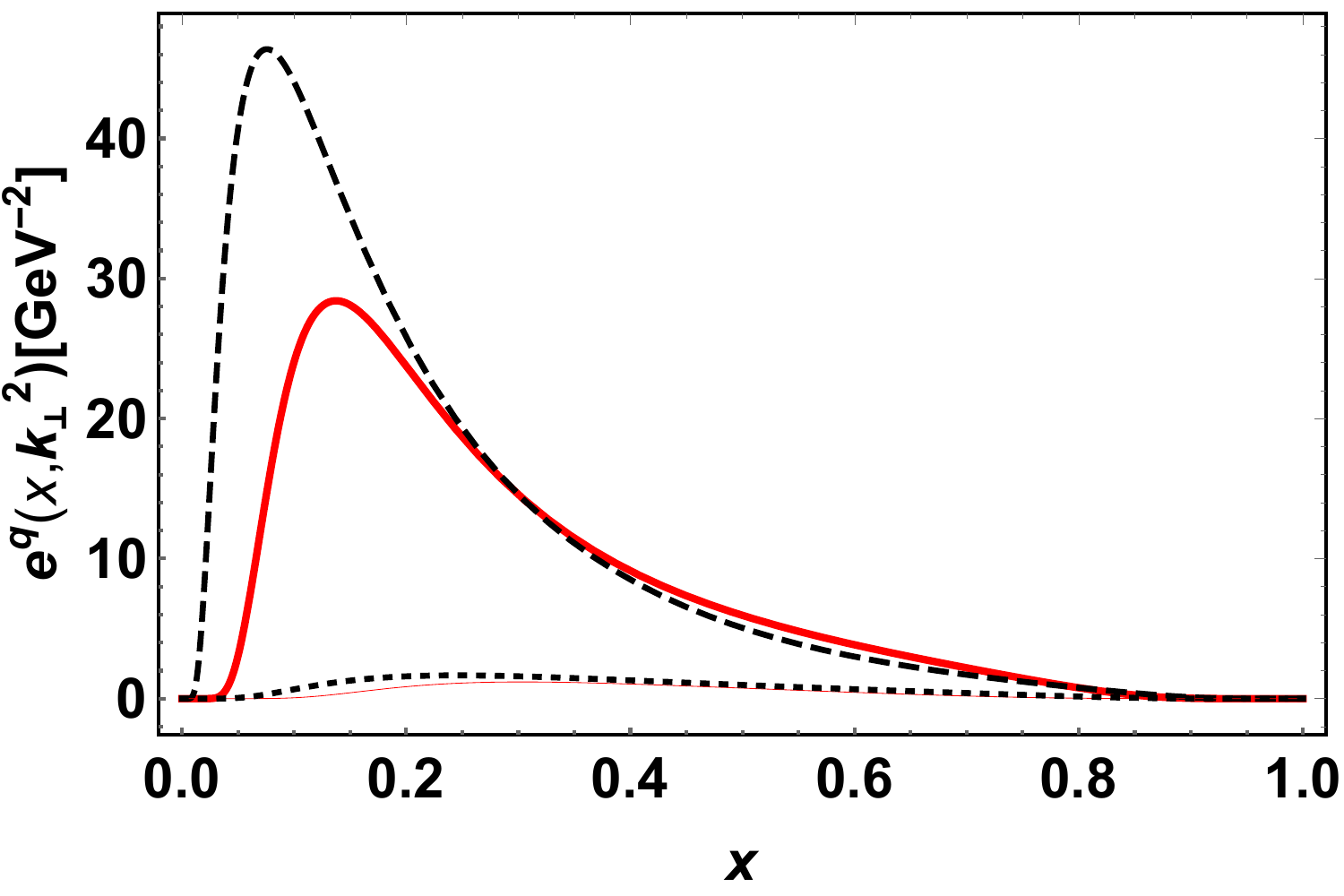}}%
\hfill \\
(c){\label{4figs-c43} \includegraphics[width=0.45\textwidth]{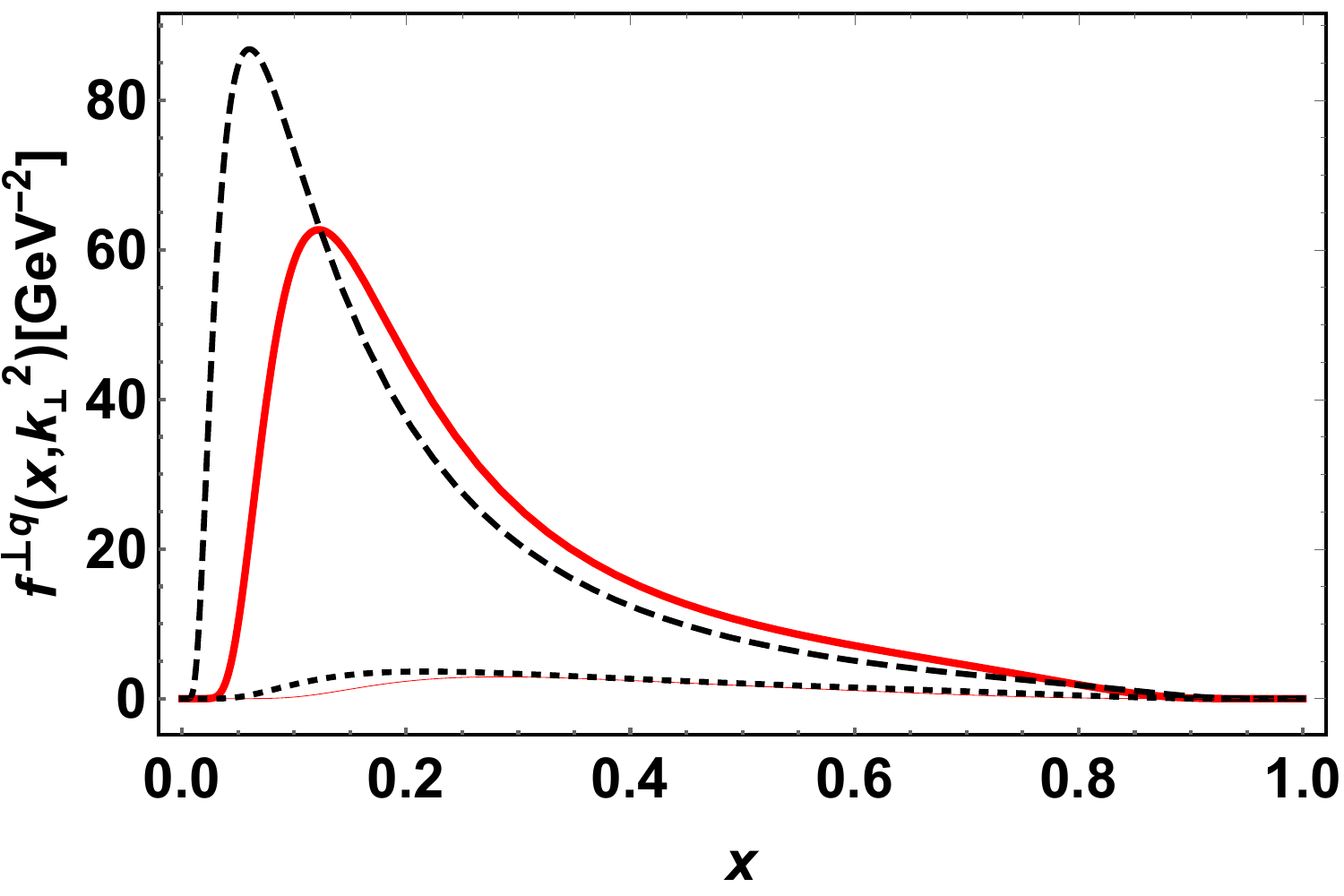}}%
\hfill
(d){\label{4figs-d44} \includegraphics[width=0.48\textwidth]{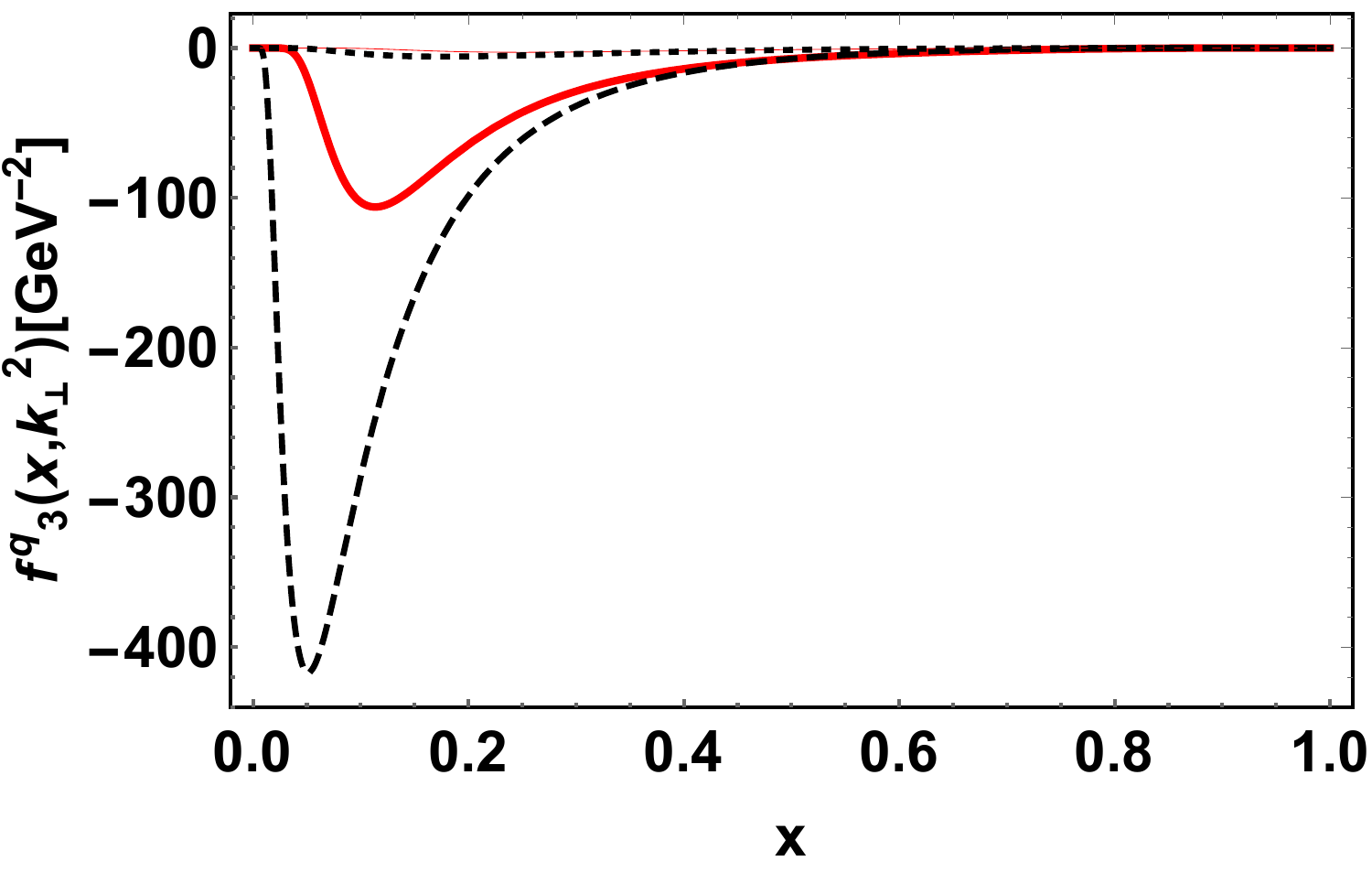}}%
\caption{Comparison of T-even TMDs with respect to longitudinal momentum fraction $(x)$ in both the LCQM and LFHM for $\bar s$-quark of kaon. The solid thick and thin red lines are for the case in LFHM, while dashed and dotted black lines are in the LCQM.}
\label{4figs10}
\end{figure}
\begin{figure}
\centering
(a){\label{4figs-a45} \includegraphics[width=0.45\textwidth]{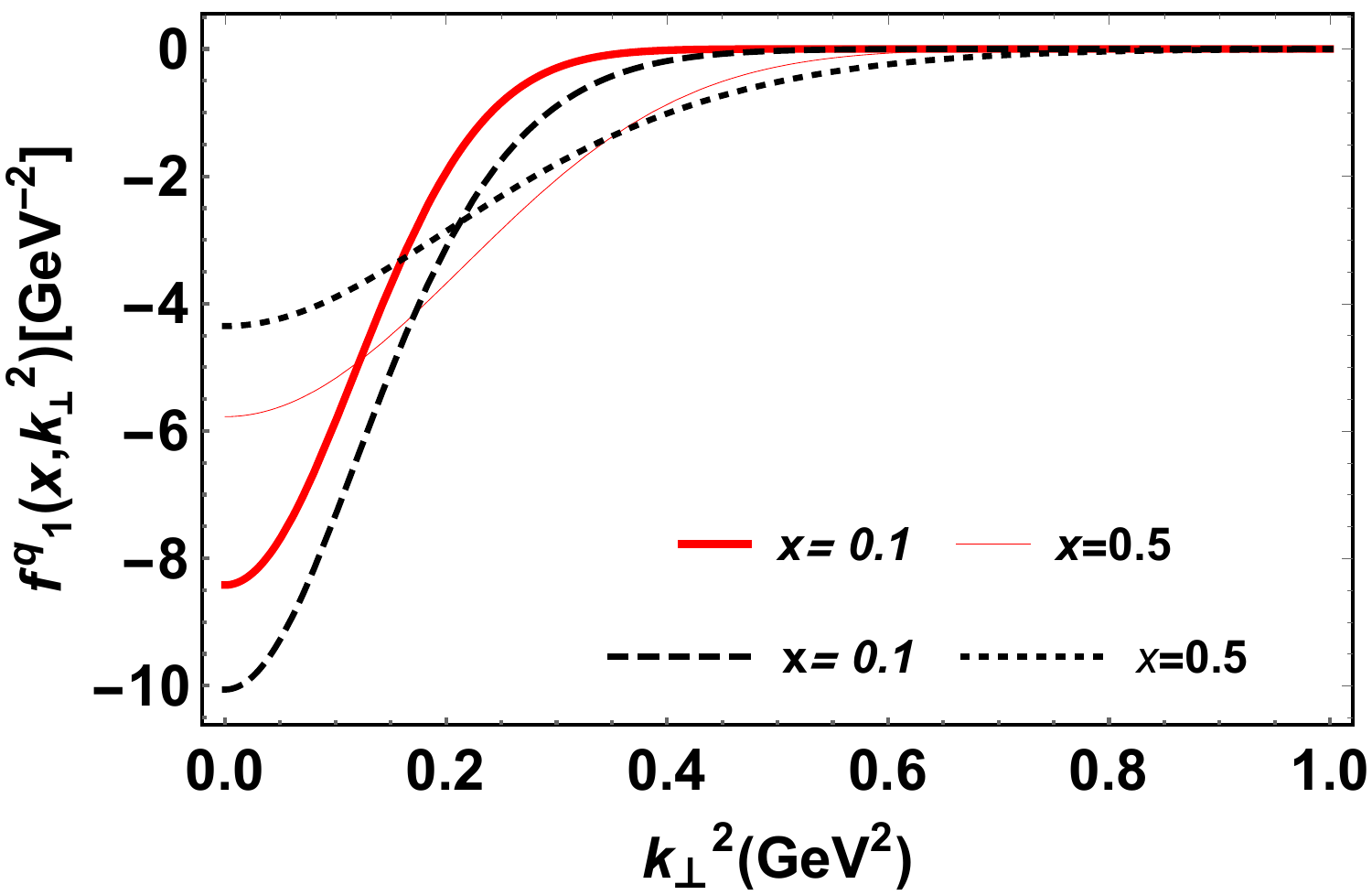}}
\hfill
(b){\label{4figs-b46} \includegraphics[width=0.45\textwidth]{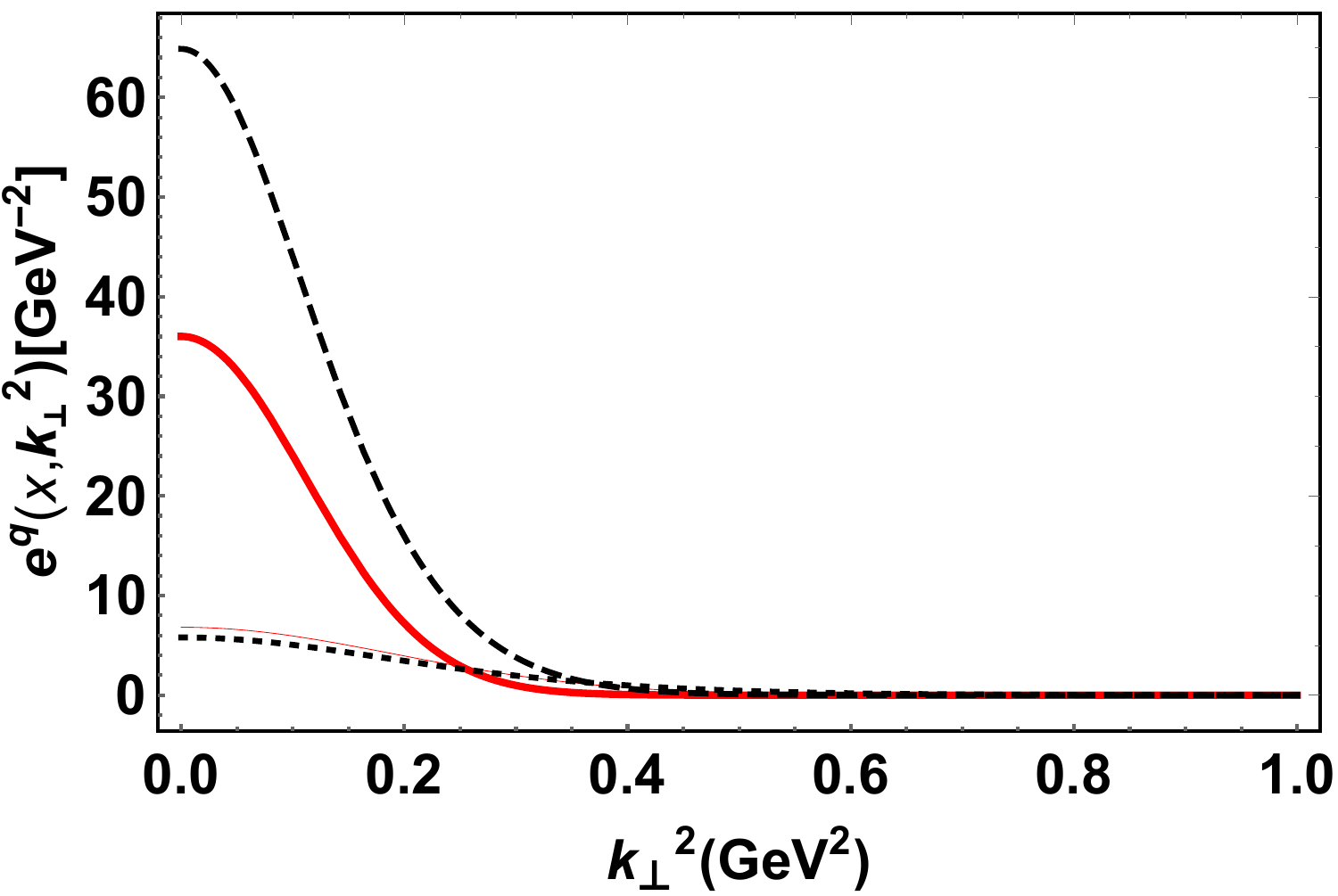}}%
\hfill \\
(c){\label{4figs-c47} \includegraphics[width=0.45\textwidth]{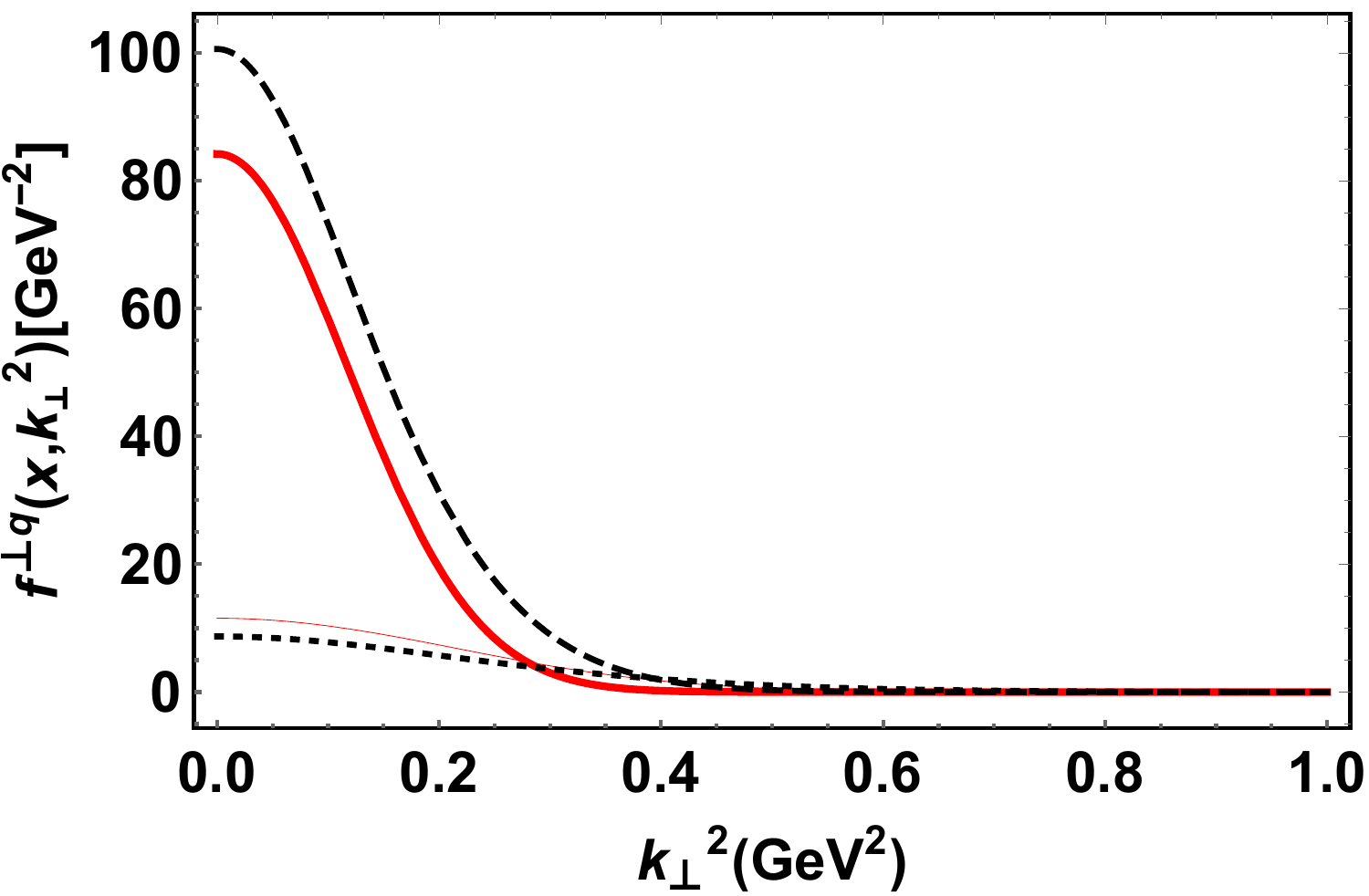}}%
\hfill
(d){\label{4figs-d48} \includegraphics[width=0.45\textwidth]{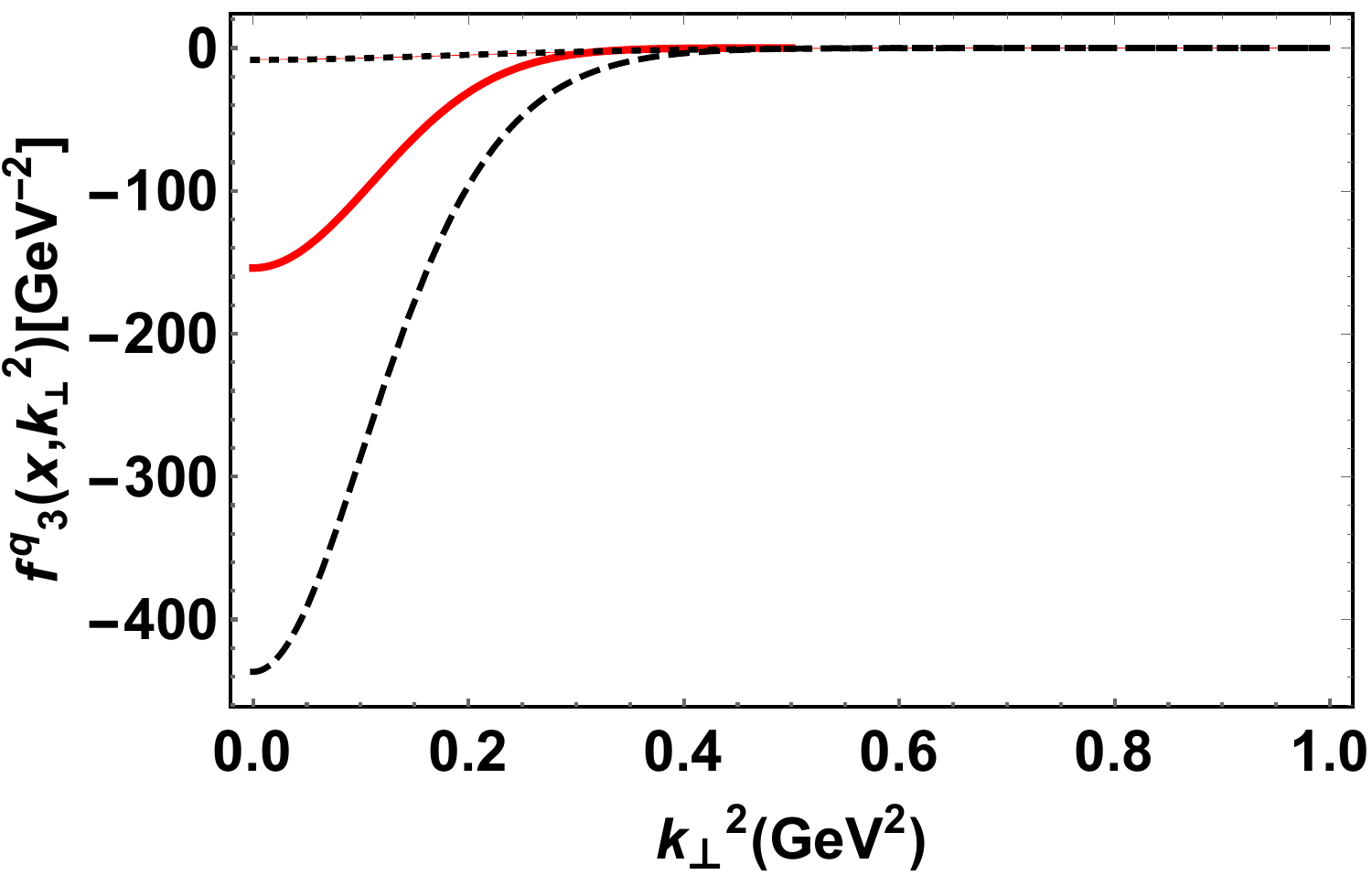}}%
\caption{Comparison of T-even TMDs with respect to $\bfk^2(GeV^2)$ in both the LCQM and LFHM for $\bar s$-quark of kaon. The solid thick and thin red lines are for the case in LFHM, while dashed and dotted black lines are in the LCQM.}
\label{4figs11}
\end{figure}
\begin{figure}
\centering
(a){\label{4figs-a53} \includegraphics[width=0.45\textwidth]{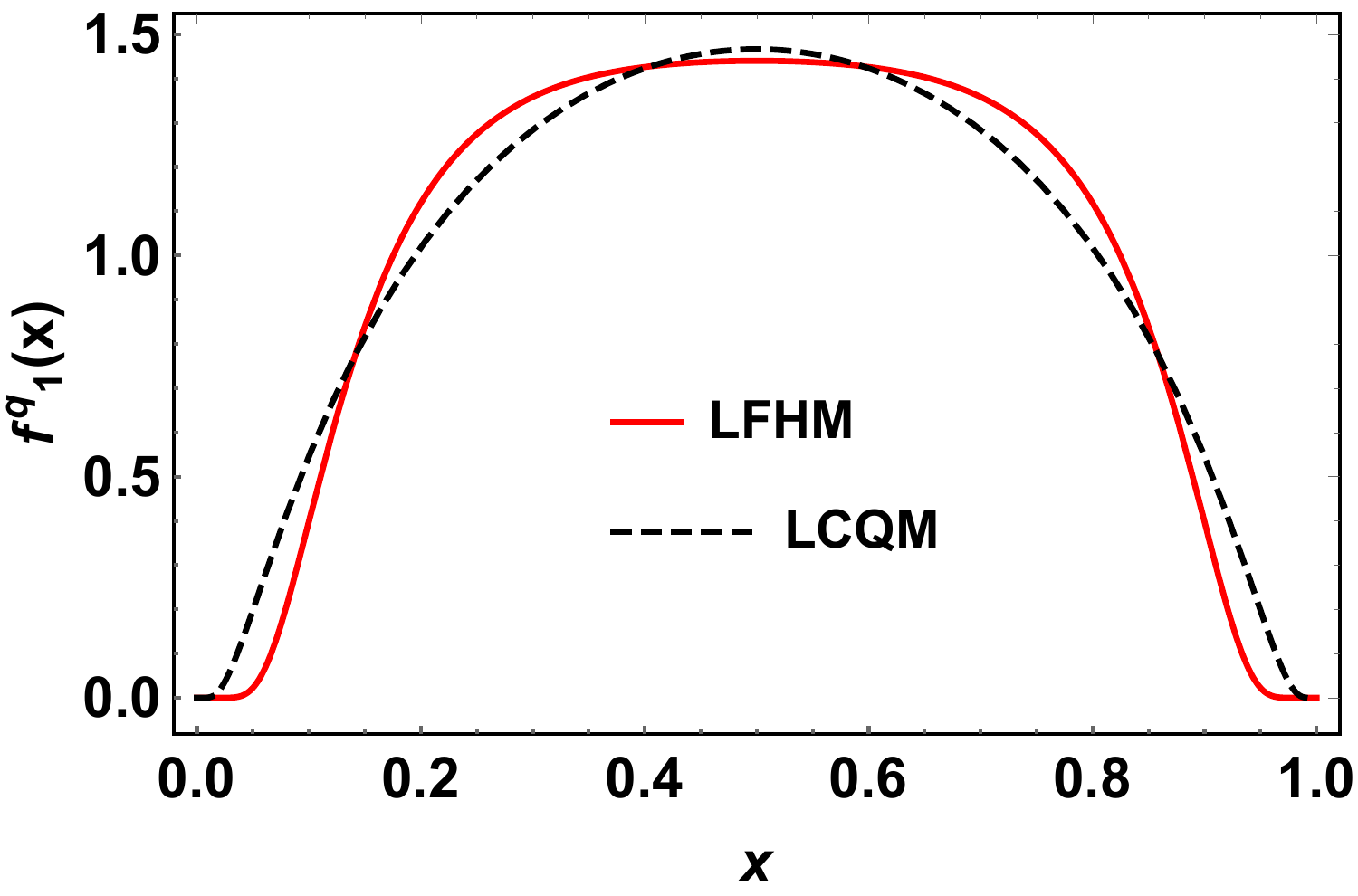}}
\hfill
(b){\label{4figs-b54} \includegraphics[width=0.45\textwidth]{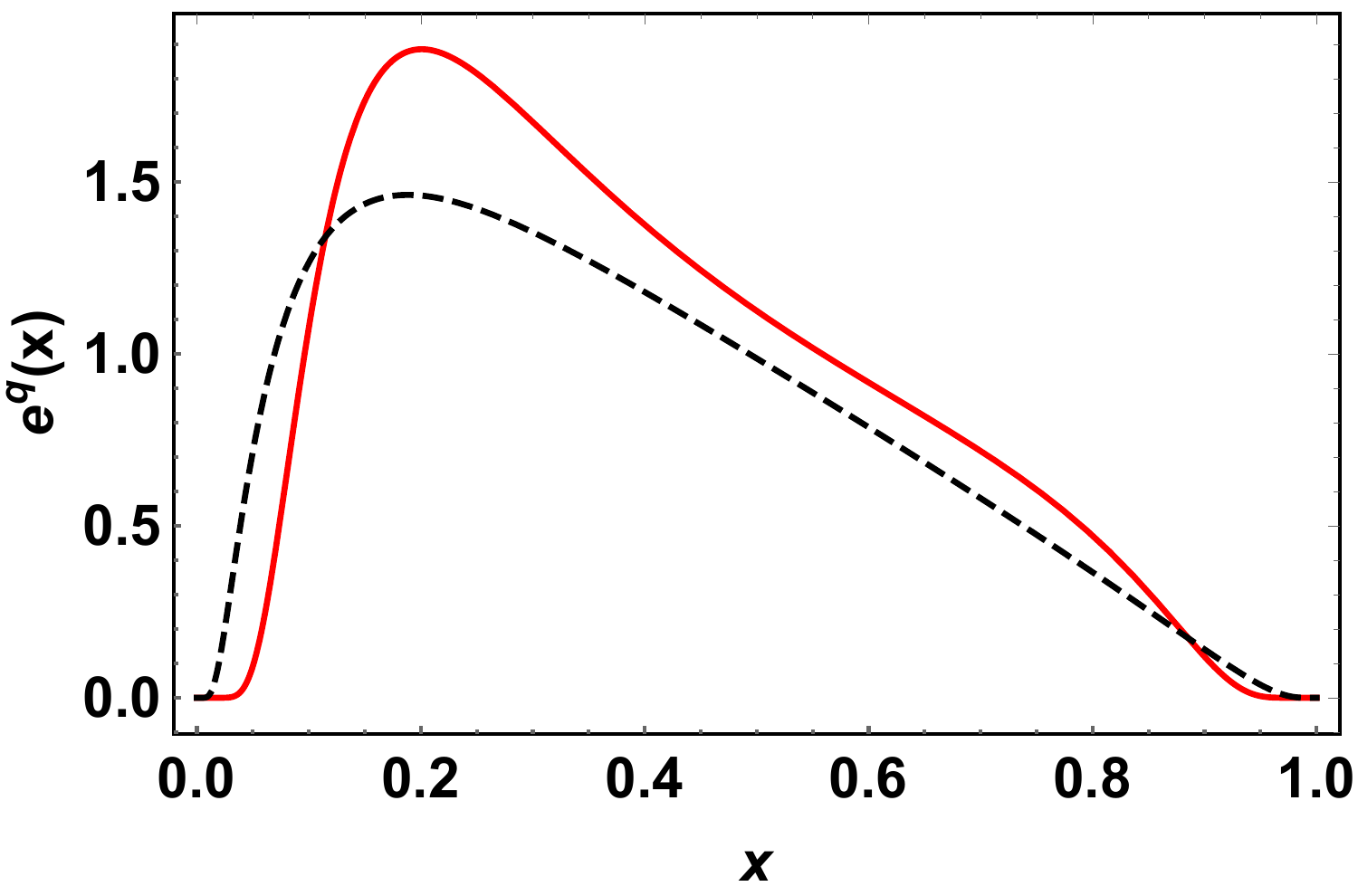}}%
\hfill \\
(c){\label{4figs-c55} \includegraphics[width=0.45\textwidth]{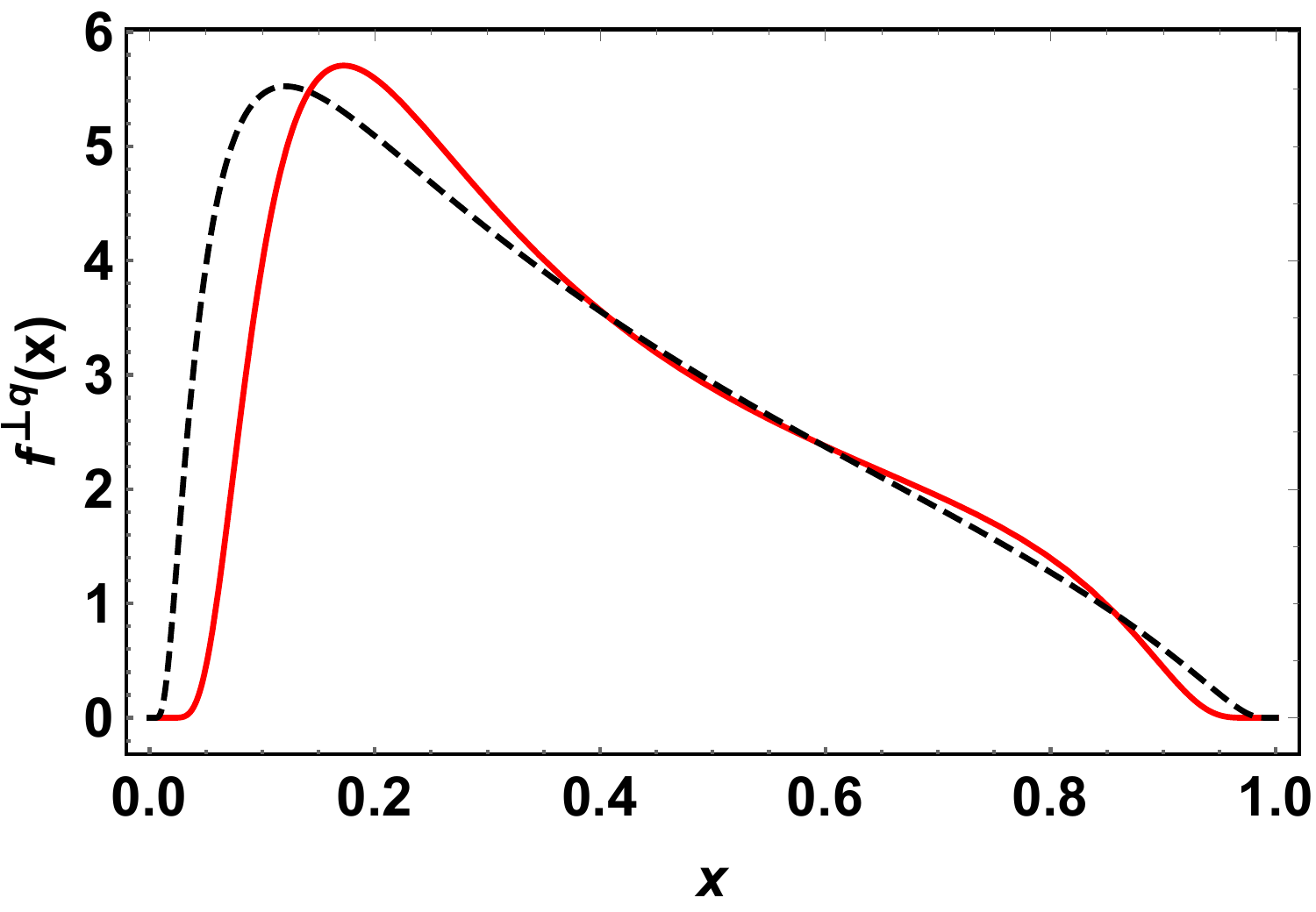}}%
\hfill
(d){\label{4figs-d56} \includegraphics[width=0.45\textwidth]{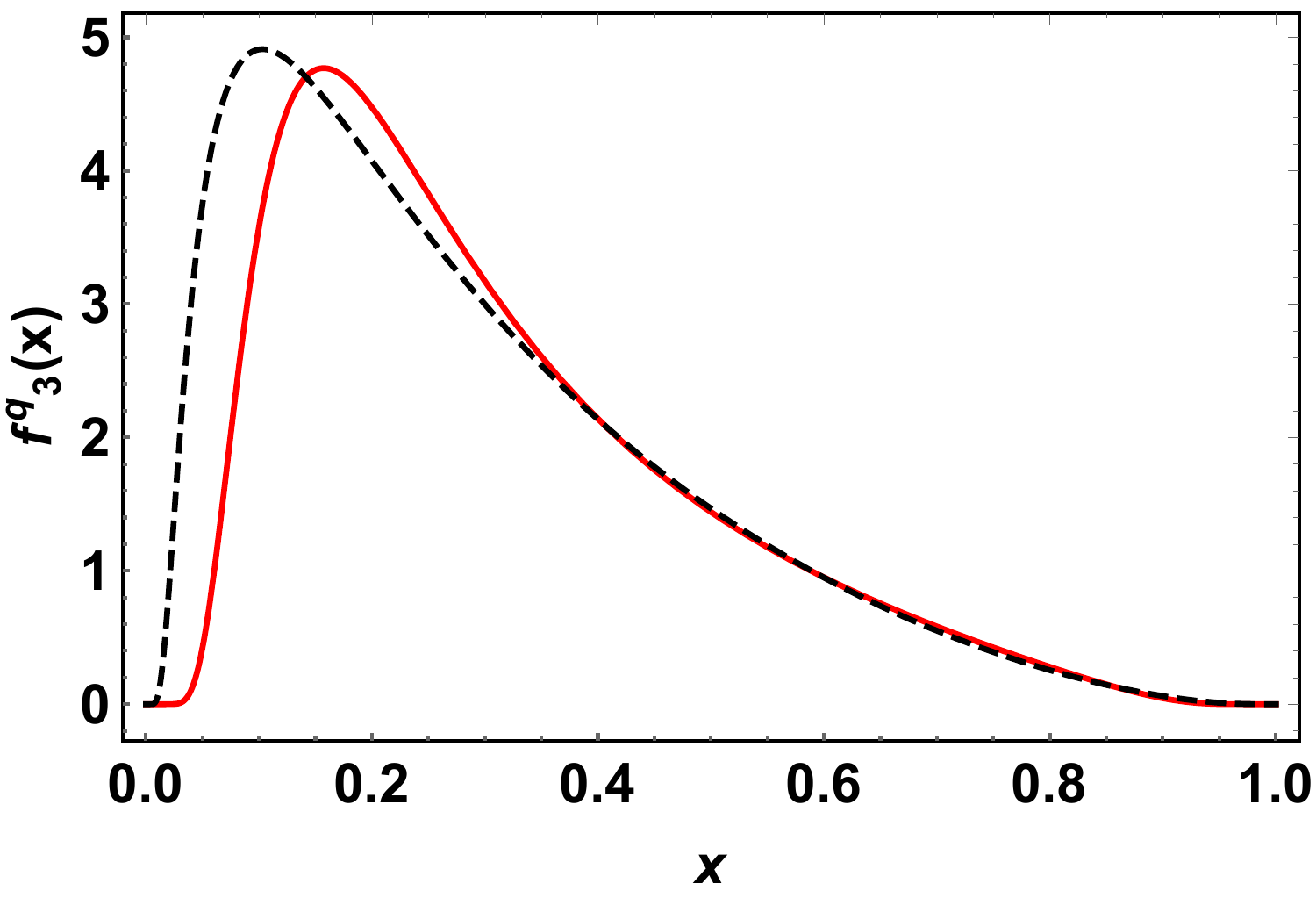}}%
\caption{Comparison of PDFs with respect to longitudinal momentum fraction $(x)$ in both the LCQM and LFHM for \textit{u}-quark of pion. The solid red line is for the case in LFHM and dashed black line is in the LCQM.}
\label{4figs12}
\end{figure}

\begin{figure}
\centering
(a){\label{4figs-a53} \includegraphics[width=0.45\textwidth]{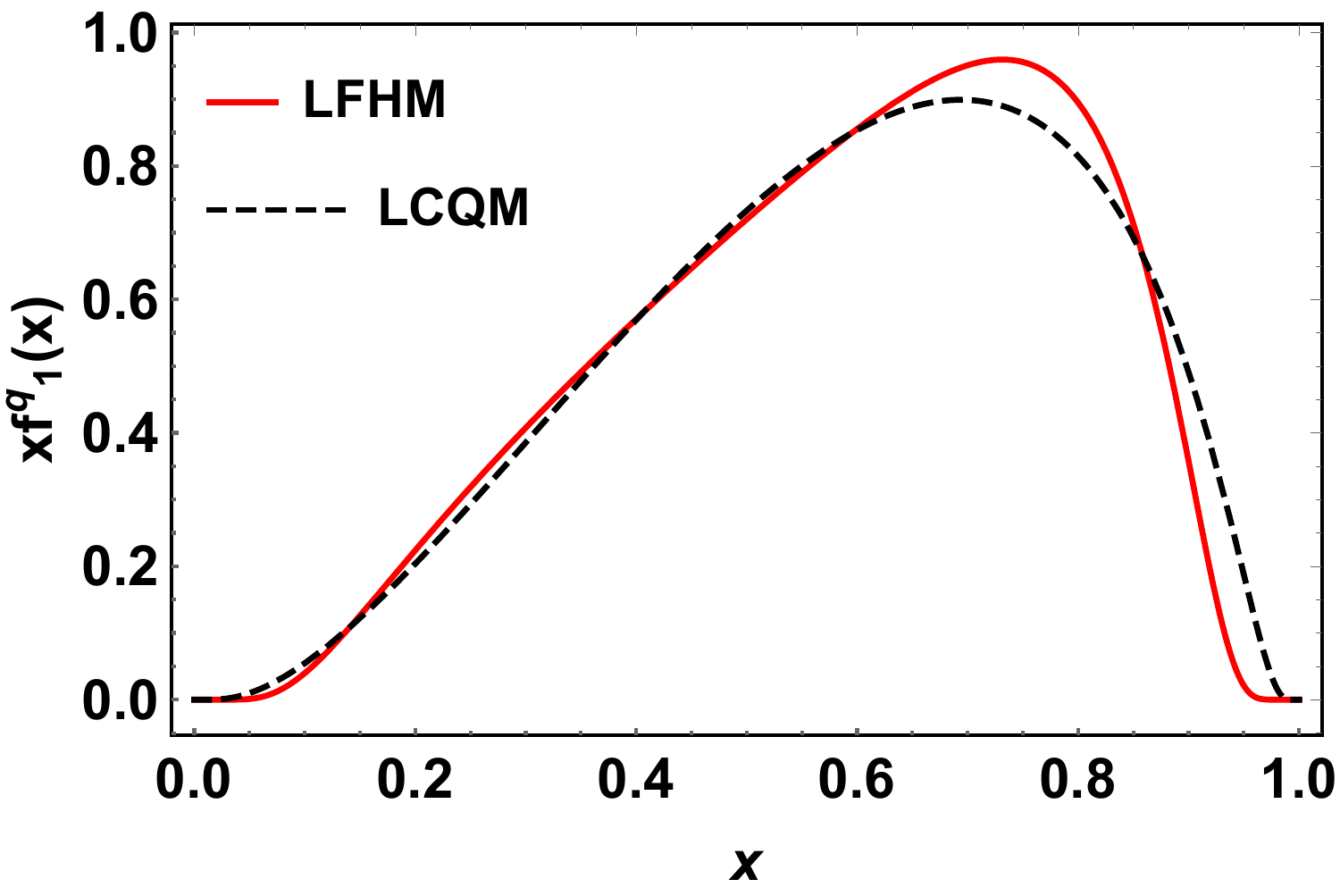}}
\hfill
(b){\label{4figs-b54} \includegraphics[width=0.45\textwidth]{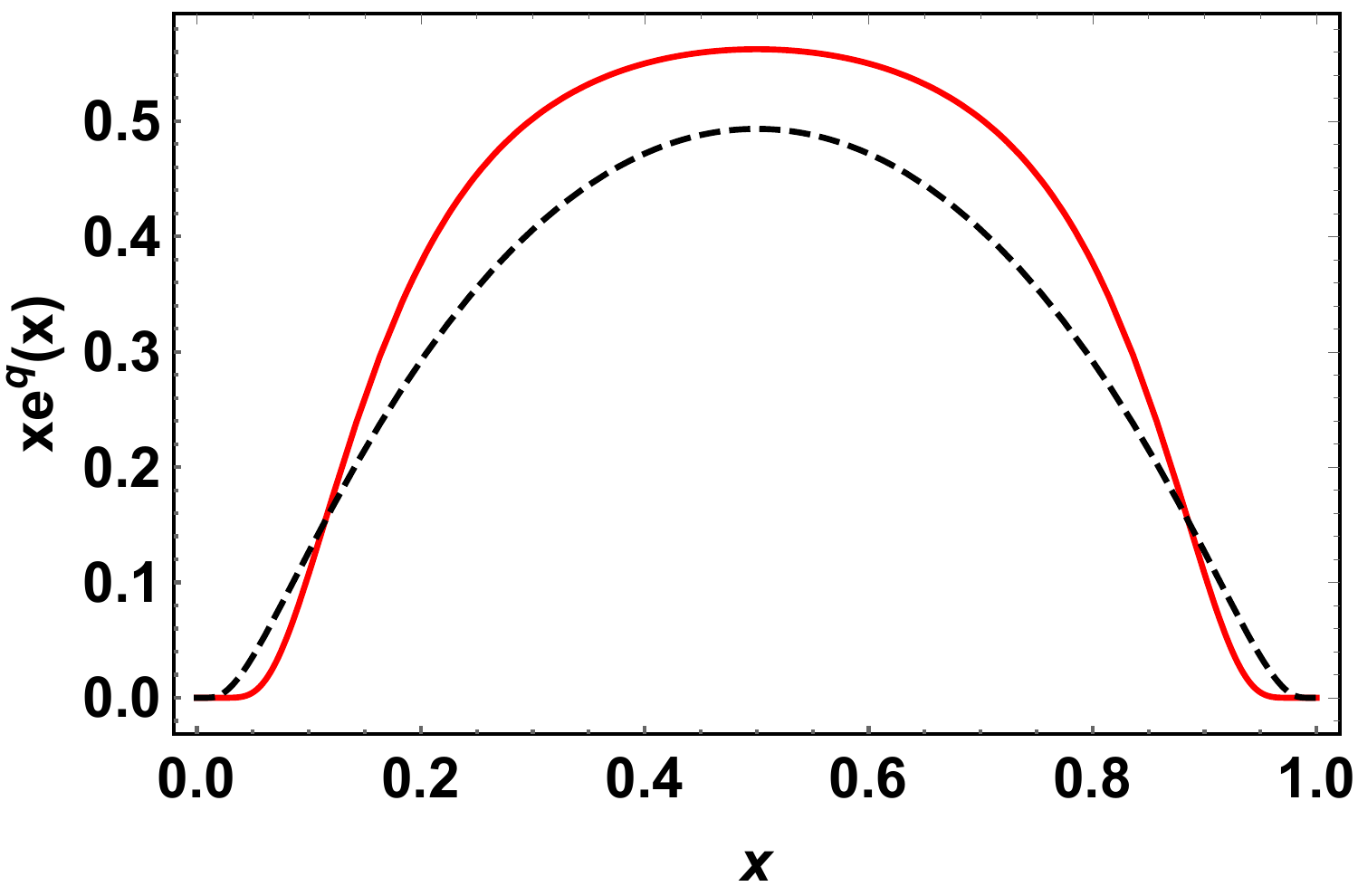}}%
\hfill \\
(c){\label{4figs-c55} \includegraphics[width=0.45\textwidth]{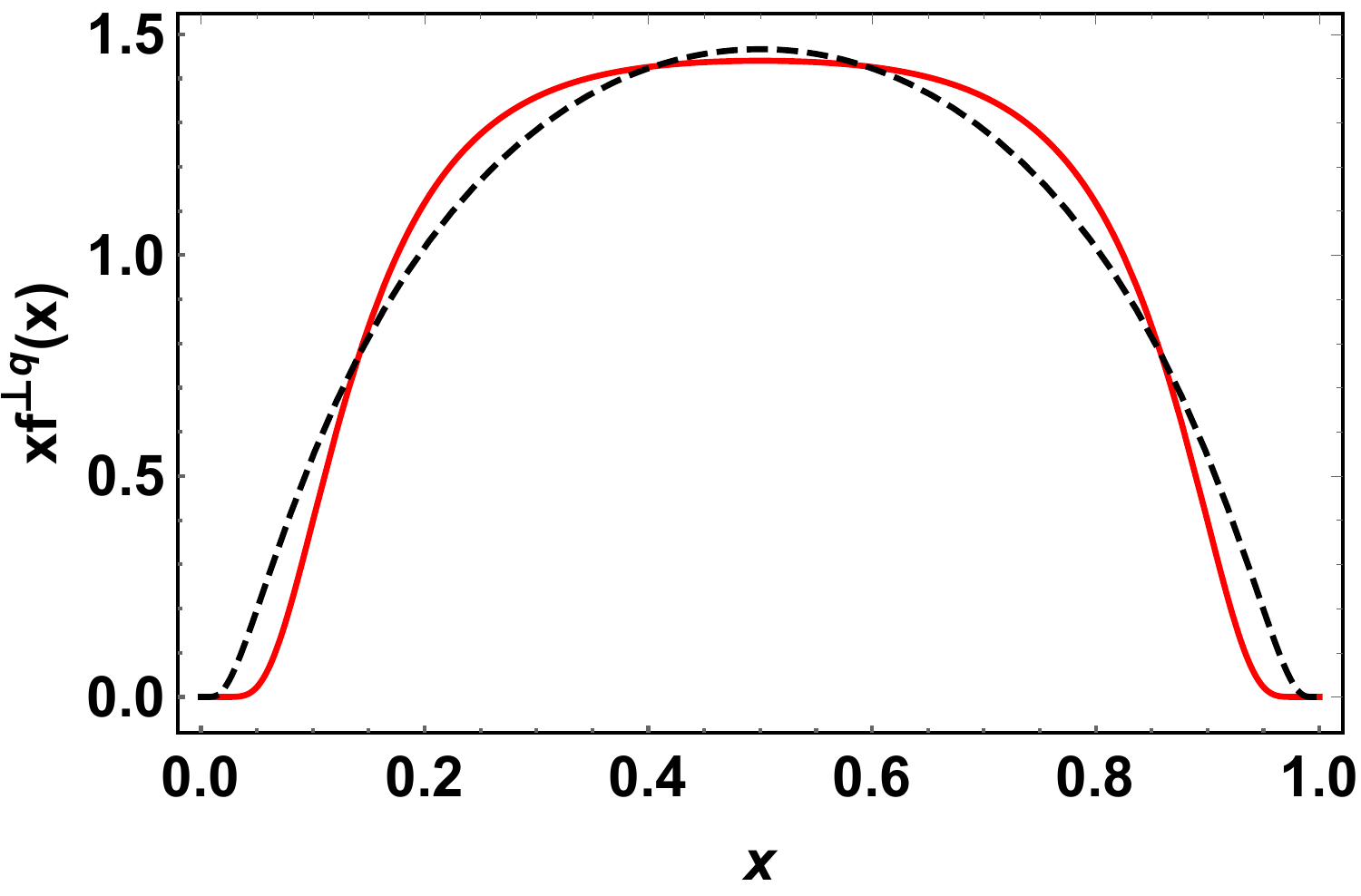}}%
\hfill
(d){\label{4figs-d56} \includegraphics[width=0.45\textwidth]{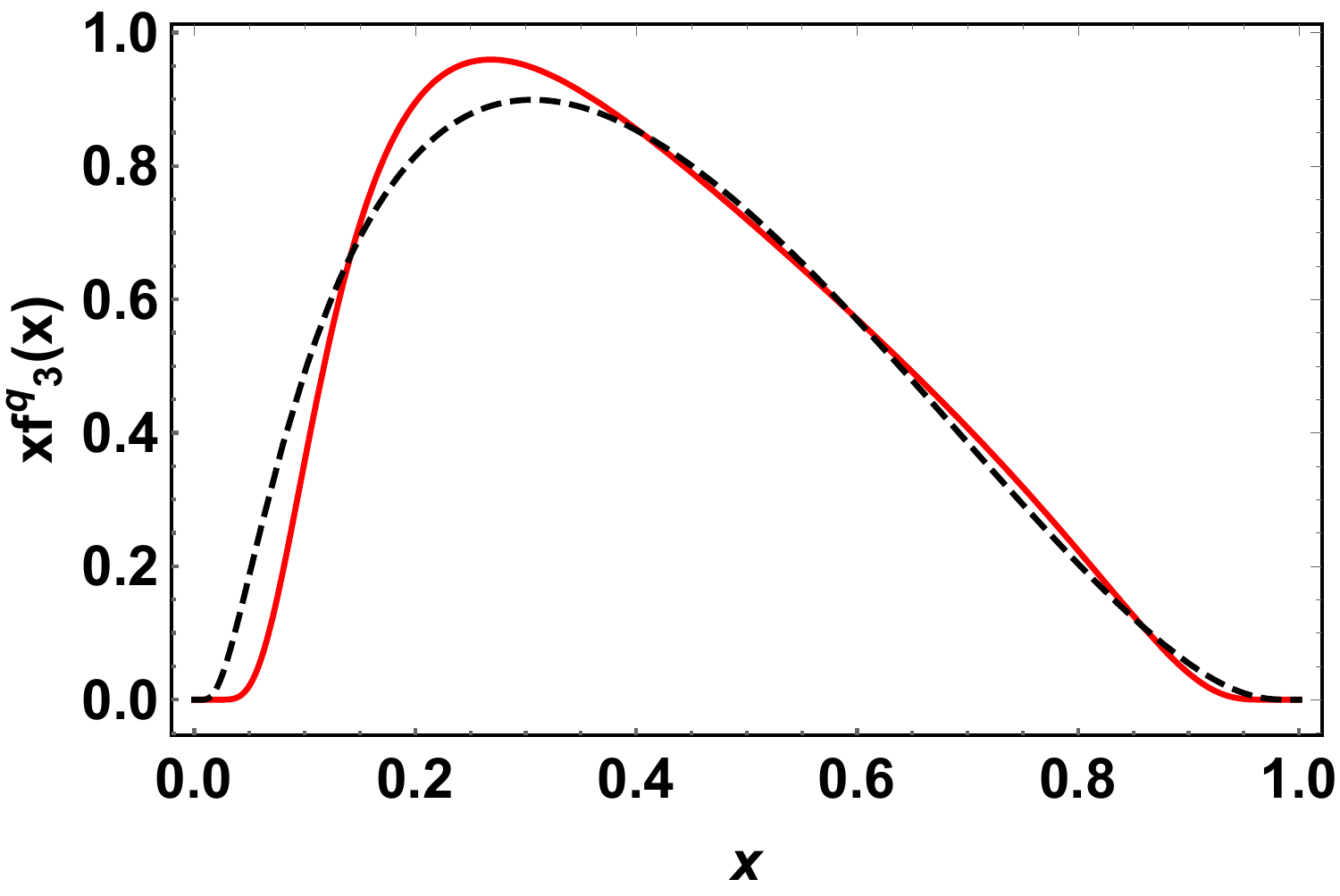}}%
\caption{Comparison of PDFs with respect to longitudinal momentum fraction $(x)$ in both the LCQM and LFHM for \textit{u}-quark of pion. The solid red line is for the case in LFHM and dashed black line is in the LCQM.}
\label{4figs13}
\end{figure}
\begin{figure}
\centering
(a){\label{4figs-a57} \includegraphics[width=0.45\textwidth]{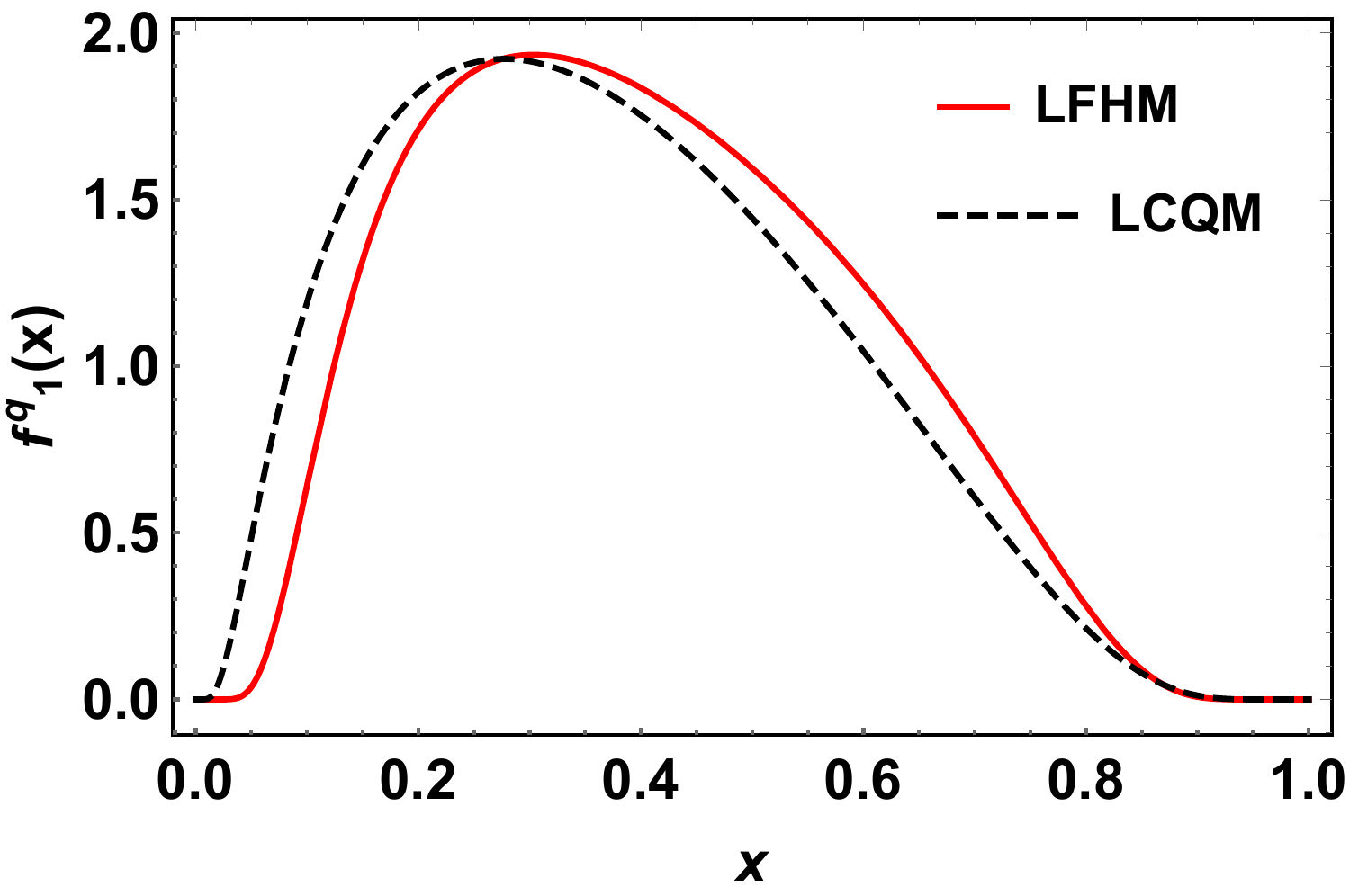}}
\hfill
(b){\label{4figs-b58} \includegraphics[width=0.45\textwidth]{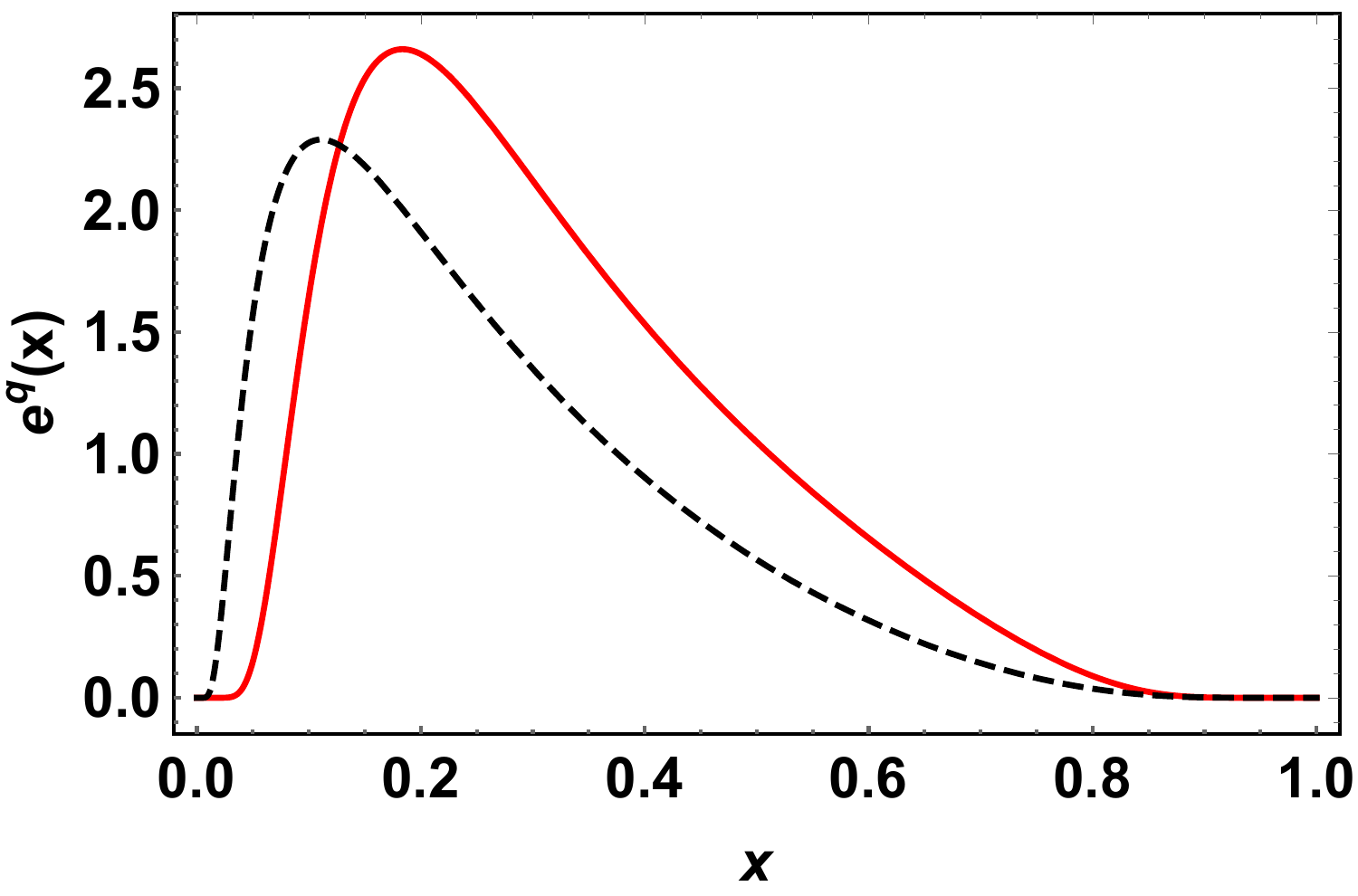}}%
\hfill \\
(c){\label{4figs-c59} \includegraphics[width=0.45\textwidth]{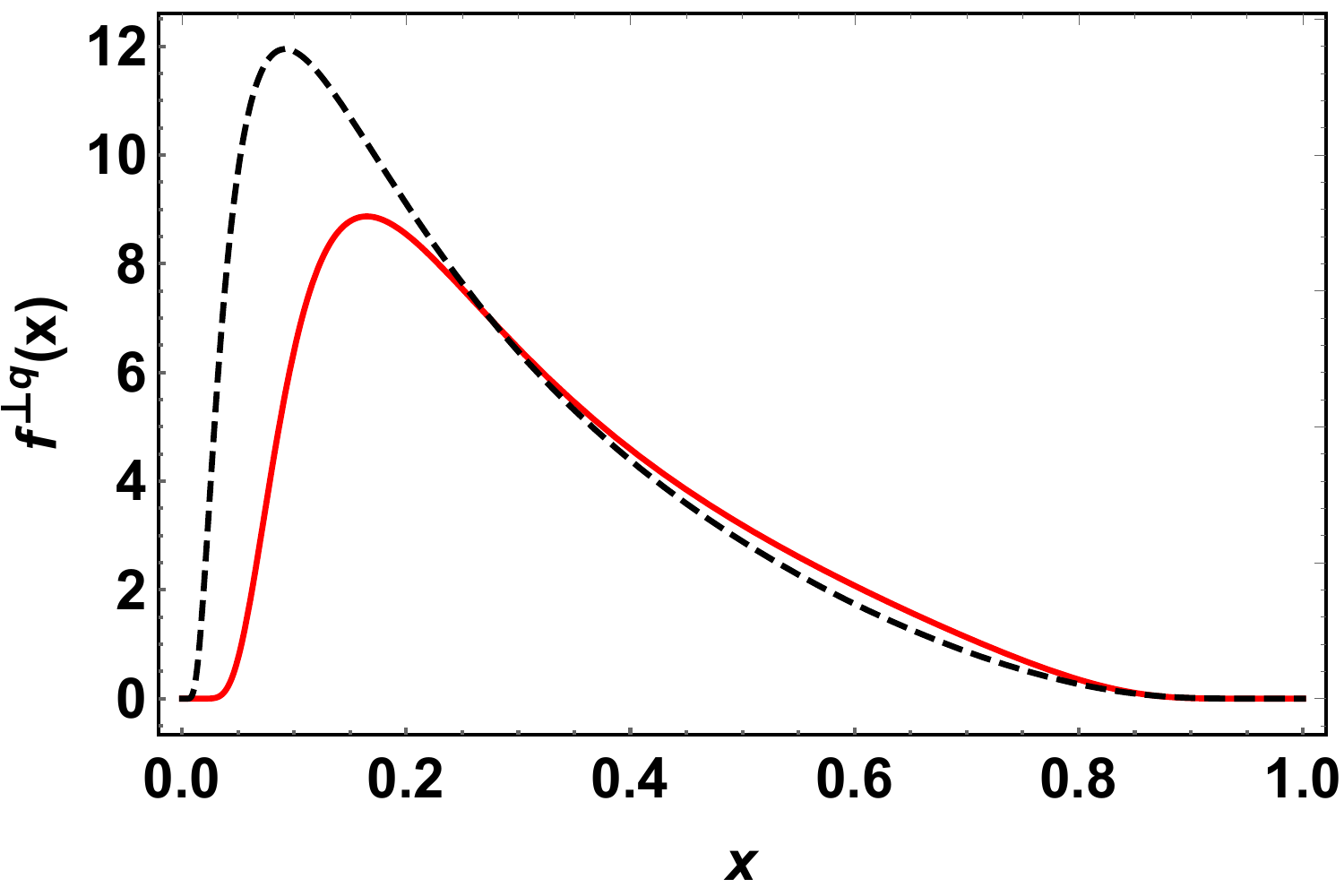}}%
\hfill
(d){\label{4figs-d60} \includegraphics[width=0.45\textwidth]{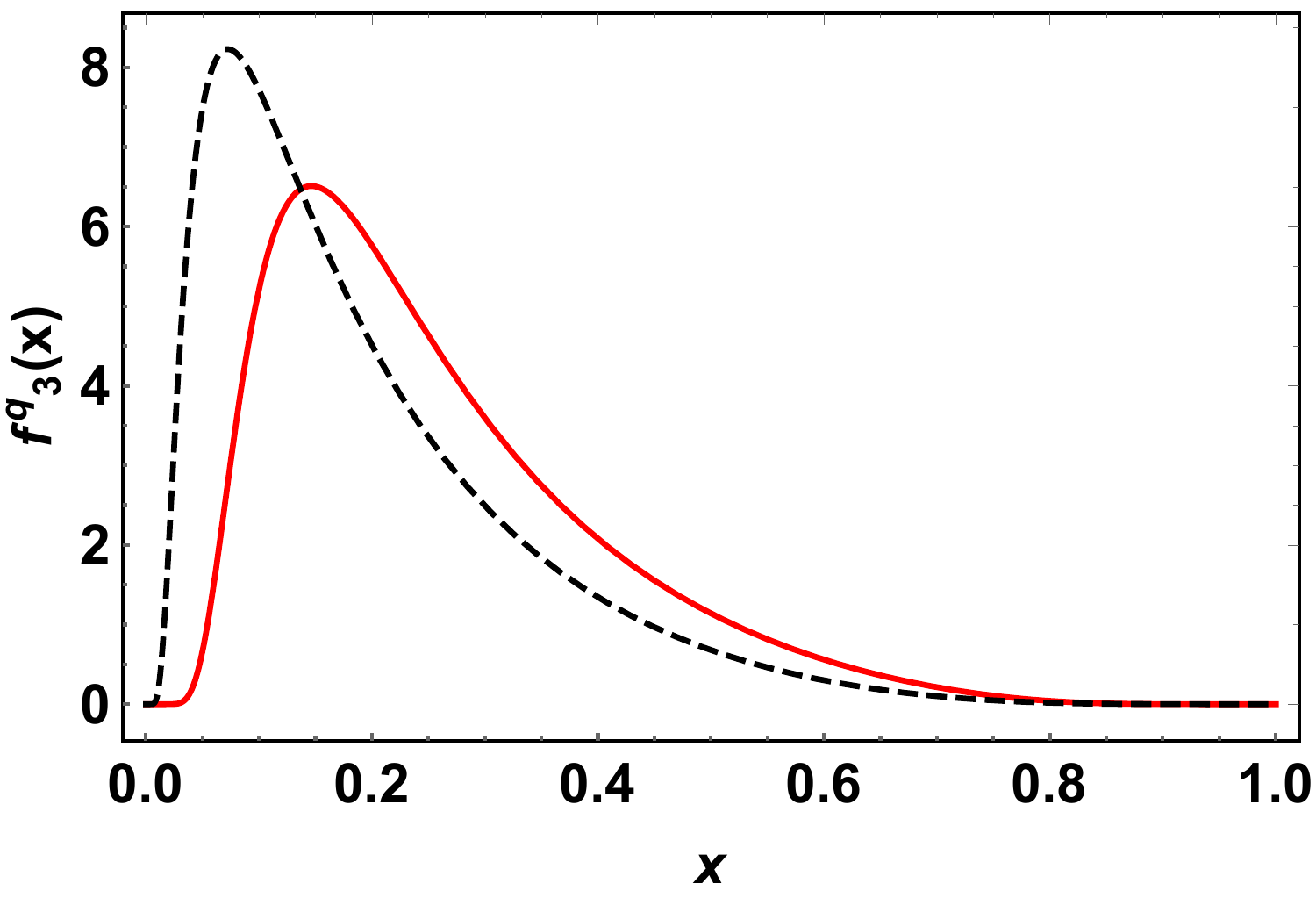}}%
\caption{Comparison of PDFs with respect to longitudinal momentum fraction $(x)$ in both the LCQM and LFHM for \textit{u}-quark of kaon. The solid red line is for the case in LFHM and dashed black line is in the LCQM.}
\label{4figs14}
\end{figure}

\begin{figure}
\centering
(a){\label{4figs-a61} \includegraphics[width=0.45\textwidth]{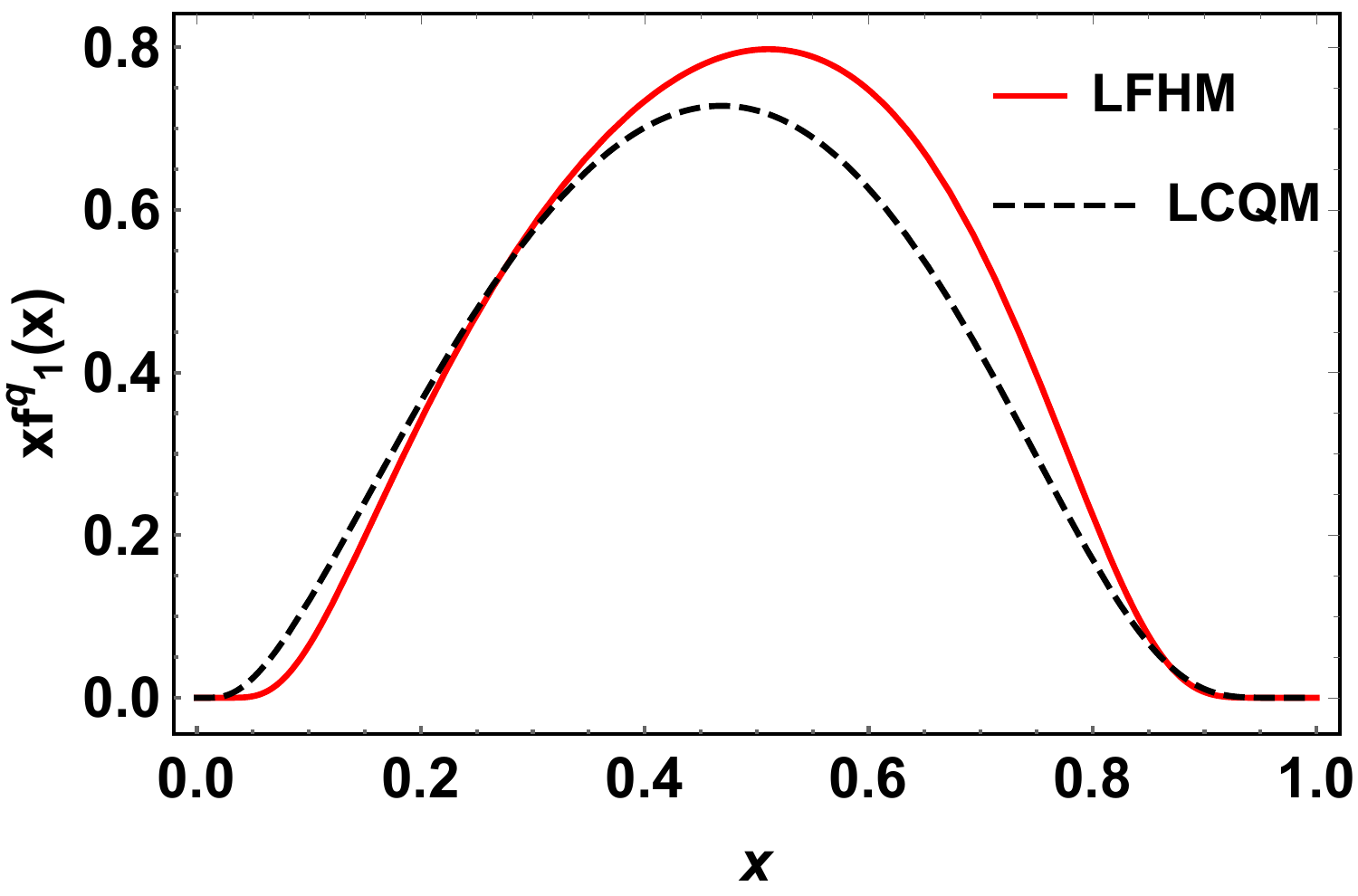}}
\hfill
(b){\label{4figs-b62} \includegraphics[width=0.45\textwidth]{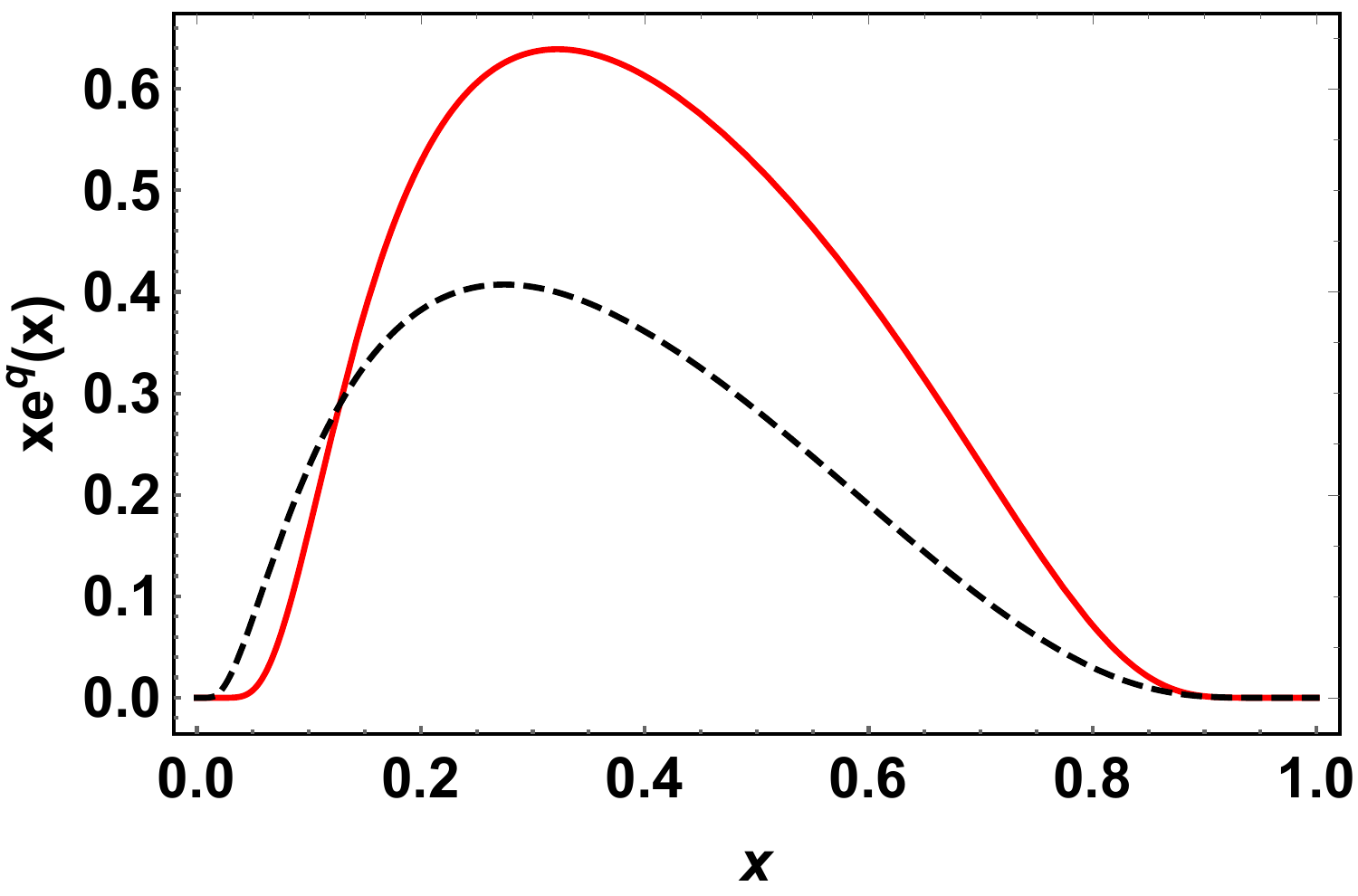}}%
\hfill \\
(c){\label{4figs-c63} \includegraphics[width=0.45\textwidth]{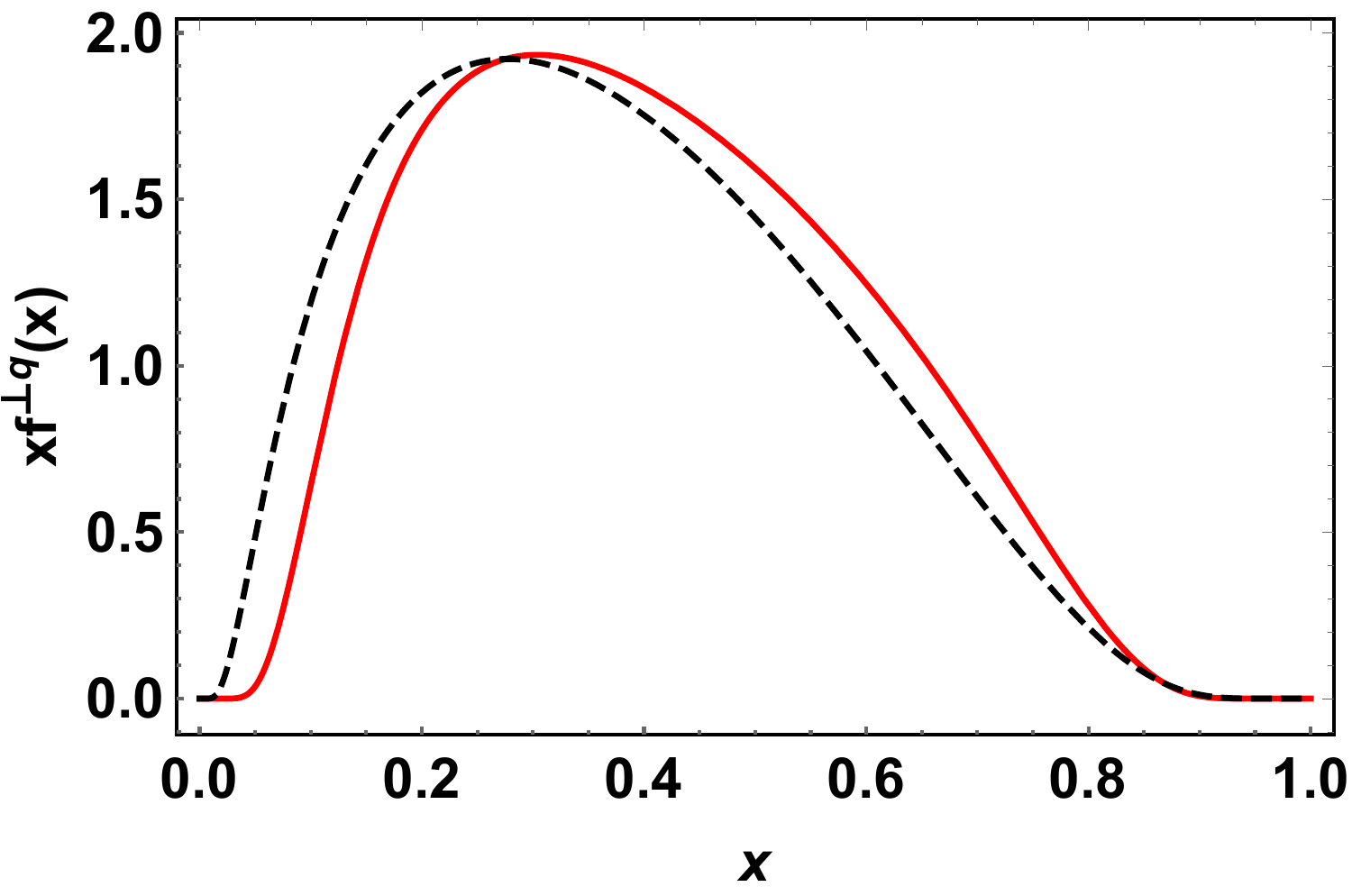}}%
\hfill
(d){\label{4figs-d64} \includegraphics[width=0.45\textwidth]{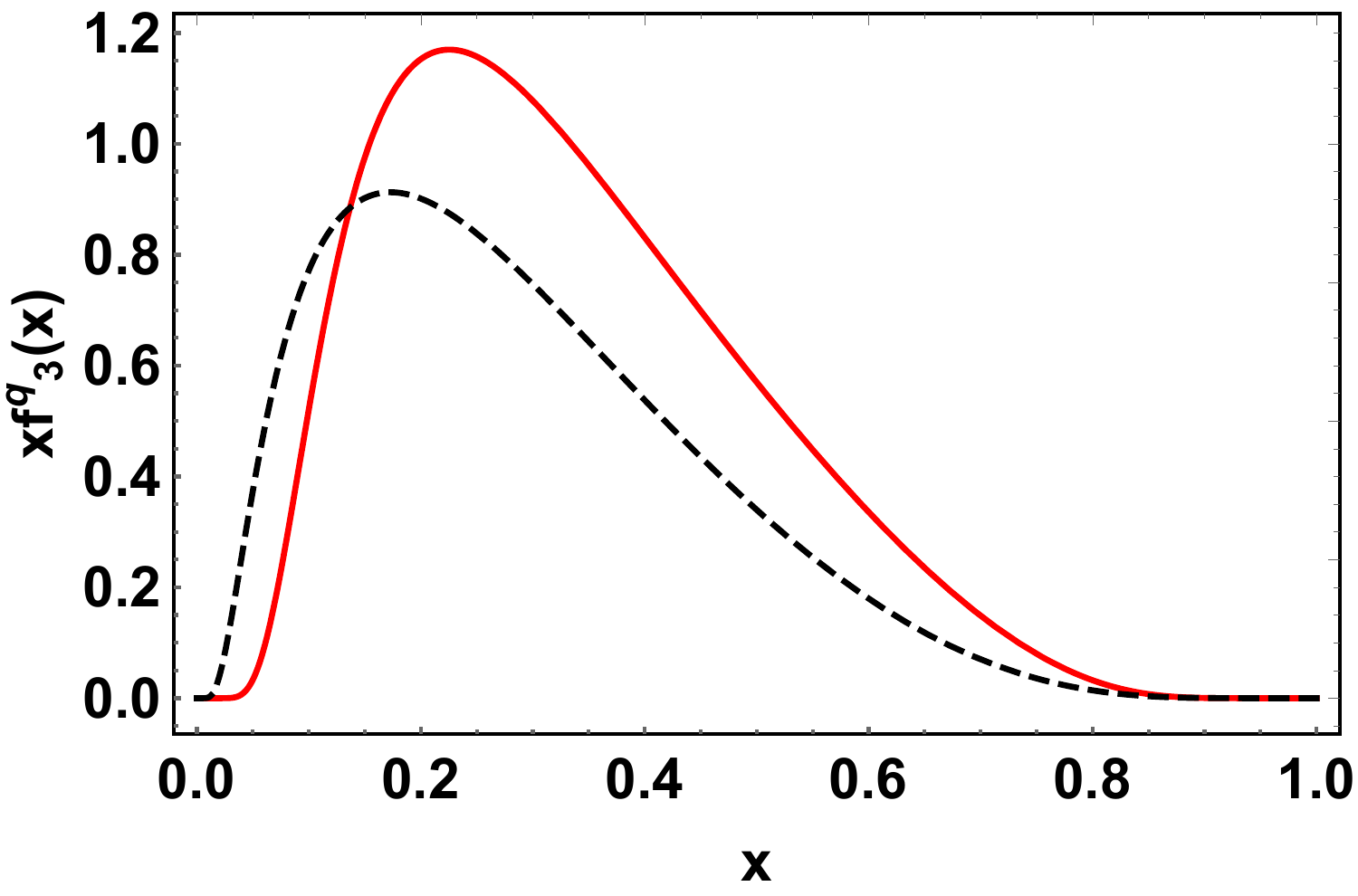}}%
\caption{Comparison of PDFs with respect to longitudinal momentum fraction $(x)$ in both the LCQM and LFHM for $\bar s$- quark of kaon. The solid red line is for the case in LFHM and dashed black line is in the LCQM.}
\label{4figs15}
\end{figure}

\begin{figure}
\centering
(a){\label{4figs-a65} \includegraphics[width=0.45\textwidth]{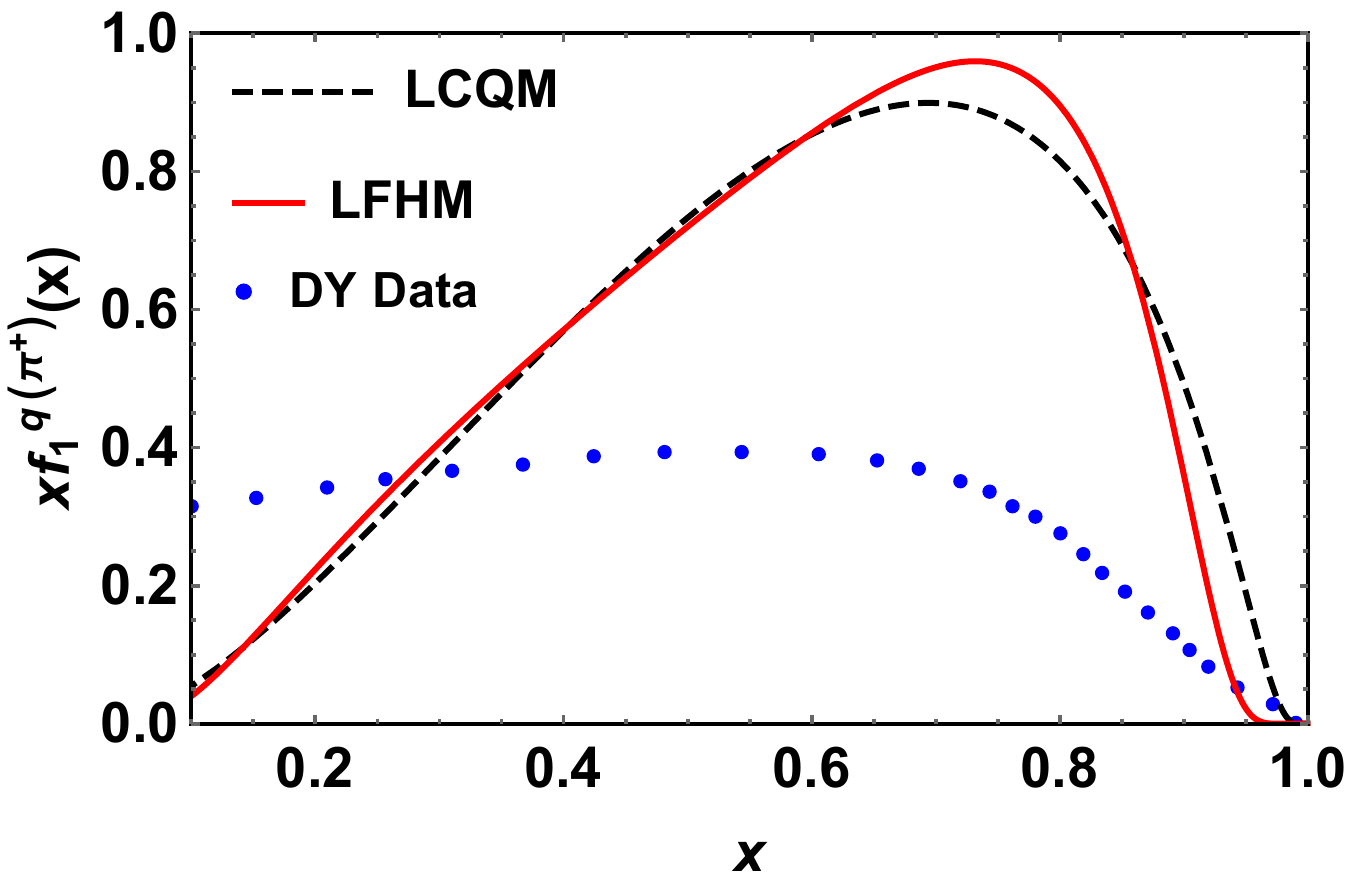}}
\hfill
(b){\label{4figs-b66} \includegraphics[width=0.45\textwidth]{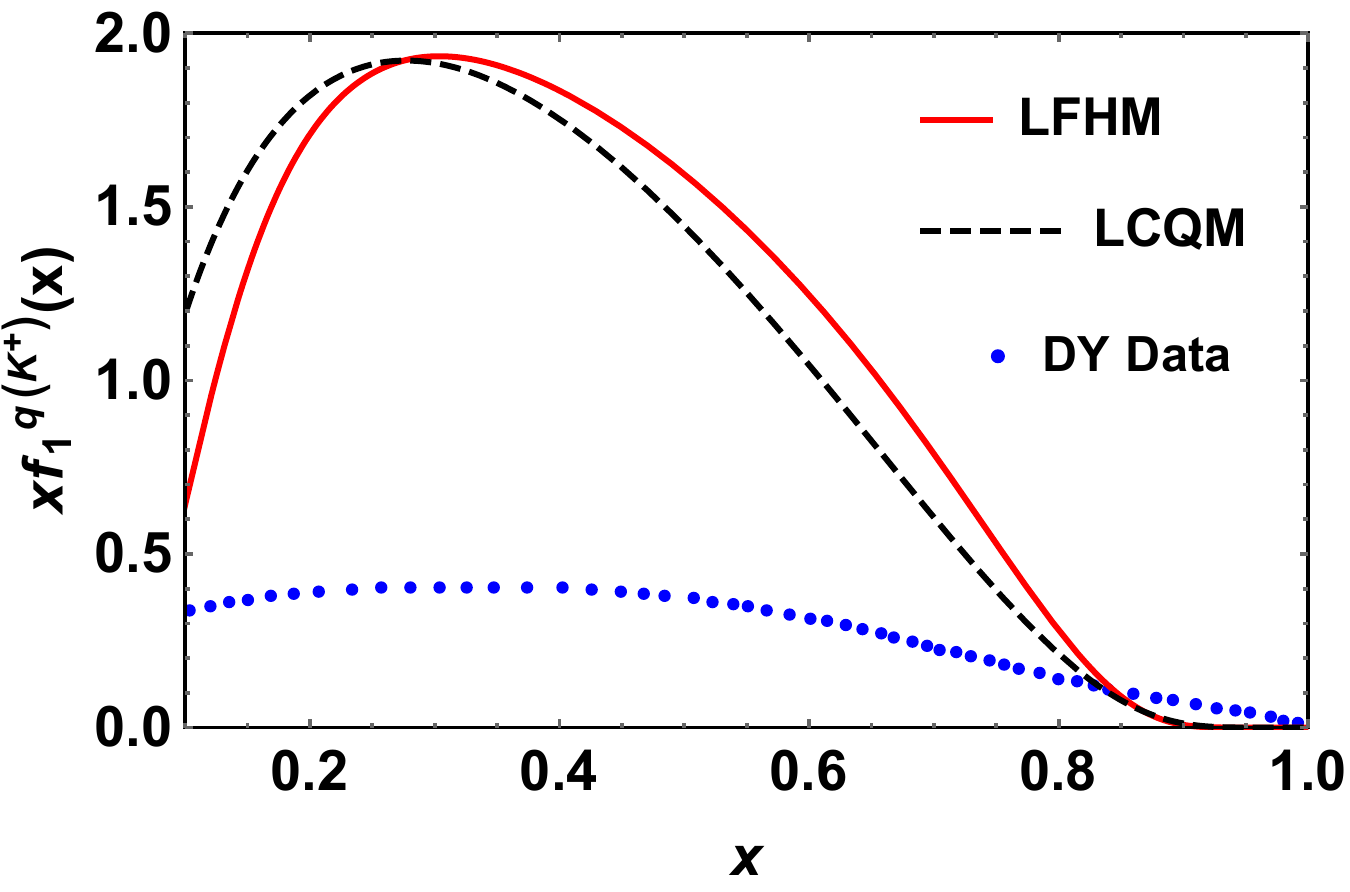}}%
\hfill\\
(c){\label{4figs-c67} \includegraphics[width=0.45\textwidth]{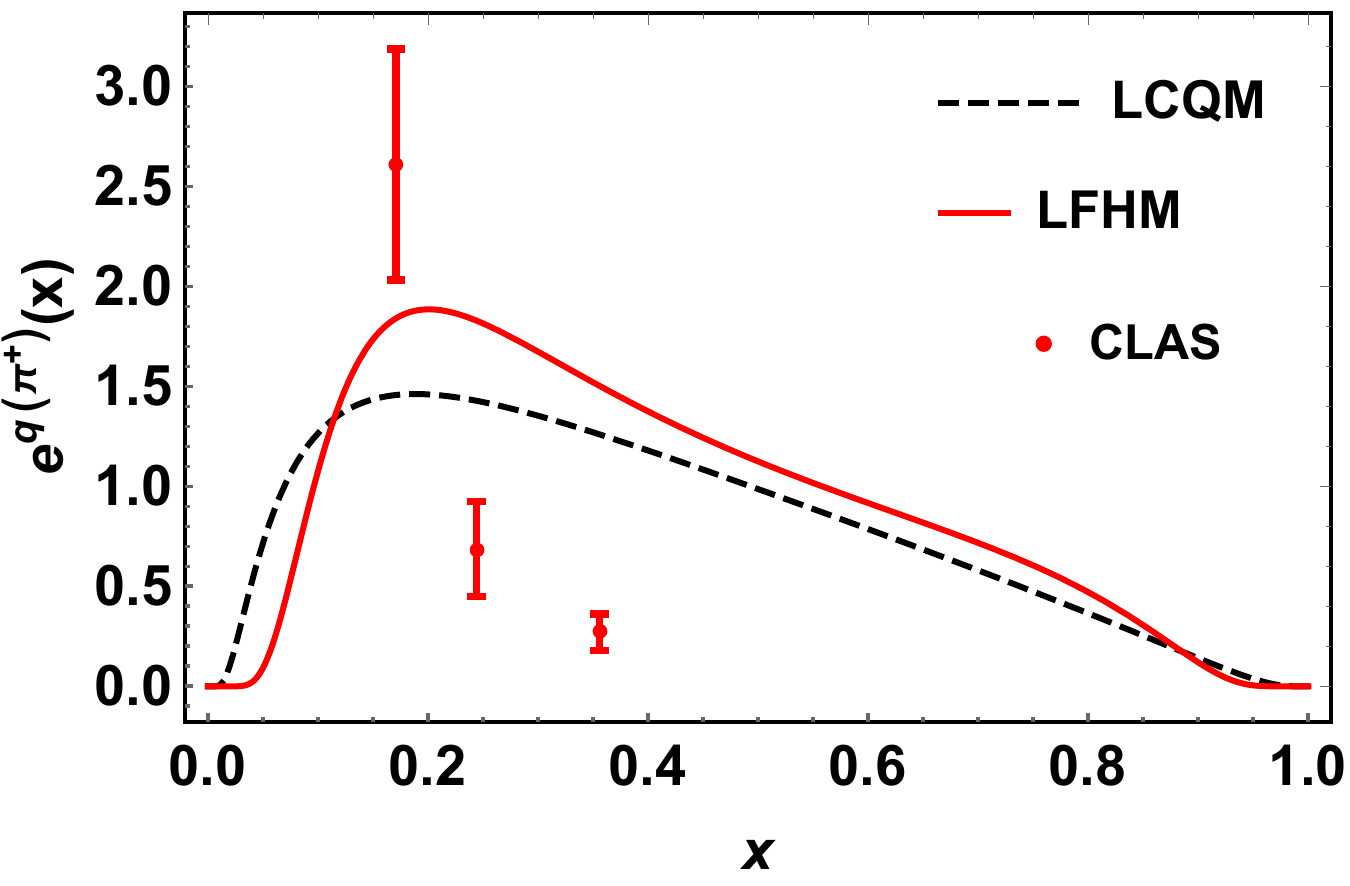}}%
\hfill
(d){\label{4figs-d68} \includegraphics[width=0.45\textwidth]{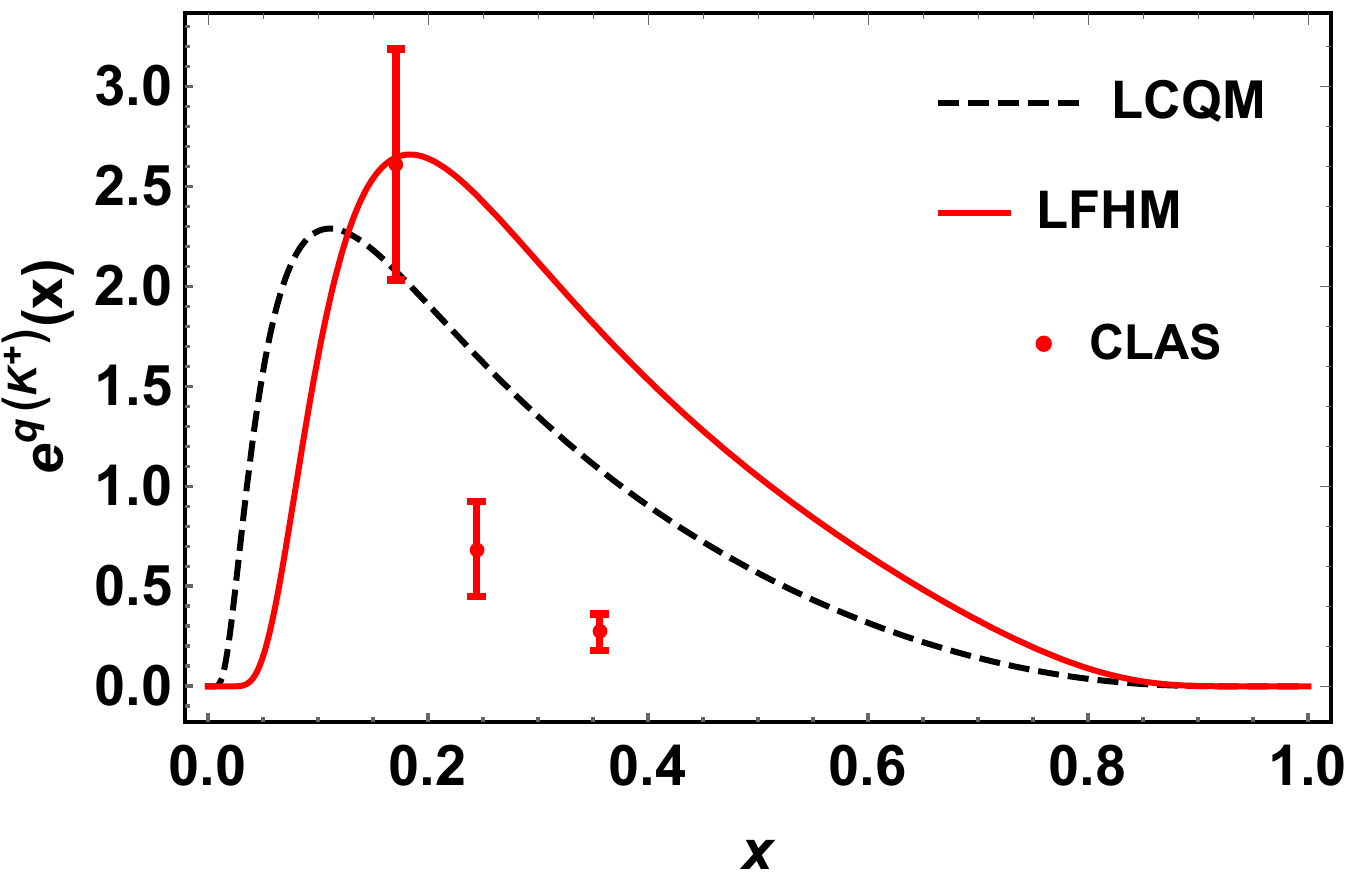}}%
\caption{PDFs comparison of our models with available pion-induced Drell-Yan, $J/\psi$ production \cite{Bourrely:2023yzi} and CLAS data \cite{Courtoy:2014ixa}.The $xf^q_1(x)$ PDF in the left two figures are compared with DY data. The right two figures are $e^q(x)$ plots compared with available nucleon CLAS data.}
\label{4figs16}
\end{figure}

\begin{eqnarray}
|M(P)\rangle_{L_z=0}  &=&
\int \frac{d^2\bfk}{16\pi^3} \, \frac{dx}{\sqrt{6 x (1-x)}} \,
\Psi_{\pi(K)}^{(0)}(x, \bfk) \, \sum_{a=1}^{3} \,
\Big[b^{\dagger a}_{\uparrow}(1) d^{\dagger a}_{\downarrow}(2)
-    b^{\dagger a}_{\downarrow}(1)
     d^{\dagger a}_{\uparrow}(2)\Big] \, |0\rangle, \\
|M(P)\rangle_{|L_z|=1}   &=&
\int \frac{d^2\bfk}{16\pi^3} \, \frac{dx}{\sqrt{2 x (1-x)}} \,
\Psi_{\pi(K)}^{(1)}(x, \bfk) \nonumber\\
&\times&  \, \Big(\frac{\bfk^+}{\sqrt{3}} \, \sum_{a=1}^{3} \,
     b^{\dagger a}_{\uparrow}(1)
     d^{\dagger a}_{\uparrow}(2) \, |0\rangle\,
+  \frac{\bfk^-}{\sqrt{3}} \, \sum_{a=1}^{3} \,
    b^{\dagger a}_{\downarrow}(1)
     d^{\dagger a}_{\downarrow}(2)  \, |0\rangle\,\Big),
\end{eqnarray}
where $(1)=(x, \textbf{k}_\perp)$ and $(2)=((1-x), - \textbf{k}_\perp)$ are the quark and antiquark coordinates. $\Psi_{\pi(K)}^{(0)}(x, \bfk)$ and $\Psi_{\pi(K)}^{(1)}(x, \bfk)$ are the pion (kaon) LF wave functions having the QOAM $|L_z|=0$ and $|L_z|=1$ respectively.   $b^{\dagger a}_\lambda(b^a_\lambda)$ and $d^{\dagger a}_\lambda(d^a_\lambda)$ are respectively the quark and anti-quark creation (anhilation) operators. Subscript $\lambda$ and superscript $a$ is used to denote helicity and color respectively. The $\bfk^{\pm}$ is the shorthand notation for $\bfk^\pm = \textbf{k}_{\perp x} \pm i \textbf{k}_{\perp y}$. The LF wave function $\Psi_{\pi(K)}^{(L_z)}(x, \bfk)$ is expressed as a product of wave functions of spin wave and momentum space as \cite{Kaur:2019jfa}
\begin{eqnarray}
    \Psi_{\pi(K)}^{(0)}(x,\bfk)&=& - \dfrac{m_{q(\bar q)} + M_{\pi(K)}x(1-x)}{\sqrt{2}x(1-x)} \, \psi^{(2)}_{\pi(K)}(x,\bfk),
\end{eqnarray}
\begin{eqnarray}
\Psi_{\pi(K)}^{(1)}(x,\bfk)&=&  - \dfrac{1}{\sqrt{2}x(1-x)} \, \psi^{(2)}_{\pi(K)}(x,\bfk).\label{space1}
\end{eqnarray}
The superscript $2$ in momentum wave function $\psi_{\pi(K)}(x,\bfk)$ in the above equations is used to distinguish LFHM from LCQM. The momentum space wave function $\psi^{(2)}_{\pi(K)}(x,\bfk)$, for $\pi$ and $K$ can be expressed as
\begin{eqnarray}
     \psi^{(2)}_{\pi} (x,\bfk) = \frac{4 \pi N_{\pi}}{\kappa_{\pi} \sqrt{x (1-x)}}   \exp{ \left[ -\frac{\bfk^2 + m^2}{2 \kappa^2_{\pi} x(1-x)} \right]}, 
 \label{LFHMpion}
\end{eqnarray}
\begin{eqnarray}
     \psi^{(2)}_{K} (x,\bfk) = \frac{4 \pi N_{K}}{\kappa_{K} \sqrt{x (1-x)}}   \exp{ \left[ -\frac{(\bfk^2 +(1-x) m^2_{q}+ x m^2_{\bar q}) }{2 \kappa^2_{K} x(1-x)} \right]}. 
 \label{LFHMkaon}
\end{eqnarray}
$N_{\pi (K)}$ and $\kappa_{\pi (K)}$ are respectively the normalization constants and model scale parameters for pion (kaon) wave function. The AdS/QCD scale $\kappa_{\pi (K)}$, which fixes the holographic Gaussian width, is extracted from mass spectroscopic data through Regge trajectory. For pion, $\kappa_{\pi}= 0.523$ GeV \cite{Kaur:2019jfa,mohmmad} and for kaon, $\kappa_{K}= 0.524$ GeV \cite{Vega:2009zb}.
\par The LF wave functions of unpolarized TMD is found to be 
\begin{eqnarray}
f_{1}^{q(\pi(K))}(x,\bfk^{2})&=& \dfrac{1}{(32 \pi^3)}\big[ \vert \Psi^{(0)}_\pi(x,\bfk)\vert^{2}+\bfk^{2} \, \vert\Psi^{(1)}_\pi(x,\bfk)\vert^{2}
\big].
\end{eqnarray}
The T-even TMDs for pion in LFHM are given by using Eqs. (\ref{Eq:eom-e1}), (\ref{Eq:eom-fperp1}) and (\ref{Eq:eom-f41})
\begin{eqnarray}
    f_{1}^{q(\pi)}(x,\bfk^{2})&=& \frac{N_{\pi}^{2}}{4 \pi }\, \frac{[\bfk^2+(m+x(1-x)M_{\pi})^2]}{\kappa_{\pi}^2 \, x^3 \,(1-x)^3 }\, \exp\bigg[-\frac{\bfk^2 + m^2)}{\kappa^2_{\pi} x (1-x)}\bigg],
\label{result-of-T-even-tmd}\\
e^{q(\pi)}(x,\bfk^{2})&=& \frac{N_{\pi}^{2}}{4 \pi }\, \frac{m [\bfk^2+(m+x(1-x)M_{\pi})^2]}{M_{\pi}\kappa_{\pi}^2 \, x^4 \,(1-x)^3 }\, \exp\bigg[-\frac{\bfk^2 + m^2)}{\kappa^2_{\pi} x (1-x)}\bigg],
\label{eqLFHM}\\
 f^{\perp q(\pi)}(x,\bfk^{2})&=& \frac{N_{\pi}^{2}}{4 \pi }\, \frac{[\bfk^2+(m+x(1-x)M_{\pi})^2]}{\kappa_{\pi}^2 \, x^4 \,(1-x)^3 }\, \exp\bigg[-\frac{\bfk^2 + m^2)}{\kappa^2_{\pi} x (1-x)}\bigg],
 \label{fperpLFHM}\\
  f_{3}^{q(\pi)}(x,\bfk^{2})&=& \frac{N_{\pi}^{2}}{4 \pi }\, \frac{(\bfk^2+m^2)[\bfk^2+(m+x(1-x)M_{\pi})^2]}{M^2_{\pi} \kappa_{\pi}^2 \, x^5 \,(1-x)^3 }\, \exp\bigg[-\frac{\bfk^2 + m^2)}{\kappa^2_{\pi} x (1-x)}\bigg].
\end{eqnarray}
Similarly, the T-even TMDs for kaon are found to be
\begin{eqnarray}
   f_{1}^{q(K)}(x,\bfk^{2})&=&  \frac{N_{K}^{2}}{4 \pi }\, \frac{[\bfk^2+(m_{q}+x(1-x)M_{K})^2]}{\kappa_{K}^2 \, x^3 \,(1-x)^3 }\, \exp{ \left[ -\frac{(\bfk^2 +(1-x) m^2_{q}+ x m^2_{\bar q}) }{\kappa^2_{K} x(1-x)} \right]},
   \label{flfhmk}\\
   e^{q(K)}(x,\bfk^{2})&=& \frac{N_{K}^{2}}{4 \pi }\, \frac{m_q[\bfk^2+(m_{q}+x(1-x)M_{K})^2]}{ M_{K}\kappa_{K}^2 \, x^4 \,(1-x)^3 }\, \exp{ \left[ -\frac{(\bfk^2 +(1-x) m^2_{q}+ x m^2_{\bar q}) }{\kappa^2_{K} x(1-x)} \right]},
   \label{elfhmk}\\
    f^{\perp q(K)}(x,\bfk^{2})&=&  \frac{N_{K}^{2}}{4 \pi }\, \frac{[\bfk^2+(m_{q}+x(1-x)M_{K})^2]}{\kappa_{K}^2 \, x^4 \,(1-x)^3 }\, \exp{ \left[ -\frac{(\bfk^2 +(1-x) m^2_{q}+ x m^2_{\bar q}) }{\kappa^2_{K} x(1-x)} \right]},
    \label{fpelfhm}\\
     f_{3}^{q(K)}(x,\bfk^{2})&=&  \frac{N_{K}^{2}}{4 \pi }\, \frac{(\bfk^2+m^2_{q})[\bfk^2+(m_q+x(1-x)M_{K})^2]}{M^2_{K}\kappa_{K}^2 \, x^5 \,(1-x)^3 }\, \exp\bigg[-\frac{\bfk^2 + m^2)}{\kappa^2_{\pi} x (1-x)}\bigg].
     \label{f3lfhm1}
   \end{eqnarray}
   
\section{Numerical Results}\label{puhan}
For numerical predictions in both LFHM and LCQM, the adopted input parameters are presented in Table \ref{input parameters}. These input parameters are fixed with experimental data by Regge trajectory. These parameters have been widely used for describing various aspects of meson structure like distribution amplitudes (DAs), form factors (FFs) and PDFs. 
\begin{table}
\centering
\begin{tabular}{|c|c|c|c|c|c|c|c|c|}
\hline 
 & \multicolumn{4}{c|}{LCQM} &\multicolumn{4}{c|}{LFHM}\\
\cline{2-9}
& $m$ (GeV) & $m_q$ (GeV) & $m_{\bar q}$ (GeV) & $\beta$ (GeV) & $m$ (GeV) &  $m_q$ (GeV) & $m_{\bar q}$ (GeV) & $\kappa$ (GeV) \\
\hline
$\pi^+ (u \bar d)$ & 0.2 & - & - & 0.410 & 0.33 & - & - & 0.523\\
\hline
$K^+ (u \bar s)$ & - & 0.2 & 0.556 & 0.405 & - & 0.33 & 0.5&0.524 \\
\hline
\end{tabular}
\caption{In case of LCQM, the pion and kaon parameters are taken from Ref. \cite{Kaur:2020vkq}. While for LFHM, the kaon and pion input parameters are taken from Ref. \cite{Vega:2009zb} and \cite{Kaur:2019jfa}.}
\label{input parameters}
\end{table}
To perform the comparative analysis of the momentum space wave function for $\pi$ and $K$ in LCQM (LFHM), we have plotted it in Figs. \ref{Tom} (\ref{jerry}) with respect to longitudinal momentum fraction $x$, while transverse momenta $\bfk^2$ is fixed at $0.2$GeV$^2$. The 3D quark distributions for pion have been presented in Fig. \ref{4figs3} with respect to longitudinal momentum fraction $x$ and quark transverse momenta $\bfk^2$ in LCQM and LFHM respectively. $f^{q}_1(x,\bfk^2)$ TMD in both the models show a symmetry about $x=0.5$ but in the case of LFHM, it shows two different peaks around $x=0.2$ and $0.8$. We can clearly see that $e^q(x,\bfk^2)$, $f^{\perp q}(x,\bfk^2)$ and $f_3^{q}(x,\bfk^2)$ TMDs in Figs. \ref{4figs} and \ref{4figs3} show similar behavior in both the models which are in agreement with the results found in Ref. \cite{Lorce:2016ugb}. The $f^{\perp q}(x,\bfk^2)$ and $f_3^{q}(x,\bfk^2)$ TMDs have nearly equal peak amplitudes and show a similar behavior in both the models. The smaller quark masses of pion and kaon enhance the amplitudes of $f^{\perp q}(x,\bfk^2)$ and $f_3^{q}(x,\bfk^2)$ TMDs than other TMDs. Similar results can be seen for the case of kaon in Figs. \ref{4figs1} and \ref{4figs4}. Except for $f^{q}_1(x,\bfk^2)$ TMD for pion, other TMDs are not symmetric but they have distribution around $x \le 0.5$. For kaon, we have plotted the $\bar s$-quark distribution in Figs. \ref{4figs5} and \ref{4figs2} in both the models. In case of $\bar s$-quark, the $f^{q}_1(x,\bfk^2)$ and $f_3^{q}(x,\bfk^2)$ TMDs show negative distributions, whereas $f^{\perp q}(x,\bfk^2)$ and $e^q(x,\bfk^2)$ TMDs show positive distributions. The $f_3^{q}(x,\bfk^2)$ TMD for $\bar s$-quark in LCQM shows a higher negative peak as compared to the other TMDs. It is observed that most of the TMDs have a distribution towards the left of $x=0.5$. Another point to notice is that all the TMDs vanish at $\bfk^2\ge 0.4$ Ge$V^2$. In Figs. \ref{4figs6} and \ref{4figs8}, we have plotted the $\textit{u}$-quark distribution comparison in both models for kaon and pion at fixed $\bfk^2=0.1$ and $0.4$ GeV$^2$ respectively. It is observed that the quark distribution at $\bfk^2=0.4$ Ge$V^2$ is very low compared to $\bfk^2=0.1$ Ge$V^2$. While in Figs. \ref{4figs7} and \ref{4figs9}, we have compared the $\textit{u}$-quark distribution in both the models at fixed longitudinal momentum fraction $x=0.1$ and $0.5$ for pion and kaon respectively. The $\bar s$-quark distribution comparison has been shown in Figs. \ref{4figs10} and \ref{4figs11} for kaon with respect to $\bfk^2(GeV^2)$ and $x$ in both the models respectively. Both the models show almost similar behavior for both the particles. This kind of behavior has also been observed in the light-front constituent model (LFCM) \cite{Lorce:2016ugb} for the case of pion. However, looking into basis light-front quantization (BLFQ) results \cite{Zhu:2023lst}, there is a slight deviation with their distributions for the case of pion. As there has not been any higher twist theoretical quark distributions work reported for the case of kaon, we cannot compare our results with other models as well as with experimental data.
\par The PDFs for pion have been plotted with respect to $x$ in Fig. \ref{4figs12} and compared in both the models. Both the model results come out similar to each other and have exactly similar behavior with LFCM \cite{Lorce:2016ugb}. In Fig. \ref{4figs13}, we have plotted the $x$PDFs with longitudinal momentum fraction $x$. The $xf^q_1(x)$ PDF follow the same trend as BLFQ \cite{Zhu:2023lst}, whereas $xe^q(x)$ and $xf^{\perp q}(x)$ PDFs are slightly different from BLFQ which may be due to the gluon contributions introduced in BLFQ which has not been included in our work. For the case of kaon, the PDFs and $\textbf{x}$PDFs have been plotted with respect to $x$ in Figs. \ref{4figs14} and \ref{4figs15} respectively.
\par The higher order PDFs have not been studied through experiments for pseudo-scalar mesons. While the pion-induced DY and $J/\psi$ production data have provided the complete information of the $f^q_1(x)$ PDFs for pion and kaon \cite{Bourrely:2023yzi} at $\sqrt{s}\ge 4$ GeV, the $e^q(x)$ PDF has been studied by CLAS experiment for nucleons at $\sqrt{s}=6$ GeV \cite{Courtoy:2014ixa}. We have compared the $xf^q_1(x)$ PDF of pion and kaon results of both models with experimental results in Fig.\ref{4figs16}. Our model scale parameters are $0.19$ GeV$^2$ and $0.20$ GeV$^2$ for LCQM and LFHM respectively, but the experimental results are at $16$ GeV$^2$. The evolution for $xf_1^q(x)$ PDF for pion has been done in the Ref. \cite{Kaur:2020vkq} in LCQM and the result matches with the FNAL-E615 experimental data \cite{Conway:1989fs} and modified FNAL-E615 data \cite{Aicher:2010cb} . Since there is no experimental data for higher twist quark distributions of mesons, we have compared the pion and kaon $e^q(x)$ PDF with nucleon CLAS experimental result \cite{Courtoy:2014ixa} in Fig. \ref{4figs16}.
\par In the LF dynamics, it is also possible to calculate the inverse momenta \cite{Lorce:2016ugb} which can be expressed as 
\begin{eqnarray}
    \langle x^{-1}\rangle_q = \int_0^1 dx\;\frac{f_1^{ \pi (K)}(x)}{x}. \label{inverse}
\end{eqnarray}
The inverse moments for both particles in both models has been given in Table. \ref{inversemomenta}. In case of LFCM, the inverse moment is found to be $2.82 N_q $ for pion \cite{Lorce:2016ugb}. These inverse moments play an important role in describing the sum rules.
\begin{table}
\centering
\begin{tabular}{|c|c|c|}
\hline 
$ \langle x^{-1}\rangle $ & LCQM & LFHM\\
\hline
$\pi^+ (u \bar d)$ & 2.79 & 2.62 \\
\hline
$K^+ (u \bar s)$ & 3.92 & 3.20 \\
\hline
\end{tabular}
\caption{The inverse moments of pion and kaon in both the models.}
\label{inversemomenta}
\end{table}
It was seen that, the inverse moments are higher for the case of kaon than the pion in both the models.
\subsubsection{Transverse dependence of TMDs}
In this subsection, we will discuss the transverse properties like average momenta, Gaussian transverse dependence and spin densities of T-even TMDs. The $x$-dependent mean transverse momenta $(n=1)$ and mean squared transverse momenta ($n=2$) can be  given as \cite{Lorce:2016ugb,Zhu:2023lst}
\begin{eqnarray}
    \langle \bfk^n\rangle = \frac{\int dx \int d^2\bfk \bfk^n TMDs} {\int dx\int d^2\bfk TMDs}.
\end{eqnarray}
One can also find the Gaussian transverse dependence ratio of these TMDs to find the extent upto which the model supports Gaussian  $\bfk$ behavior \cite{Boffi:2009sh,Lorce:2016ugb}. The ratio is expressed as 
\begin{eqnarray}
    	R_G = \frac{2}{\sqrt{\pi}} \frac{\langle \bfk \rangle}{\langle \bfk^2\rangle^{1/2}}.
\end{eqnarray}
The $R_G$ has also been observed phenomenologically in many DIS experiments \cite{DAlesio:2007bjf,Schweitzer:2010tt}. In Table \ref{trans}, we compare our predictions for $\langle \bfk \rangle$, $\langle \bfk^2 \rangle^{\frac{1}{2}}$ and $R_G$ of our models with available LFCM \cite{Lorce:2016ugb}, BLFQ \cite{Zhu:2023lst} and statistical model (SM) \cite{Bourrely:2023yzi} data for both the particles. While comparing our computations with the available data, we have found that the $\langle \bfk \rangle$ values for pion and kaon are comparable. In the case of LFHM, the $R_G$ values are nearly equal to $1$, but for LCQM, it is slightly less than $1$. While looking into $\bar s$-quark, the $\langle \bfk \rangle$ and $\langle \bfk^2 \rangle^{\frac{1}{2}}$ value are compared with SM data for the case of kaon as given in Table \ref{inversemomenta11}.
\begin{table}
\centering
\begin{tabular}{|c|c|c|c|c|c|c|c|c|c|c|c|c|c|c|c|c|c|c|c|}
\hline 
\begin{turn}{-90}$u$-quark \end{turn} & \multicolumn{6}{c|}{LCQM (This work)} &\multicolumn{6}{c|}{LFHM (This work)}  &\multicolumn{3}{c|}{LFCM \cite{Lorce:2016ugb}}  &\multicolumn{2}{c|}{BLFQ \cite{Zhu:2023lst}} &\multicolumn{2}{c|}{SM data \cite{Bourrely:2023yzi}} \\
\cline{2-20}
 & \multicolumn{3}{c|}{$\pi^+(u \bar d)$}  & \multicolumn{3}{c|}{$K^+(u \bar s)$} & \multicolumn{3}{c|}{$\pi^+(u \bar d)$}  & \multicolumn{3}{c|}{$K^+(u \bar s)$}& \multicolumn{3}{c|}{$\pi^+(u \bar d)$}& \multicolumn{2}{c|}{$\pi^+(u \bar d)$} & $\pi^+(u \bar d)$ & $K^+(u \bar s)$ \\
 \cline{2-20}
& $\langle \bfk \rangle$ &  ${\langle \bfk \rangle}^{\frac{1}{2}}$ & $R_G$& $\langle \bfk \rangle$ &  ${\langle \bfk \rangle}^{\frac{1}{2}}$ & $R_G$& $\langle \bfk \rangle$ &  ${\langle \bfk \rangle}^{\frac{1}{2}}$ & $R_G$ & $\langle \bfk \rangle$ &  ${\langle \bfk \rangle}^{\frac{1}{2}}$ & $R_G$  & $\langle \bfk \rangle$ &  ${\langle \bfk \rangle}^{\frac{1}{2}}$ & $R_G$  & $\langle \bfk \rangle$ &  ${\langle \bfk \rangle}^{\frac{1}{2}}$ & $\langle \bfk \rangle$ & $\langle \bfk \rangle$ \\
\hline
$f_1$ & 0.22 & 0.26 & 0.96 & 0.26 & 0.30 & 0.98 & 0.24 & 0.27 & 1.00 & 0.24 & 0.27 & 1.00 & 0.28 & 0.32 & 0.99 & 0.26 &0.30 &0.24 & 0.22\\
\hline
$e$ & 0.18 & 0.22 & 0.95 & 0.21 & 0.25 & 0.96 & 0.21 &0.24 & 0.99 & 0.21 & 0.24 & 0.99 & 0.26 & 0.30 & 0.99 &0.26 & 0.30 &-&-\\
\hline
$f_\perp$ & 0.21 & 0.25 & 0.96 & 0.23 & 0.27 & 0.96 & 0.23 & 0.26 & 0.99 & 0.22 & 0.25 & 0.99 & 0.26 & 0.30 & 0.99 &0.25 &0.29&-&- \\
\hline
$f_3$ & 0.21 & 0.24 & 0.95 & 0.23 & 0.27 & 0.96 & 0.22 & 0.25 & 0.99 & 0.22 & 0.25 & 0.99 & 0.30 & 0.33 & 0.98& -&-&-&- \\
\hline
\end{tabular}
\caption{The average transverse momenta of \textit{u}-quark of both LCQM and LFHM for both the particles with Gaussian transverse momenta ratio $R_G$. Our results have been compared with LFCM, BLFQ and SM data. The SM data is at $\sqrt{s}=10$ GeV while our data at their respective model scale. }
\label{trans}
\end{table}

\begin{table}
\centering
\begin{tabular}{|p{1.2cm}|p{1.8cm}|p{1.8cm}|p{1.8cm}| p{1.8cm}|p{1.8cm}|p{1.8cm}|}
\hline 
$\bar s$-quark & \multicolumn{2}{c|}{LCQM (This work)} & \multicolumn{2}{c|}{LFHM (This work)} & \multicolumn{2}{c|}{SM Data at $10$ GeV$^2$ \cite{Bourrely:2023yzi}}\\
\cline{2-7}
&$\langle \bfk \rangle$ & $\langle \bfk \rangle^{\frac{1}{2}}$ &$\langle \bfk \rangle$ & $\langle \bfk \rangle^{\frac{1}{2}}$ &$\langle \bfk \rangle$ & $\langle \bfk \rangle^{\frac{1}{2}}$\\
\hline
~~~$f_1$ & 0.276 & 0.321 & 0.227 & 0.258 &0.276 &-\\
\hline
~~~$e$ & 0.217 & 0.257 & 0.202 & 0.231 &- &-\\
\hline
~~~$f^\perp$ & 0.238 & 0.282 & 0.211 & 0.241 & - &-\\
\hline
~~~$f_3$ & 0.179 & 0.214 & 0.192 & 0.221 & -&- \\
\hline
\end{tabular}
\caption{The average transverse momenta of $\bar s$-quark of both the LCQM and LFHM compared with only available SM data for kaon. The SM data is at $\sqrt{s}=10$GeV while our data  at model scale $\le 1$ GeV$^2$.}
\label{inversemomenta11}
\end{table}

\par
\begin{figure}
\centering
(a){\label{4figs-a68} \includegraphics[width=0.45\textwidth]{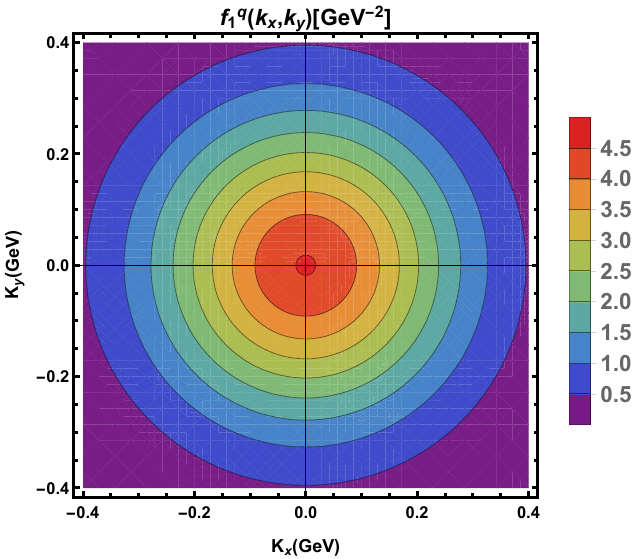}}
\hfill
(b){\label{4figs-b69} \includegraphics[width=0.45\textwidth]{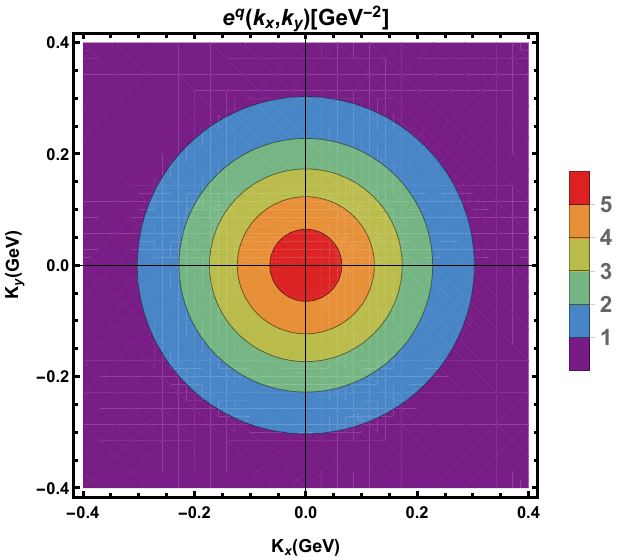}}%
\hfill \\
(c){\label{4figs-c70} \includegraphics[width=0.45\textwidth]{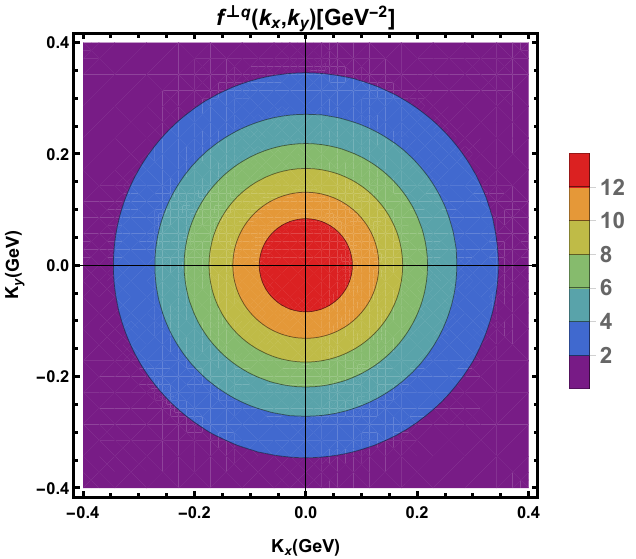}}%
\hfill
(d){\label{4figs-d71} \includegraphics[width=0.45\textwidth]{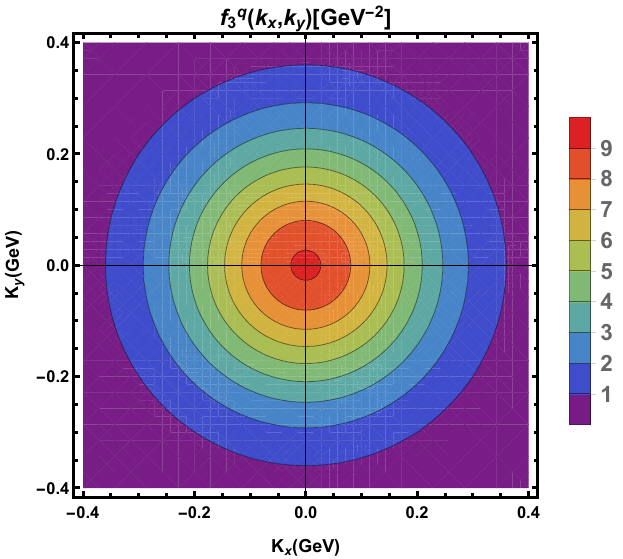}}%
\caption{Transverse structure of \textit{u}-quark pion TMDs in the LFHM at model scale 0.20 GeV$^2$.}
\label{4figs17}
\end{figure}

\begin{figure}
\centering
(a){\label{4figs-a72} \includegraphics[width=0.45\textwidth]{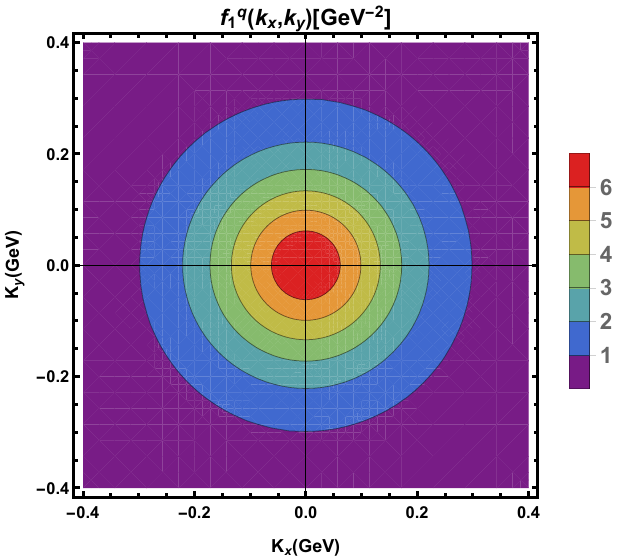}}
\hfill
(b){\label{4figs-b73} \includegraphics[width=0.45\textwidth]{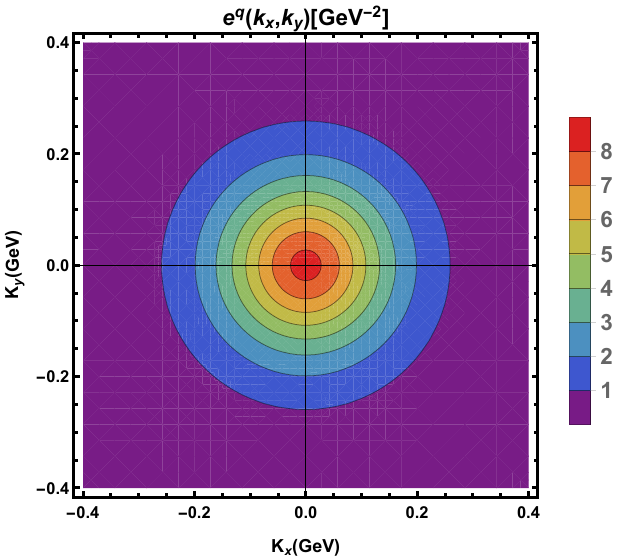}}%
\hfill \\
(c){\label{4figs-c74} \includegraphics[width=0.45\textwidth]{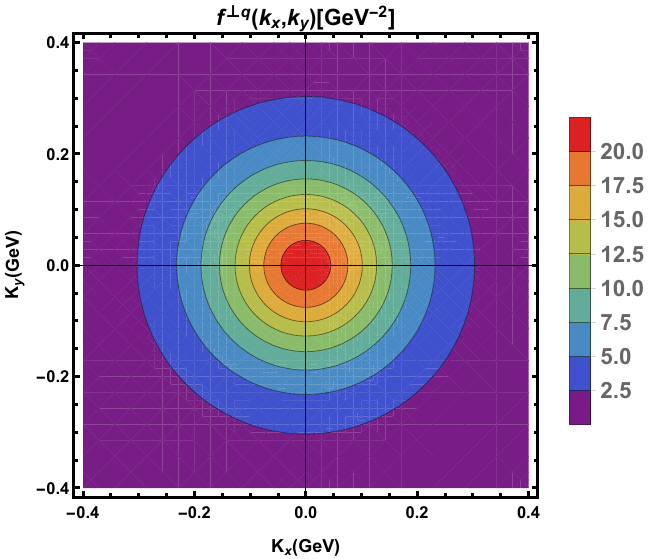}}%
\hfill
(d){\label{4figs-d75} \includegraphics[width=0.45\textwidth]{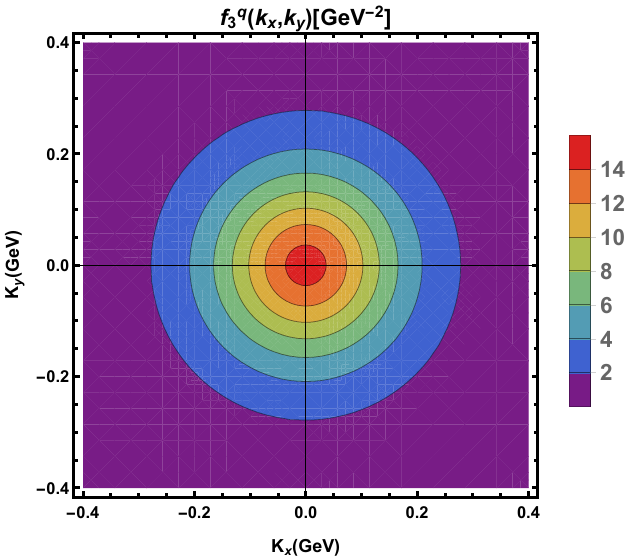}}%
\caption{Transverse structure of \textit{u}-quark pion TMDs in the LCQM at model scale 0.19 GeV$^2$.}
\label{4figs18}
\end{figure}

\begin{figure}
\centering
(a){\label{4figs-a76} \includegraphics[width=0.45\textwidth]{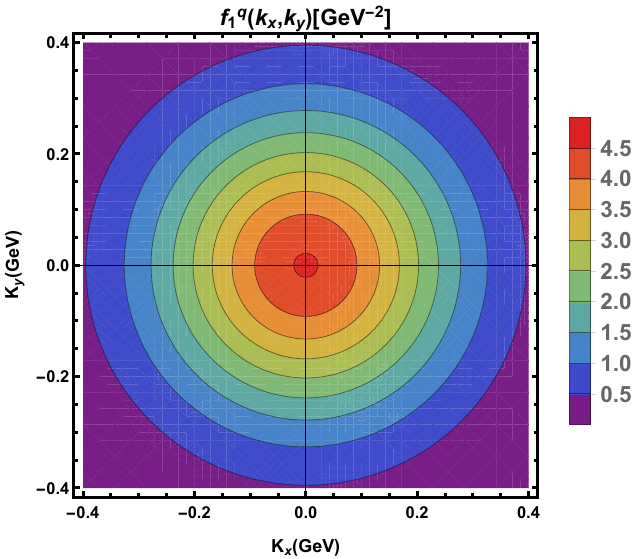}}
\hfill
(b){\label{4figs-b77} \includegraphics[width=0.45\textwidth]{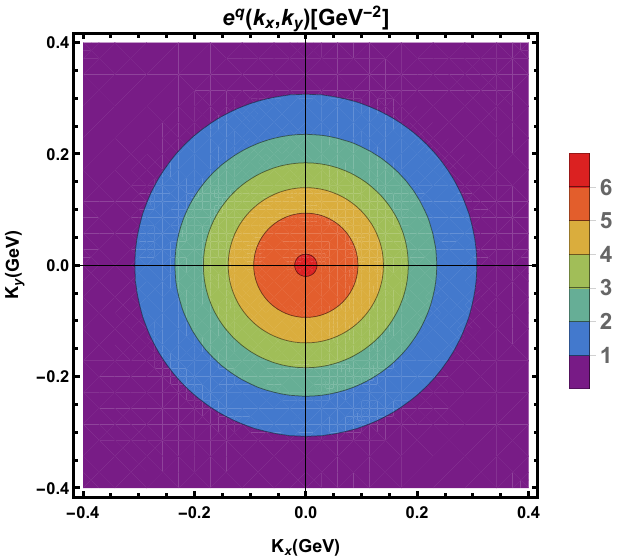}}%
\hfill \\
(c){\label{4figs-c78} \includegraphics[width=0.45\textwidth]{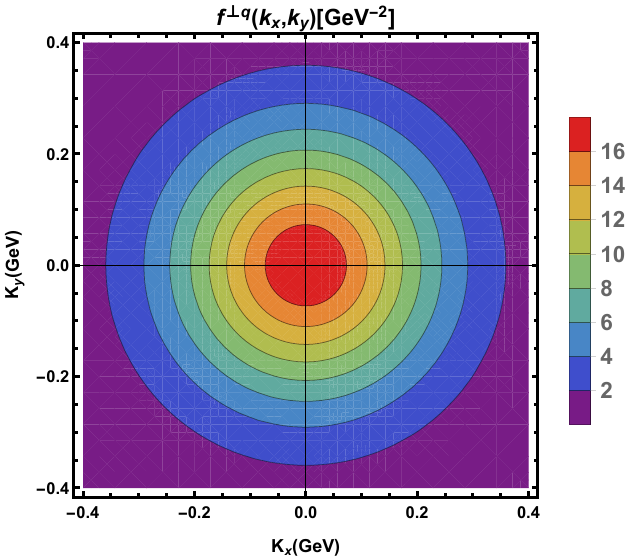}}%
\hfill
(d){\label{4figs-d79} \includegraphics[width=0.45\textwidth]{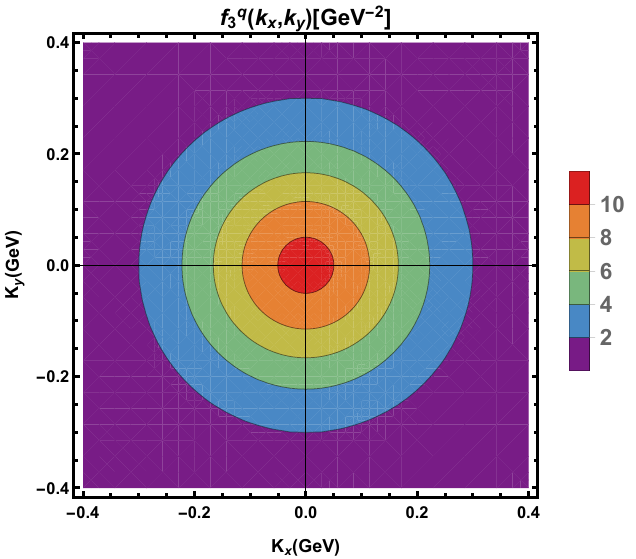}}%
\caption{Transverse structure of \textit{u}-quark kaon TMDs in the LFHM at model scale 0.20 GeV$^2$.}
\label{4figs19}
\end{figure}

\begin{figure}
\centering
(a){\label{4figs-a80} \includegraphics[width=0.45\textwidth]{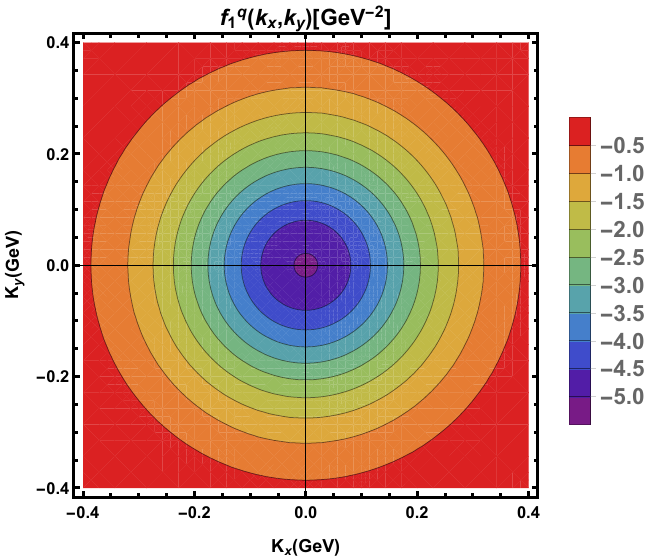}}
\hfill
(b){\label{4figs-b81} \includegraphics[width=0.45\textwidth]{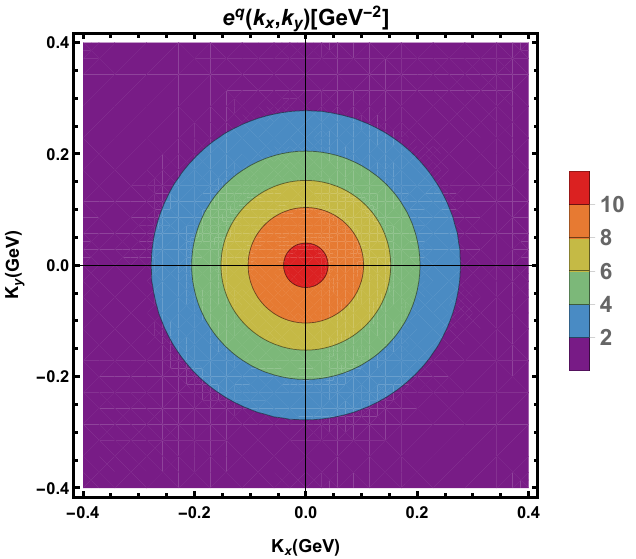}}%
\hfill \\
(c){\label{4figs-c82} \includegraphics[width=0.45\textwidth]{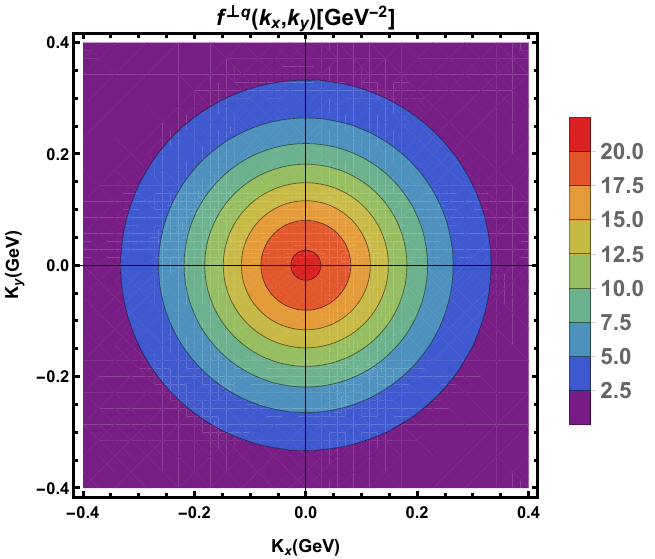}}%
\hfill
(d){\label{4figs-d83} \includegraphics[width=0.45\textwidth]{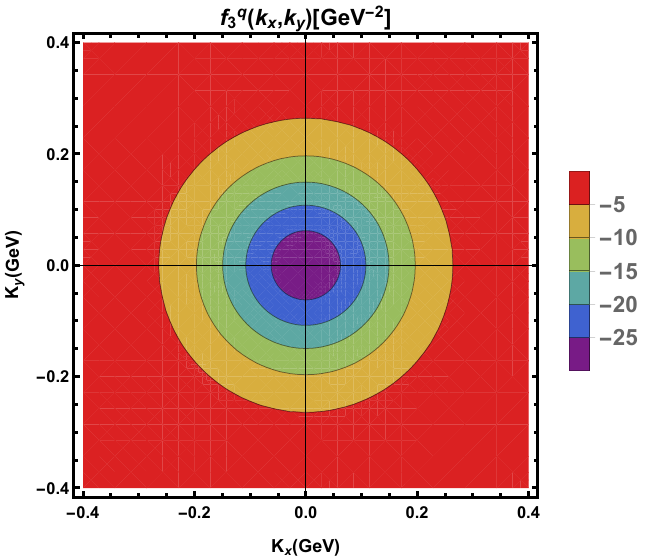}}%
\caption{Transverse structure of $\bar s$-quark kaon TMDs in the LFHM at model scale 0.20 GeV$^2$.}
\label{4figs20}
\end{figure}

\begin{figure}
\centering
(a){\label{4figs-a84} \includegraphics[width=0.45\textwidth]{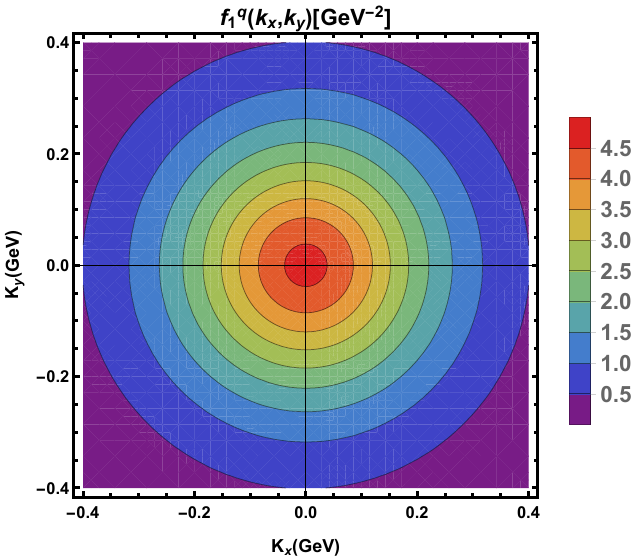}}
\hfill
(b){\label{4figs-b85} \includegraphics[width=0.45\textwidth]{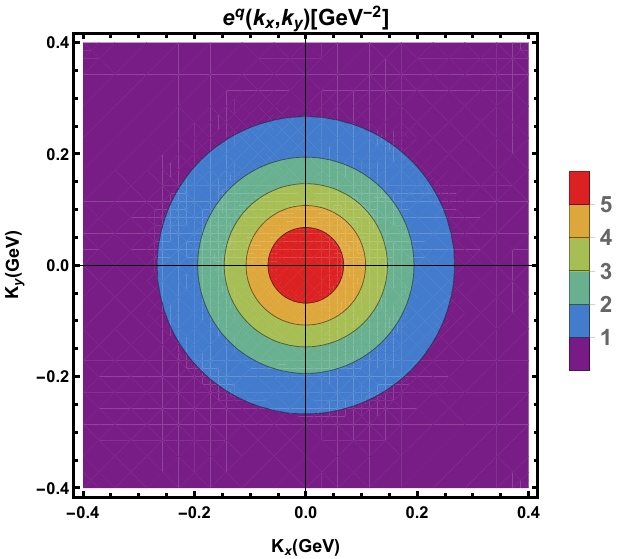}}%
\hfill \\
(c){\label{4figs-c86} \includegraphics[width=0.45\textwidth]{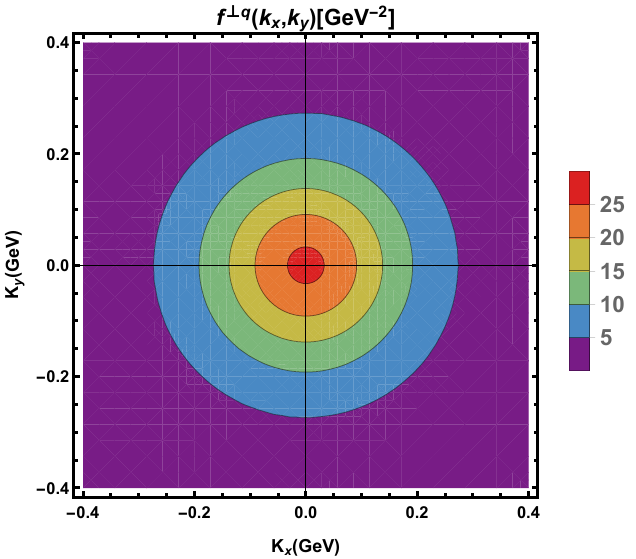}}%
\hfill
(d){\label{4figs-d87} \includegraphics[width=0.45\textwidth]{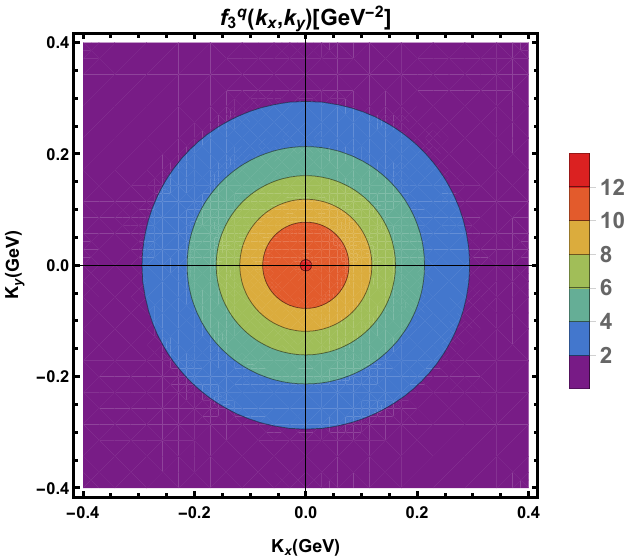}}%
\caption{Transverse structure of \textit{u}-quark kaon TMDs in the LCQM at model scale 0.19 GeV$^2$.}
\label{4figs21}
\end{figure}

\begin{figure}
\centering
(a){\label{4figs-a88} \includegraphics[width=0.45\textwidth]{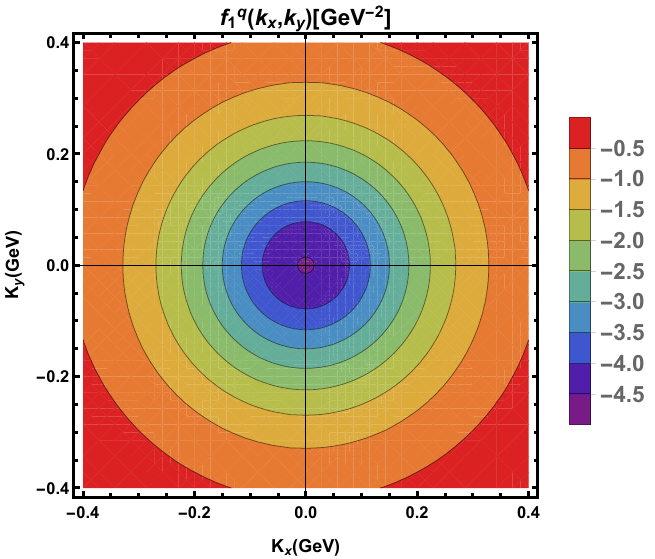}}
\hfill
(b){\label{4figs-b89} \includegraphics[width=0.45\textwidth]{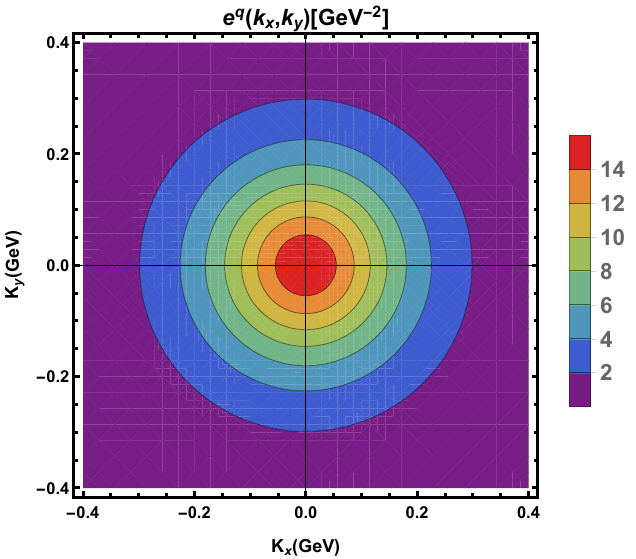}}%
\hfill \\
(c){\label{4figs-c90} \includegraphics[width=0.45\textwidth]{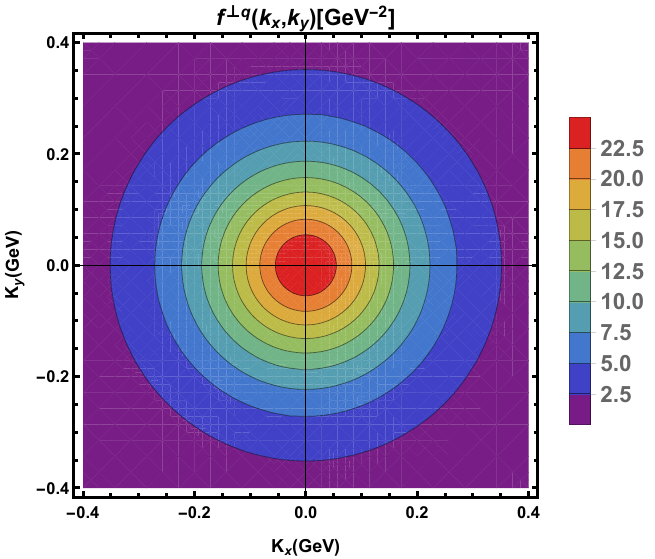}}%
\hfill
(d){\label{4figs-d891} \includegraphics[width=0.45\textwidth]{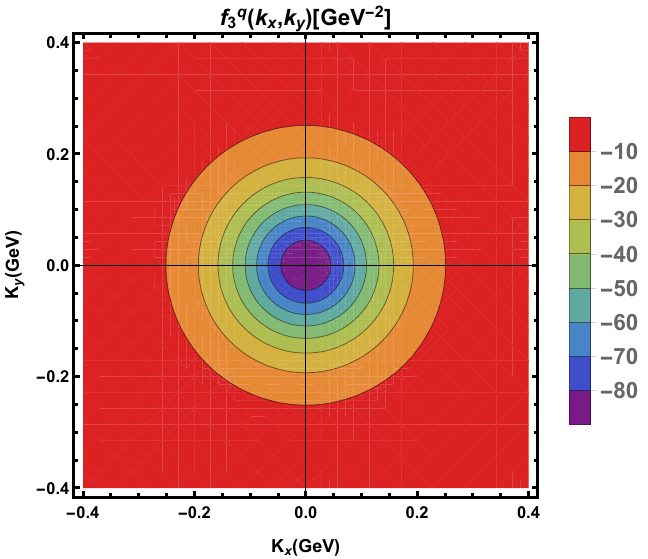}}%
\caption{Transverse structure of $\bar s$-quark kaon TMDs in the LCQM at model scale 0.19 GeV$^2$.}
\label{4figs22}
\end{figure}

\begin{figure}[!htb]
\centering
\includegraphics[width=.5\linewidth]{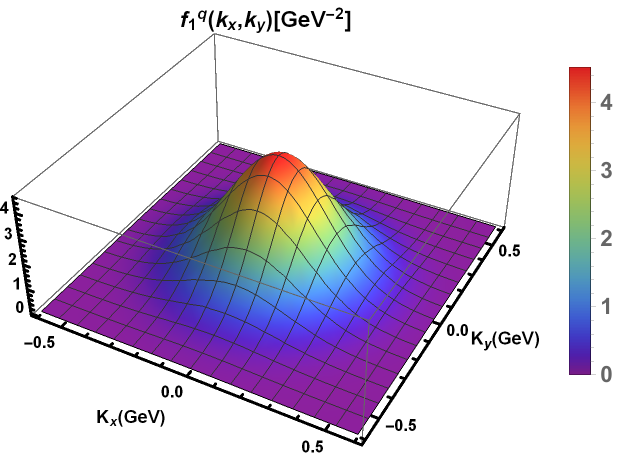}
    \caption{The transverse structure of $f_1^u (x, \bfk^2)$ TMD of pion in momentum space in the LFHM at model scale $0.2$ GeV$^2$ }\label{3d transverse}
\end{figure}

 The quark densities with transverse momenta $\bfk (\textbf{k}_x,\textbf{k}_y)$ and transverse spin $\textbf{s}_\perp$ can be obtained by integrating all the T-even TMDs over longitudinal momentum fraction $x$ and expressed as
\begin{eqnarray}
    \textit{H}^q (\textbf{k}_x,\textbf{k}_y,\textbf{s}_\perp)=\int_{0}^{1} TMD \,dx .
\end{eqnarray}
Here, $\textit{H}$ is used for $f_1$, $e$, $f^\perp$ and $f_3$. As we are dealing with the unpolarized TMDs, the transverse spin $\textbf{s}_\perp$ will be $0$. The quark densities have been plotted for all the TMDs with respect to $\textbf{k}_x$ and $\textbf{k}_y$ in Figs. \ref{4figs17} and \ref{4figs18} for \textit{u}-quark of pion and kaon in LFHM and LCQM respectively. In Figs. \ref{4figs19} and \ref{4figs20}, we have plotted the \textit{u} and $\bar s$-quark densities for kaon in LFHM. Similarly, in Figs. \ref{4figs21} and \ref{4figs22}, the kaon quark densities have been plotted in LCQM. For a more detailed study on quark densities, we have plotted $3$D $f_1^\textit{u}(\textbf{k}_x,\textbf{k}_y)$ for pion with respect to momentum space $(\textbf{k}_x,\textbf{k}_y)$ in Fig. \ref{3d transverse}. It is found that all these unpolarized T-even TMDs have a sharp peak at the center of the particle ($\bfk=0$). All these quark density distributions are symmetric and localized near the center about $\bfk=0$. The width of the quark transverse momenta decreases as $x$ increases. On LF dynamics, the higher the value of the longitudinal momentum fraction, the lesser the kinetic energy carried by the quark. The transverse momentum becomes broader to carry a large amount of kinetic energy, as the total energy is limited in the distribution.  One can get the spin densities by Fourier transformation of $\textit{H}^q (\textbf{k}_x,\textbf{k}_y)$. The spin densities can relate TMDs with GPDs of pseudo-scalar mesons.

\section{Conclusion}\label{conli}
In this work, we have presented twist-$2$, twist-$3$ and twist-$4$ T-even valence quark TMDs for pion and kaon in the light-cone quark model and the light-front holographic model. Both the models have been introduced with different dynamic spin effects. By solving the quark-quark correlation function for spin-$0$ mesons, we get the TMDs in the overlap form of light-front wave functions for both the models. The relations between the $e^q (x, \bfk^2)$, $f^{\perp q}(x, \bfk^2)$ and $f_3^{q}(x, \bfk^2)$ TMDs with $f_1^{q}(x, \bfk^2)$ TMD, have been presented. We have then discussed the positivity inequalities among these TMDs. PDFs corresponding to these TMDs have also been calculated along with their sum rules for both models.
\par We have studied the behavior of pion and kaon momentum wave functions in both the models with respect to $x$ at transverse momenta $\bfk^2=0.2$ GeV$^2$. The properties of all T-even TMDs have been studied using $3$D and $2$D plots for \textit{u} and $\bar s$-quarks with respect to $x$ and $\bfk$ in both the models for both the particles. It is found that only $f_1^q (x,\bfk^2)$ TMD shows symmetry about $x \longrightarrow (1-x)$ while other TMDs do not show any symmetry. It is observed that all the TMDs amplitude shifts toward $x \le 0.5$ when we move from the leading twist to the higher twist. All the T-even TMDs vanish at $\bfk^2 \ge 0.4$ GeV$^2$. In the case of kaon, $f_1^s (x, \bfk^2)$ and $f_3^s (x, \bfk^2)$ TMDs show negative distributions, while the other two TMDs show positive distributions. It is also observed that higher twist TMDs have higher peak values than the leading twist $f_1^q (x, \bfk^2)$ TMD. The behavior of these quark TMDs is similar in both the models and follows the same trend when compared with the light-front constituent model. Both the models satisfy the sum rules for PDFs and positivity inequalities for TMDs. We have studied the PDFs corresponding to each of the TMDs and plotted them with respect to $x$. The $f_1^q (x)$ PDF shows symmetry and peak value around $x=0.5$. While other three PDFs $e^q (x)$, $f^{\perp q} (x)$ and $f_3^q (x)$ have peak value around $x=0.2$ with no symmetry. All the PDFs have the same trend in our models along with LFCM. But the PDF results are slightly different from the BLFQ results except the $x f_1^{q}(x, \bfk^2)$ PDF.  We have also compared our PDF result with pion-induced DY, $J/\psi$ production data and nucleon $e^q$ PDF CLAS data.
\par We have also calculated the inverse transverse moments of $f_1 (x)$ PDF for both the particles in our models. The inverse moments calculation may lead to the sum rule of PDFs in the near future. We have also calculated the $x$-dependent mean transverse momenta $\langle \bfk \rangle$, mean squared transverse momenta $\langle \bfk \rangle^{\frac{1}{2}}$ and Gaussian transverse ratio $R_G$ in our work. We have compared our data with available LFQM, BLFQ and SM data for \textit{u}-quark. In the case of $\bar s$-quark of kaon, we have compared our data with only available SM data. The $\langle \bfk \rangle$ and $\langle \bfk \rangle^{\frac{1}{2}}$ values of our models are coming very close to other models data. The kaon TMDs and PDFs work has not been done in any theoretical models yet for the higher twist. The $\langle \bfk \rangle$ data for \textit{u}-quark leading twist TMDs for kaon is coming very close to SM data. The $\bar s$-quark $\langle \bfk \rangle$ value equal to SM data. We have also discussed the quark densities in momentum space $(\textbf{k}_{\perp x},\textbf{k}_{\perp y})$ for both the particles in both models. As all these TMDs are unpolarized T-even, all the plots follow symmetry and have peaks about $\bfk=0$.
\par The COMPASS program will give important insights into the leading and sub-leading structure of pion and kaon in future experiments. Electron-ion collider (EIC) and JLab experiments may provide some wealth of information for TMDs and PDFs shortly.
\par We plan to further add gluon contributions to the higher order TMDs along with sea quark contributions in our future work. We are currently focusing on the T-odd TMDs for pseudo-scalar mesons to have a complete structure of TMDs up to twist-4.  
%
%
%
%
%
%
%
%
%
%
%
%
%
%
%
%
%
%
%
%
%
%
%
%
%
%
%
%
%

\end{document}